\documentclass[twocolumn,natbib]{svjour3}         
\smartqed  
\usepackage{graphicx}
\begin{document}

\title{Global Properties of Solar Flares
}


\author{Hugh S. Hudson        
}


\institute{H.~S. Hudson \at
              SSL/UC Berkeley \\
              Tel.: +1 510-643-0333\\
              \email{hhudson@ssl.berkeley.edu}           
}

\date{Received: date / Accepted: date}

\maketitle

\setcounter{tocdepth}{3}
\tableofcontents

\bigskip
\begin{abstract}
This article broadly reviews our knowledge of solar flares.
There is a particular focus on their global properties, as opposed to the microphysics such as that needed for magnetic reconnection or particle acceleration as such.
Indeed solar flares will always remain in the domain of remote sensing, so we cannot observe the microscales directly and must understand the basic physics entirely via the global properties plus theoretical inference.
The global observables include the general energetics -- radiation in flares and mass loss in coronal mass ejections (CMEs) -- and the formation of different kinds of ejection and global wave disturbance: the type~II radio-burst exciter, the Moreton wave, the EIT ``wave,'' and the ``sunquake'' acoustic waves in the solar interior.
Flare radiation and CME kinetic energy can have comparable magnitudes, of order $10^{32}$~erg each for an X-class event, with the bulk of the radiant energy in the visible-UV continuum.
We argue that the impulsive phase of the flare dominates the energetics of all of these manifestations, and also point out that energy and momentum in this phase largely reside in the electromagnetic field, not in the observable plasma.

\end{abstract}

\section{Introduction}\label{sec:intro}
\cite{1859MNRAs..20...13C} first reported the occurrence of a solar flare, a manifestation seen as he observed sunspots in  ``white light'' through a small telescope.
One could immediately conclude from this chance observation that the disturbance of the solar atmosphere was compact, brief, and extremely energetic.
Carrington's fellow amateur observer \cite{1859MNRAS..20...15H} confirmed the observation and  likened the brilliance of the display to that of the bright A0 star $\alpha$~Lyrae.
Further evidence of the significance of the event lay in its effects seen in terrestrial compasses, both prompt and delayed \citep[e.g.,][]{chapman-bartels} 

The idea that a solar disturbance could affect a terrestrial instrument such as a compass
seemed highly improbable at the time, but it turned out to be indeed a correct association, so this first observed flare served to suggest immediately the capability for a such a solar event to have widely felt influences.
This article briefly reviews general flare physics and discusses its problems from the point of view of large-scale effects such as the generation of global waves, and of ejecta.
The idea is to ask what we can learn about the fundamental processes in the flare by observations of its large-scale effects.
We must bear in mind the other aspect of the flare of September~1, 1859\footnote{Henceforth we adopt the IAU naming convention, which for this flare would be SOL1859-09-01T11:18; see DOI 10.1007/s11207-010-9553-0.}, namely that its effects seemed to originate in compact and short-lived features in the deep atmosphere.
Coronal mass ejections (CMEs), on the other hand, can become huge and reach the scale of the solar system.
Thus the physics involved in a solar flare require consideration of multiple scales.

This article discusses the global properties of a flare, and by this we mean the phenomena extending well beyond the chromosphere and the coronal loops that define most of the radiative effects (the flare proper, in a strict sense).
To introduce the discussion of these global properties, and to have a framework for them, we begin in 
Section~\ref{sec:background} with an overview of key observational and theoretical facts and ideas about solar flares.
Knowledgeable readers should be able to skip this material, which goes over general occurrence properties and then summarizes what appear to be the four main phases of a flare, mainly as identified via X-ray signatures.
Then Sections~\ref{sec:glob} (Global effects) and Section~\ref{sec:ergs} (Energetics) discuss various global properties, including CMEs.
The newest global facts, from the past decade or so, come from many spacecraft (RHESSI, SOHO, TRACE, Hinode, SORCE, STEREO for example) and  from radio and optical observatories on the ground.
Sections~\ref{sec:waves} and~\ref{sec:synth} in particular tackle the global waves, which have advanced observationally to become an important new guide to flare and CME physics.

\section{Background}
\label{sec:background}

\subsection{Flare morphology}
\label{sec:morph}

As noted by Carrington, a powerful solar flare can locally increase the intensity of the photosphere by an detectable factor.
He could visually see his white-light flare relatively easily; in his words "...the brilliancy was fully equal to that of direct sun-light.''
They occurred in small patches near a sunspot group.
His description means that the flare was at least as bright as sunspots are dark, although the flare brightenings were much smaller in area and of course transient in nature, lasting only a few minutes.
His colleague Hodgson described the flare as ``much brighter'' than the photosphere and ``most dazzling to the protected eye.''
In the quiet Sun, the background intensity fluctuations, for reasonable telescopic angular resolution, have RMS magnitudes of a few percent \citep[e.g.,][]{1988ARA&A..26..473H}.
These are the result of the convective motions (granulation) seen in the quiet Sun, and the large image contrasts within active regions.
These image contrasts convert to time-series fluctuations at low angular resolution and in bad seeing conditions.
Because a flare detectable in the visible continuum needs to overcome these observational hurdles, there were only some 56~known white-light flares in the century and a quarter following Carrington \citep{1983STIN...8424521N} until recently; these have generally been the most energetic events (GOES X-class flares).
Nowadays space-based observations with no seeing limitation \citep{1992PASJ...44L..77H,2003A&A...409.1107M,2006SoPh..234...79H,2009RAA.....9..127W}  make it possible to detect much weaker and therefore more numerous events.
\cite{2006SoPh..234...79H} report a GOES~C1.3  event observed by TRACE, and \cite{2008ApJ...688L.119J} found a GOES~C2.0 event even with ground-based observations.

In the chromosphere flare detectability becomes far easier, and spectroscopic observations in (e.g.) H$\alpha$ defined flare physics for many decades.
Continuing the historical progression, the development of radio astronomy and then UV and X-ray astronomy made coronal observations possible even in front of the solar disk.
At these extreme wavelengths the photosphere becomes dark (for the short wavelengths) or elevated in altitude (for radio waves), and flare effects become dominant.

The coronal parts of a solar flare have many loop-like features, which rather clearly represent striations along the magnetic field.
These flare loops (or sometimes confusingly called ``post-flare loops'' or ``post-eruption arcades'')  appear first  in soft X-rays, at temperatures of order 2~$\times$~10$^7$~K.
This morphology is consistent with the chromospheric structure, which in major events usually has the ``two-ribbon'' pattern.
The two ribbons appear in the two polarities of the photospheric line-of-sight magnetic field, and the coronal loops connect them across the polarity inversion line.

\begin{figure}
\centering
\includegraphics[width=0.49\textwidth]{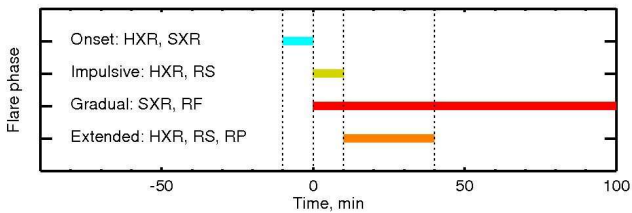}
\caption{Time ranges associated with the different phases of a flare.
The dominant X-ray signatures are designated by HXR (hard X-rays, above $\sim$20~keV) and SXR (soft X-rays, characterized by temperatures below $\sim$30~MK).
The dominant radio emission mechanisms are designated by RS (gyrosynchrotron), RF (free-free), and RP (plasma-frequency mechanisms).
Gamma rays typically accompany HXR when detectable \citep[e.g.,][]{1994ApJ...426..767C,2009ApJ...698L.152S}.
}
\label{fig:phases}       
\end{figure}

Different phases of a flare have characteristic evolutionary patterns.
In this review we discuss three relatively well-defined phases, namely the \textit{impulsive}, \textit{gradual}, and \textit{extended flare} phases, in Section~\ref{sec:xray}.
These phases are distinguishable via their X-ray signatures, but in fact there is ambiguity and overlap.
Recent observations also point to an early non-thermal phase, prior to the impulsive phase, that may also have distinguishing characteristics.
RHESSI observations of SOL2002-07-23T00:35 gave a first clear example of this \citep{2003ApJ...595L..69L}.
Figure~\ref{fig:phases} illustrates the timing of the four identified phases schematically

\subsection{Flare dynamics}

The temporal evolution of a flare also has characteristic dynamical patterns.
The impulsive phase marks the epoch of most intense energy release, which adds mass to the corona by expelling it from the chromosphere. 
The newly ``evaporated'' material flows upward into the corona and becomes visible in soft X-rays because the gas pressure has risen dramatically.
In major two-ribbon flares multiple X-ray loops appear in a roughly cylindrical ``arcade'' formation.
The newly formed high-temperature coronal plasma then gradually cools through the EUV, UV, and optical ranges, where it forms the H$\alpha$ post-flare loops.
These were known historically as ``loop prominence systems.''

During the impulsive phase the flare emits hard X-ray bremsstrahlung and microwave gyrosynchrotron radiation; the non-thermal electrons implied by these radiations appear to receive the bulk of the flare energy release, but not all of it.
The microwave emission requires electrons at energies of 0.1-1~MeV and contains negligible total energy, but during the impulsive phase a large fraction of the total energy also may appear in $\gamma$-ray-emitting energetic ions \citep[e.g.,][]{1995ApJ...455L.193R}.
The evaporation of chromospheric material into the coronal magnetic loops corresponds to the \cite{1968ApJ...153L..59N} effect, the dominant temporal behavior pattern:
the coronal manifestations of a flare essentially integrate the impulsive-phase energy release, owing to the relatively slow cooling times of coronal material.
The white-light flare continuum, which together with its UV extension is about two orders of
magnitude more important the flare X-rays \citep{2005JGRA..11011103E,2010arXiv1001.1005H}, also occurs in the impulsive phase, along with the powerful electron acceleration.
See Sections~\ref{sec:morph} and \ref{sec:sed} for further comments.

Mass motions often occur simultaneously with the flare brightening. 
The most powerful events have a one-to-one association with coronal mass ejections (CMEs), which have the clear appearance of the expansion of the coronal field and the creation of new ``open'' field that can support solar-wind flow.
Sometimes the CME appears to entrain much or all of a pre-existing solar prominence.
A CME is the spectacular ejection of mass outward past the occulting disk of a coronagraph observing the corona in Thomson-scattered white light.
One can use the intensity to estimate the mass, often more than $10^{15}$~g, and the radial motion (often more than $10^3$~km/s) shows that the kinetic energy may exceed $10^{32}$~erg for major events.
The departure of coronal mass also produces the soft X-ray dimming \citep{1976SoPh...48..381R,1996ApJ...470..629H} signature of a flare.

Flares also frequently produce jets of material apparently along the field; X-ray jets \citep[e.g.,][]{1996PASJ...48..123S} frequently occur in association with a microflare brightening in closed fields, even though the jet flow itself occurs on large-scale or even open fields.
Flare mass motions thus include both flows perpendicular to the field, identifiable with CMEs, and flows parallel to it: sprays and surges at chromospheric wavelengths, and jets at X-ray and EUV wavelengths.
The CMEs occur preferentially in the most energetic events; \cite{2006ApJ...650L.143Y} find that virtually all X-class flares have accompanying CMEs.
The energy threshold for near one-to-one correspondence appears to be at about the GOES X2~level, some multiple of $10^{32}$~erg total energy.
Jets can happen in tiny events, even at the network level; they accompany microflares but can also occur in the quiet Sun, for example in the polar regions \citep[e.g.,][]{1996PASJ...48..123S}.

The close association of perpendicular plasma flows in flares is not surprising, since the field must restructure itself globally to release energy in the form of an implosion \citep{2000ApJ...531L..75H}.
What is surprising is that such flows are hard to detect except in the most energetic events, which have CME associations; even more surprising is that the CME eruption appears to \textit{add} energy to the global field by creating a large-scale current sheet (Section~\ref{sec:ergs}).
Recent observations have provided some evidence for implosive motions leading during or just prior to the impulsive phase of a flare \citep[e.g.,][]{2004ApJ...612..546S,2006A&A...446..675V}
This is consistent with the observation that ribbon expansion typically does not develop clearly until the later phase of a flare \citep[e.g.,][]{2009ApJ...693..132Y}, consistent
with a chromospheric interpretation of these early motions as a contraction \citep[e.g.,][]{2007ApJ...660..893J}.

\subsection{Flare and CME occurrence patterns}

\subsubsection{Flares and microflares}

\begin{figure}
\centering
\includegraphics[width=0.49\textwidth]{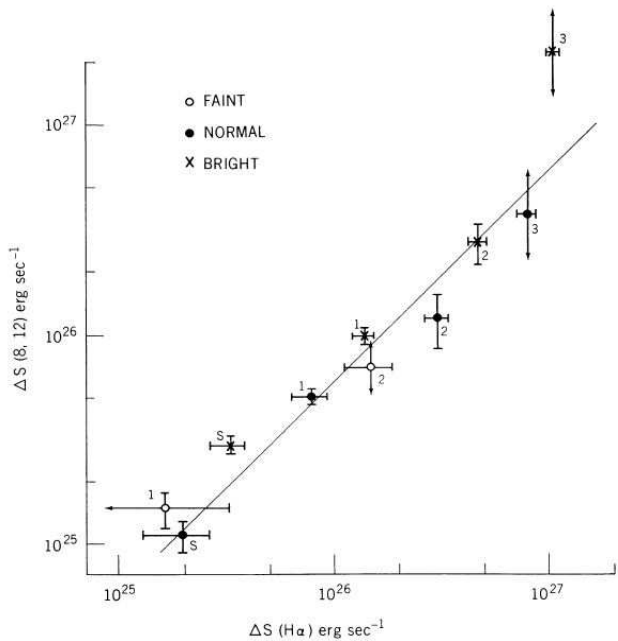}
\caption{An early example of Kahler's ``big flare syndrome'': the tight correlation found between soft X-rays and~H$\alpha$ emission by \cite{1971SoPh...16..431T}.
}
\label{fig:thomas-teske}       
\end{figure}

Flares have complicated images and time histories, but there are simplifiying patterns.
Normal classifications in common use include the soft X-ray energy flux in the standard GOES spectral bands of 1-8\AA~and 0.5-4\AA.
The more traditional H$\alpha$~classification the flare importance by area and visual brightness.
Table~\ref{tab:imp} gives a quick overview of these classifications.
It should be noted that most of the extensive properties of flare scale roughly together, with a scatter
of some fraction of one order of magnitude, over several decades of range.
This is one interpretation of the ``big flare syndrome'' \citep{1982JGR....87.3439K}, which we illustrate in Figure~\ref{fig:thomas-teske} \citep{1971SoPh...16..431T}.
This reveals a tight correlation between two quite different flare properties: the emission from the corona in soft X-rays at about 10$^7$~K, and the emission from the chromosphere in H$\alpha$~at about 10$^4$~K. 
A successful model of this relationship would require some sort of regulatory mechanism -- part of the big-flare syndrome -- to avoid the parameter dependence that obviously is not there.

Flares tend to occur in isolation, localized in space and time but with strong correlations; typically one active region will produce dozens of flares, especially during periods of flux emergence (often near the beginning of the lifetime of a given region, but not always).
The most powerful events usually occur in active regions.

In a time series such as the GOES soft X-ray data there will often be long intervals of slow variation in between the discrete flare events. 
This strongly implies the existence of two distinct sources of coronal energy release: flare-like and steady.
The flare-like component consists of readily identifiable outbursts whose occurrence distribution function in peak flux follows a power law.
Generally the distribution function dN/dR~$\propto$~R$^{-\alpha}$ has a slope ${\alpha} < 2.0$ for a given observable parameter~R such as the GOES peak flux, implying that the smaller events contribute relatively less than the larger events \cite[e.g.,][]{1991SoPh..133..357H}.
Calibrating the observable~R against the total energy of an event (the big-flare syndrome), we conclude that microflares cannot ``heat'' the corona although the integrated energy is reasonably close \citep[e.g.,][]{1995PASJ...47..251S}.
Note that this concept makes little sense in any case, since the corona is extensive and relatively steady, whereas the flares are compact and transient; RHESSI for example observes no microflares outside active regions \citep[e.g.,][]{2008ApJ...677..704H}.

\begin{table}
\caption{GOES and H$\alpha$ classifications}
\label{tab:imp}      
\begin{tabular}{l l l l}
\hline\noalign{\smallskip}
GOES & H$\alpha$ class & H$\alpha$ area & Emission measure  \\
            &                                & Sq. degrees & cm$^{-3}$ \\
\noalign{\smallskip}\hline\noalign{\smallskip}
X10& 4 & 24.7 & 10$^{51}$ \\
X    & 3 & 12.4 &10$^{50}$  \\
M   & 2 & 5.1 &10$^{49}$ \\
C   & 2 & 2.0  &10$^{48}$  \\
B   & S  & $<$2.0 & 10$^{47}$ \\
A   & S & $<$2.0 & 10$^{46}$  \\
\noalign{\smallskip}\hline
\end{tabular}
\end{table}

Parker's (1988) concept of ``nanoflares'' could help.
It seems entirely reasonable that frequent small-scale, non-thermal energy releases could 
provide a sufficiently smooth energy input to maintain steady coronal heating.
In Parker's view this would result from the dynamics of small-scale current sheets that would 
inevitably form in the coronal plasma.
\nocite{1988ApJ...330..474P}
Observationally, this is a difficult problem since the nanoflares would be undetectable
individually almost by definition; the events can run together and confuse their time series to the
point where they could not be distinguished individually.
If nanoflares do explain coronal heating \citep[e.g.,][]{1994ApJ...422..381C}, they must have different occurrence patterns from the flares or microflares.

\subsubsection{Flares and CMEs}

The most energetic CMEs occur in close association with powerful flares (see Sections~\ref{sec:cmeless} and~\ref{sec:role}).
Nevertheless large-scale CMEs do occur in the absence of major flares, even though these tend to be slower and less energetic.
These involve flare-like events in quiet regions of the Sun, with large-scale ribbon features seen in chromospheric lines \citep{1986STP.....2..198H} and the appearance of large-scale soft X-ray arcades.  
Such quiet-Sun CMEs frequently coincide with filament eruptions (e.g. Hanaoka et al., 1994).
 \nocite{1994PASJ...46..205H}
The physics of the filament channel and magnetic flux rope is thereby strongly implicated in the formation of a CME, and this may also apply to active-region flare/CME occurrences but on a smaller scale and with greater intensity of energy release.
These events in the quiet Sun are slower, cooler, and fainter than active-region flares \cite[e.g.,][]{1995JGR...100.3473H}.
The term ``impulsive phase'' might not seem to apply to them, since they are generally not detectable with current instrumentation in hard X-rays, $\gamma$-rays, or gyrosynchrotron emission, but this may simply reflect the limited sensitivity of the observations.
Finally we note the well-known property of CME initiation, namely that it may significantly precede the flare impulsive phase \citep[e.g.][]{2010JASTP..72..643M}.

\subsection{X-ray signatures}\label{sec:xray}

\begin{figure}
\centering
\includegraphics[width=0.49\textwidth]{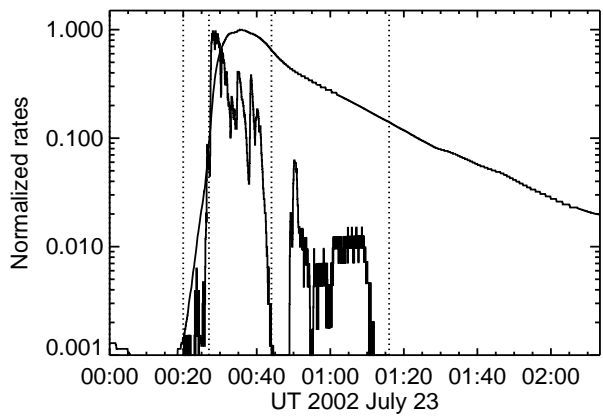}
\caption{The four different phases illustrated for the RHESSI $\gamma$-ray flare SOL2002-07-23T00:35.
The smooth line shows the GOES 1-8\AA~soft X-ray band, and the histogram the RHESSI 100-200~keV
counting rate, each normalized to its maximum.
The three dotted lines separate the four phases shown in Figure~\ref{fig:phases} schematically.
In this event the extended flare continues for almost an hour.
For this famous event, RHESSI obtained the first $\gamma$-ray images of a flare in the 2.223~MeV
line during the impulsive phase, 00:27:20-00:34:40~UT \citep{2003ApJ...595L..77H}.
}
\label{fig:real_phases}       
\end{figure}

Figure~\ref{fig:real_phases} illustrates the four flare phases we discuss (see Figure~\ref{fig:phases} for a schematic view), using the event SOL2002-07-23T00:35 as an example.

\subsubsection{Early phase}

Often the GOES light curve will have a precursor increase prior to the impulsive phase of a flare.
Although this is often interpreted as ``preheating,'' soft X-ray images show that in many cases the precursor, though a part of the flare, does not coincide with the structure supporting the impulsive phase
\cite{1998SoPh..183..339F}, hence not reflecting the plasma conditions at the site of major energy release, but instead
shows preflare dynamics associated with flux emergence \citep[e.g.,][]{2007A&A...472..967C}.

Hard X-rays also show interesting features prior to the impulsive phase, as discovered in the event of Figure~\ref{fig:real_phases} \citep{2003ApJ...595L..69L}.
These observations reveal a coronal source with a steep non-thermal spectrum and weak footpoint emissions.
The observed emissions require a major part of the flare energy, even though the impulsive-phase acceleration has not yet begun.
\cite{2009A&A...498..891B} suggest that this phase proceeds via con\-duction-driven evaporation as a 
response to this energy input, which is presently not understood.
We note that observations of source motions now are beginning to suggest the contraction of the coronal
magnetic field in this phase \citep{2000ApJ...531L..75H}; see also the discussion below in Section~\ref{sec:ergs} regarding flare energetics.

\subsubsection{Impulsive phase}\label{sec:imp}

Spiky bursts of hard X-rays and microwave-millimeter wave radiation characterize the impulsive phase of a flare, due respectively to the bremsstrahlung and gyrosynchrotron emission from non-thermal electrons, inferred to have been accelerated into a continuous power-law distribution that may extend to the MeV range.
The time variations during this phase exhibit the ``soft-hard-soft'' pattern, with a close negative correlation between the hard X-ray flux and its power-law spectral index $\gamma$ as determined by a fit to $j(\varepsilon) \propto \varepsilon^{-\gamma}$ \citep{1970ApJ...162.1003K,2004A&A...426.1093G}.
Such bursts sometimes exhibit quasi-periodic pulsations \citep{1971SoPh...16..186P}.

\begin{figure}
\centering
\includegraphics[width=0.45\textwidth]{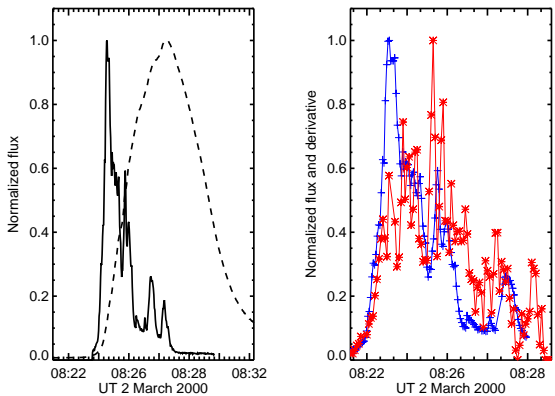}
\caption{\textit{Left:} the Neupert effect as seen in X-rays.
Solid line, the X-ray fluxes for a compact flare of SOL2000-03-02T08:22, as observed by Yohkoh/HXT
in its 33-53~keV band (solid) and GOES 1-8\AA~soft X-rays (dashed), both normalized to their
peak values.
\textit{Right:} a comparison between the GOES derivative (red) and the Yohkoh/HXT hard X-ray counting rate (blue), 33-53~keV.
}
\label{fig:neupert}       
\end{figure}

The new bolometric observations \citep{2006JGRA..11110S14W} show that this flux, L$_X$, amounts to about 1\% of the bolometric luminosity (Emslie et al. 2005; Kretzschmar et al., 2010; Quesnel et al. 2010) for the most energetic flares. 
\nocite{2010NatPh...6..690K}
\nocite{2010arXiv1003.4194Q}
\nocite{2005JGRA..11011103E}
We interpret these soft X-ray fluxes as described above, namely as hot plasma collecting in closed magnetic structures that extend into the corona.
These flux tubes trap plasma at high temperatures, which then cools following the Serio scaling law $n_e^2 \propto T$ \citep{1991A&A...241..197S}.
The coronal residence time is a few minutes (the reason for the Neupert effect, illustrated in Figure~\ref{fig:neupert}, since this time scale usually exceeds that of the energy release).
The growth phase of the flare soft X-ray source detected by GOES corresponds to the impulsive phase, in which hard X-ray emission appears.
The right panel of this Figure compares GOES derivative and hard X-ray fluxes, showing the effect clearly but also illustrating the absence of an exact relationship \citep{1993SoPh..146..177D,2002A&A...392..699V}.
One would not expect any particular detailed relationship in view of the complex spatial structures of a flare, which implies a range of physical conditions and presumably variations of the evaporation physics.

\begin{figure}
\centering
\includegraphics[width=0.4\textwidth]{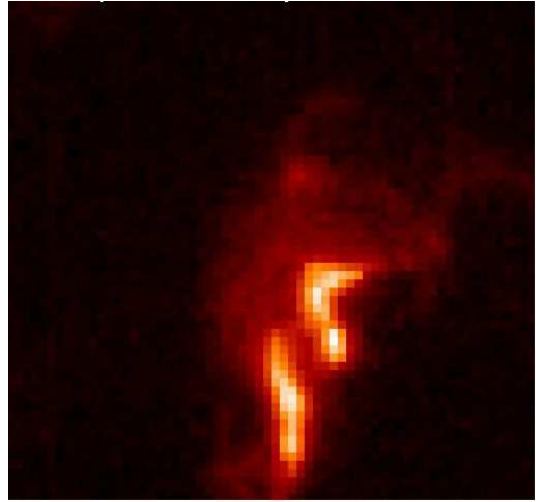}
\caption{Soft X-ray image from Yohkoh/SXT for the flare SOL2001-08-25T16:45,
studied by \cite{2003ApJ...595..483M} as a white-light flare.
The image scale can be inferred from the size of the individual SXT pixels, 2.54$''$.
At the time of this exposure the ribbon sources clearly appeared in soft X-rays, and in TRACE white light, and were crossing the sunspot umbra.
}
\label{fig:sxt_flare}       
\end{figure}

Soft X-ray images beautifully define solar flares as coronal magnetic structures.
The first observations came from sounding rockets, and then from Skylab \citep[e.g. articles in][]{1980sfsl.work.....S}, but the definitive view was produced by the SXT instrument \citep{1991SoPh..136...37T} on the Yohkoh satellite, launched in 1991.
These observations revealed the coronal structures of flares often to consist of loops, clearly defined by the magnetic field in a manner consistent with the low plasma beta (ratio of gas pressure to magnetic pressure) inferred from the observations.
In the impulsive phase of a flare, though, at the time of major energy release, the soft X-ray images of flares may not look like loops.
Figure~\ref{fig:sxt_flare} shows the example of the well-studied GOES X5.3 flare SOL2002-08-25T16:45 \citep{2002ApJ...574.1059K,2003ApJ...595..483M}.
In this image the dominant structures are not loops, but instead features close to the chromospheric and photospheric ribbon features of the flare.
These features correspond to impulsive soft X-ray emission in the chromospheric footpoints \citep{1993ApJ...416L..91M,1994ApJ...422L..25H}.
This image, not untypical of the impulsive phase, further explains why one would not expect a precise match to the Neupert effect: the flare structures are complex.

\subsubsection{Gradual phase}

\begin{figure}
 \centering
\includegraphics[width=0.4\textwidth]{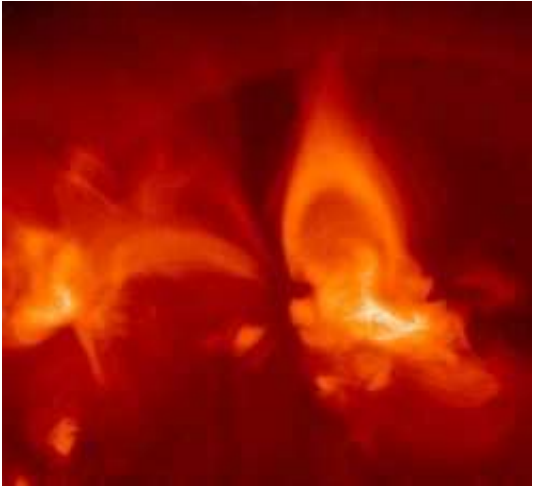}
\caption{Soft X-ray image from Yohkoh/SXT for a flare SOL2000-06-07T14:49.
The tip of the cusp strikingly resembles a coronal helmet streamer, and such a cusp configuration is often taken to indicate the presence of a large-scale current sheet but may also have a geometrical origin.
The image scale (roughly 1~R$_\odot$) can be inferred from the faint glimpse of the N~polar coronal hole in the background.
}
\label{fig:sxt_cusp}       
\end{figure}

The soft X-ray images of the gradual phase of a flare often look strikingly different from the impulsive-phase image shown in Figure~\ref{fig:sxt_flare}.
In the most powerful events, and especially those with CME association and the ``long decay event''  (LDE) property of an hours-long duration, one can often see a well-developed cusp structure in the soft X-ray images.
Figure~\ref{fig:sxt_cusp} shows an excellent example, again from Yohkoh/SXT.
The cusp is striking, but what is even more striking is the fact that it originated in a flare one day earlier (GOES X2.3). 
The image here shows the situation just before a GOES X1.2 flare in the same active region, which was observed to have no drastic effect on the cusp left over from the previous flare \citep[see e.g.,][]{2003LNP...612...58F}.
This behavior predominates; extensive surveys such as those of \cite{2004A&A...422..337T} and \cite{2009ApJ...690..347L} do not remark on exceptions.
But on occasion an apparently stable arcade resulting from a previous flare (as in the cusped arcade of Figure~\ref{fig:sxt_cusp}) can indeed blow out \citep{2009ApJ...703..757L,ruiliu}.
This implies that the increased stability of the restructured field does not mean that it cannot accumulate sufficient energy and/or helicity to erupt again \citep{2001GeoRL..28.3801N}.
As with homologous flares, this presumably means that flux emergence has extended spatial and temporal coherence.

The usual interpretation of the cusp structure is in terms of the standard magnetic-reconnection model \citep[the consensus theoretical views expressed in][]{2006SSRv..123..251F}.
This scenario most naturally explains the gradual phase of a flare as the reconnection and closing down of the field opened in a transient manner by the associated CME.
This presupposes that the CME expansion stores energy in the field, as it appears to do during the eruption.
The impulsive phase remains ill-understand in this standard picture.
The problem lies in how to evade the Aly-Sturrock theorem \citep{1984ApJ...283..349A,1991ApJ...380..655S}, which states that a global open-field configuration maximizes the stored magnetic energy for a fixed boundary field.
\cite{2006SSRv..123..251F} offer several suggestions regarding this, but in my view none of them seem convincing except for the implosion argument given in Section~\ref{sec:large-scale} below.

In any case, following the eruption, the standard model naturally explains the two-ribbon arcade configuration and even the extended heating; as the reconnection proceeds to higher and higher altitudes, new high-energy loops appear. 
These then cool by radiation and conduction -- in rough balance because they regulate each other \citep[e.g.,][]{1980sfsl.work..341M,1990ApJ...357..243F} -- appearing sequentially in emissions characteristic of lower and lower temperatures \citep[e.g.,][]{1967ApJ...149L..79N}, and
and at the same time shrinking geometrically \citep{1987SoPh..108..237S,1996ApJ...459..330F}.
The conductive losses from these loops cause their footpoints to radiate strongly in transition-region and chromospheric lines, such as the bright ribbon structures seen in both and well studied in H$\alpha$.
The ``supra-arcade downflows'' provide an interesting new wrinkle in this area \citep{1999ApJ...519L..93M,2004ApJ...616.1224S}.
The on-line movie attached to \cite{2002SoPh..210..341G}, for example, shows beautiful ``tadpoles'' drifting down with the hot spiky structure above the arcade \citep[e.g.,][]{1998SoPh..182..179S}.
\cite{2004ApJ...605L..77A} have shown that these downflows, or tadpoles, also tend to match the timing of hard X-ray emission as observed by RHESSI for the $\gamma$-ray event SOL2002-07-23T00:35.

\subsubsection{Extended flare phase}\label{sec:ext}

The ``extended flare'' phase designates events with major coronal non-thermal developments generally coinciding in time with  the gradual-phase arcade development, but not clearly related to it.
These can include meter-wave bursts of type~II and type~IV \citep[e.g.,][]{1963ARA&A...1..291W} and now hard X-ray signatures as well \citep[e.g.,][]{1986ApJ...305..920C,2008A&ARv..16..155K}.

These processes involve the long-term storage of particles in coronal magnetic fields, at altitudes ranging up to a fraction of 1~R$_\odot$.
The particles may be relativistic; pion-decay $\gamma$~rays can in some cases be detected hours after the flare injection \citep{1987ApJ...318..913C,1993A&AS...97..349K}.
The meter-wave radio signatures are complex and fascinating \citep[e.g.,][]{2008A&ARv..16....1P}, resulting from a mixture of free-free, gyrosynchrotron, and plasma radiation resulting from coupling from Langmuir turbulence excited by various processes.
Often the extended phase includes meter-wave emission of both type~II (Langmuir-wave conversion at a shock front propagating through the ambient corona) and type~IV (attributed to relativistic electrons trapped in large-scale closed fields).

The prototype hard X-ray extended event was the event SOL1969-03-30T02:47 \citep{1971ApJ...165..655F}, for which the identification of the coronal sources could be done by chance occultation \cite[see also][]{1978ApJ...224..235H}, from which one can get a height estimation by simple geometry.
The coronal sources of the extended flare generally have harder spectra, slower time variations, and lower peak microwave frequencies; the sources may move with time as in the famous ``Westward Ho!'' type IV burst \citep{1970SoPh...13..448R} SOL1969-03-01T23:00 seen at 80~MHz, or more recently detected moving hard X-ray sources \citep{2001ApJ...561L.211H,2007ApJ...669L..49K}.
There are also apparently stationary hard X-ray sources in the extended flare phase.
These sources may have bright footpoint emission as in the impulsive phase \cite{2004ApJ...603..335Q,2008A&ARv..16..155K} and exhibit spectral variations consistent with stable trapping with collisional energy losses for the lowest-energy particles.
The implied spatial scales of this trapping may be large \citep[e.g.,][]{2008A&ARv..16..155K}, and have long durations (e.g., Figure~\ref{fig:real_phases}).

The extended phase of a flare is the site of one of the major mysteries in the flare/CME/SEP environment (SEPs are ``solar energetic particles''), namely what we could call the ``Kiplinger effect'': SEPs are associated with the particular ``soft-hard-harder'' spectral evolution seen in this phase \citep{1995ApJ...453..973K}.
This distinguishes the hard X-ray behavior from that seen in the impulsive phase (Section~\ref{sec:imp}).
The implied association is mysterious because the SEPs, according to current consensus, originate mainly in diffusive shock acceleration at the CME-driven bow shock.
There should be no direct connection between this process and the corona where gradual hard X-ray sources appear; specifically no magnetic connection could exist within the standard models (see following).
The RHESSI observations have now convincingly confirmed this association \citep{2009ApJ...707.1588G}, as shown in Table~\ref{tab:kipl}.

\begin{table}
\caption{The ``Kiplinger effect''$^a$}
\label{tab:kipl}       
\begin{tabular}{l r r}
\hline\noalign{\smallskip}
& SEP & no SEP  \\
\noalign{\smallskip}\hline\noalign{\smallskip}
SHH$^b$ & 12 & 6 \\
no SHH & 0 & 19 \\
\noalign{\smallskip}\hline
\end{tabular}

$^a$\cite{2009ApJ...707.1588G}

$^b$The ``soft-hard-harder'' pattern; see text
\end{table}

\subsubsection{The standard model}\label{sec:model}

The development of a flare/CME process is extremely complicated, involving major gaps in our understanding of the microphysics in particular.
Accordingly the accepted concepts involve simple cartoons or sketches that in principle link the observables to the energy sources and mechanisms involved.\footnote{An archive of cartoons can be found on http://solarmuri .ssl.berkeley.edu/$\sim$hhudson/cartoons/}
New observations are typically described in terms of the features of the cartoon, which in principle is not the best way to discover new mechanisms that may not feature in the accepted model or any of its variants.

\begin{figure}
\includegraphics[width=0.45\textwidth]{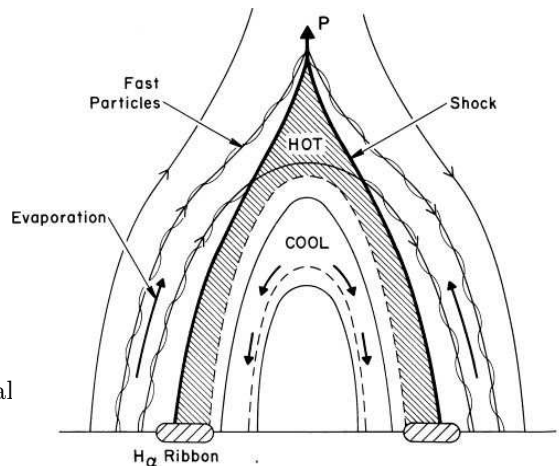}
\caption{The standard magnetic-reconnection model of a solar flare, adapted from \cite{1983ApJ...266..383C}.
Above it one may imagine the X~point or line, or current sheet, of the reconnection process, and in 3D~the plasmoid that entrains the erupting filament \citep[e.g.,][]{1974SoPh...34..323H}.
}
\label{fig:model}  
\end{figure}

Without question the present standard model for flare and CME development is the ``CSHKP'' model \citep{1992LNP...399....1S}, whose essence consists of a large-scale current sheet in which magnetic reconnection drives the flows that release energy.
An early full description of such a model was that of \cite{1974SoPh...34..323H}, in the context of filament eruption, but the initial ``C''~of CSHKP refers to \cite{1964NASSP..50..451C}.
More recent work has not resulted in any substantial modifications of this cartoon, and Figure~\ref{fig:model} shows a modern version due to \cite{1983ApJ...266..383C}.
Note that this is a 2D representation of an inherently 3D physical process; the third dimension allows for
the formation of a \textit{plasmoid} analogous to structures found in the geomagnetic tail.
This 3D view was clearly seen in earlier representations of the CSHKP model, such as that of \cite{1974SoPh...34..323H}.

The standard model does not readily lend itself to explaining the impulsive phase, since the reconnection process in principle requires the existence of a large-scale current sheet only created by the flare itself via the expansion of the field (see Section~\ref{sec:xray} for a discussion of flare phases, and Section~\ref{sec:large-scale} for a summary overview of this problem).
Another weakness of the standard model is its MHD underpinnings.
As has long been known \citep[e.g.,][]{1976SoPh...50..153L} and now abundantly confirmed (see Section~\ref{sec:ergs}) the energy of a flare or of a CME has a major component of non-thermal particle acceleration.
Since an MHD plasma has no non-thermal particles as such, the standard model cannot describe these key observations.
Nevertheless the CSHKP model captures many of the features of eruptive flares, especially the behavior of the flare ribbons.

\subsection{Flare spectroscopy}

\begin{figure}
\includegraphics[width=0.45\textwidth]{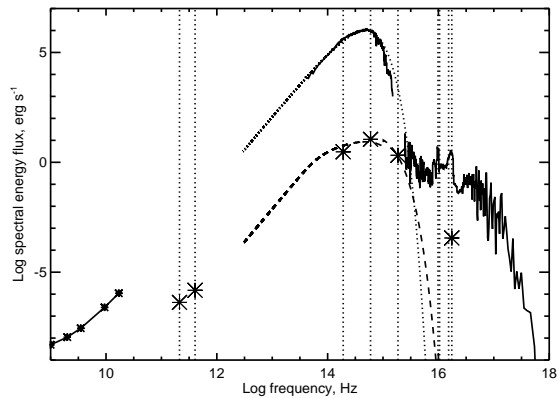}
\caption{Overview of solar flare broad-band spectrum in the impulsive phase, scaled from various 
observations by GOES class to the X1~level \citep{2010arXiv1001.1005H} and compared with the ``Sun as a star'' spectrum (curves and histogram).
The asterisks show the flare continuum at various wavelengths, with the dashed line representing a blackbody at 10$^4$~K.
}
\label{fig:over_spec}  
\end{figure}

Flares generally increase the brightness of the Sun across the entire wavelength range; remarkably making the Sun an astronomically bright object where it normally might be darker than the night sky -- in hard X-rays, for example.
The global spectrum determines the fate of the flare energy release, at least as regards radiant energy, and so it is a key part of the discussion in Section~\ref{sec:ergs}.
It also is the only tool for studying flares and similar phenomena on stars, as opposed to the Sun where we can resolve the spectral components spatially to a good degree, as described in the sections above.
Figure~\ref{fig:over_spec} gives a continuum (broad-band) overview of the flare spectrum, in comparison
with the spectrum of the quiet Sun.

The overall spectrum of the quiet Sun has a near-blackbody character at visual and near-IR wavelengths, with absorption features (lines and continua) becoming more prominent at UV wavelengths.
Excesses over a fitted blackbody function then appear at EUV and far-IR wavelengths and beyond.
One defines the effective temperature $T_{eff}$ in terms of the solar luminosity ${\mathcal L}_\odot$:
$${\mathcal L}_\odot = \sigma T_{eff}^4,
$$
 with a value of about 5777~K \citep[e.g.,][]{2000asqu.book.....C}.
 For comparison the photospheric temperature of the standard VAL-C model \citep{1981ApJS...45..635V} is 6420~K.
 This substantial difference reflects the complicated physics of the solar atmosphere, which of course a 1D~model such as VAL-C cannot fully capture.
The lowest frequency at which the quiet Sun can be detected does not extend much below 20~MHz \citep[e.g.,][]{1977SoPh...54...57E} because of competition from non-solar sources.
Interestingly the Fermi satellite has detected quiet-Sun $\gamma$-ray emission above 20~MeV \citep{2009arXiv0912.3775O}, giving a spectral range of some 15~decades!

The overall behavior of the global spectrum of the Sun during a flare is to broaden out.
At long wavelengths huge radio fluxes appear, and at short wavelengths the Sun can suddenly become a powerful source of hard X-rays and $\gamma$-rays.
The spectral range extends yet further, and solar bursts have been detected down to the plasma frequency of the solar wind at $\sim$1~AU (roughly 30~kHz).
In general the extension of the solar spectrum to the highest and lowest frequencies signals the presence on non-thermal effects secondary to the presence of high-energy particles.
Indeed, if we include solar energetic particles detected by low-latitude neutron monitors, the observable spectral range of a solar flare expands to some 20~decades.

All of the detectable wave bands provide some diagnostic information, each characteristic of the sources in the particular wave band.
As will be appreciated from the discussion above, a solar flare involves an enormous range of physical parameters in its constituent parts.
In most cases we simply cannot obtain complete data.
A further point to note is line-of-sight confusion.
In most non-radio wavebands the corona is transparent, so that foreground or background confusion can easily result.
It is clear morphologically that filaments and filament-channel magnetic fields can play an important role in CME eruptions, but to characterize the coronal plasma in a filament cavity is difficult \citep[see, e.g.,][]{2009ApJ...700L..96S}.

\section{Global effects}\label{sec:glob}

\subsection{Energy buildup and release}

For a clear discussion of the nomenclature of the various global manifestations in the corona, see the short article by \cite{2005EOSTr..86..112V}.
In general we believe that the flare and CME derive their energy from storage in the magnetic field, and so there must be a sudden restructuring involved in this energy release. 
We are interested here mainly in these sudden effects, rather than in the buildup of energy that must proceed much more slowly.
At intermediate time scales, there may be other time scales involved.
For example, the ``two-ribbon flare'' paradigm was known from early H$\alpha$ observations to involve the apparent growth of ``loop prominence systems'' \citep[e.g.,][]{1964ApJ...140..746B} or ``sporadic coronal condensations,'' to use some archaic terminology.
We now recognize this apparent growth as a cooling process, in that the flare itself creates higher-temperature plasmas initially (see Section~\ref{sec:xray}); these higher-temperature loops are not
detectable in H$\alpha$ or even at the EUV wavelengths of SOHO/EIT, for example.
The plasmas cool and become visible outside the X-ray band, eventually recombining to produce H$\alpha$ and the ``coronal rain'' downflows resulting from the draining of high-density coronal flux tubes by mass motions along the field.
The initial formation of the hot plasmas can be extended in time \citep{1979SoPh...61...69M}, a property of the standard model  \citep[e.g.,][]{1974SoPh...34..323H}.

The decisive part of the overall process, though, is the initial energy release. 
That is the main theme of this overview, and we proceed now to discuss the different global
signatures of the process.

\subsection{CMEs}\label{sec:cmes}

The CMEs of course cannot be observed directly with the flare effects, since they are defined in terms of the coronagraphic observations.
Flares, on the other hand, are observed in the low corona and below, so the observing domains are
almost always disjoint.
This is a bit awkward, since we are interested in CMEs in this review mainly as they guide us to the initial energy release.
\cite{2001JGR...10625199H} discuss other ``non-coronagraphic'' views of the CMEs, many of which allow the CME development -- within model restrictions -- to be tracked back to the lower atmosphere.
Examples of this would be EUV or radio observations, which do not have the above-the-limb limit imposed by coronagraphs.
One must always be careful, however, in identifying specific features seen in such different emission processes with true (Thomson scattering) observations of CMEs according to their definition.
For instance, a height-vs.-time plot for a CME that compares EUV and Thomson-scattering signatures will have some uncertainty related to feature identification.

The last decade or so has seen a substantial improvement in our observational understanding of CME origins.
The first of the breakthroughs was the discovery of a frequently close linkage between the impulsive phase of a powerful flare, both spatially and temporally, and the acceleration phase of the associated CME\footnote{Counterexamples to the close linkage are also frequently cited \citep[e.g.][]{1995A&A...304..585H}. 
In this review we emphasize the close relationships because of their likelihood to help us follow the energy as the flare/CME process develops.
See Section~\ref{sec:role} for further discussion of this point.}
(Dere et al. 1997; Zarro et al. 1999; Zhang et al. 2001).
\nocite{1997SoPh..175..601D,1999ApJ...520L.139Z,2001ApJ...559..452Z}.
This observation suggested a solution to one of the major problems in CME dynamics -- how can the coronal magnetic field evolve catastrophically into a higher-energy state?
The answer may be that there is a balancing act, perhaps mediated by large-scale magnetic reconnection, that allows the flaring part of the field to collapse and release energy \citep{2000ApJ...531L..75H}, part of which then goes into the CME kinetic energy (see discussion in Section~\ref{sec:large-scale}).
The best current assessment of the energetics \citep{2005JGRA..11011103E} suggests that the CME and flare require comparable amounts of energy (Section~\ref{sec:ergs}) at least for the most powerful events.

\subsection{Waves}\label{sec:waves}

The restructuring of the coronal magnetic field will naturally induce waves radiating outwards;
in particular a fast-mode MHD wave was identified early as a likely candidate \citep{1968SoPh....4...30U} for the Moreton wave (chromospheric) and the meter-wave type~II burst (coronal). 
The restructuring also can be seen as a CME, and this phenomenon is closely connected 
to the physics of global waves.
Note that the flare excitation of a large-scale wave is often described as a pressure pulse, but because of the low plasma beta the gas pressure would contribute only negligibly (Section~\ref{sec:large-scale}).
As discussed by \cite{2005EOSTr..86..112V}, flare-excited waves can propagate freely, as ``simple'' or ``blast'' waves, which will decay fairly promptly, or they can continue to gain energy from a driving ``piston'' due to other physics.
Here the important piston would be the CME eruption, which can continue into interplanetary space as an ICME, continually driving a bow shock ahead of the disturbance.
The impact of one of these ICME-driven shocks striking the Earth's magnetosphere, the ``storm sudden commencement,'' had been identified as a shock signature well before the space age
\citep[e.g.,][]{chapman-bartels}.

The interplanetary shock wave has many interesting properties; for example the theory of diffusive shock acceleration widely used in astrophysics applies here as well in supernova remnants.
A recent paper by \cite{2008AIPC.1039..111M} suggests that a major fraction (of order 10\%) of the total energy can wind up in particles accelerated by this shock front, presumably by this process.

\subsubsection{Type~II bursts}

\begin{figure}
\centering
\includegraphics[width=0.45\textwidth]{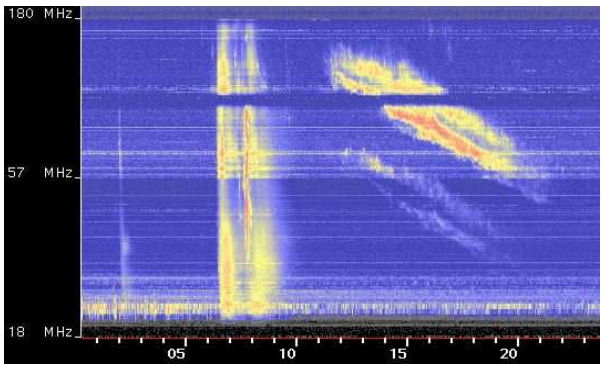}
\caption{An almost too-ideal example of type~III (fast drift) and type~II (slow drift) waves captured at
Culgoora Solar Observatory (Australia).
The negative frequency drifts show the outward motion of the exciter(s) through the corona towards lower densities.
The type~III bursts coincide with (parts of) the impulsive phase of the flare.
}
\label{fig:type_ii}  
\end{figure}

The first of the global signatures to be detected reliably was the meter-wave type~II radio burst, also called at that time the ``slow drift'' burst to distinguish it from the fast-drift type~III bursts, which have an entirely different explanation \cite[e.g.,][]{1963ARA&A...1..291W}.
Figure~\ref{fig:type_ii} shows a beautiful example.
In both cases the radio emission results from coupling between Langmuir waves generated by small-scale motions within the disturbance, coupling to electromagnetic radiation at the plasma frequency or its harmonic.
Since the plasma frequency decreases as the coronal density drops off radially ($\omega_{pe} = \sqrt{4 \pi n_e e^2 / m_e}$), the drifts strongly tend to be towards lower frequencies.
The straightforward interpretation would be that of a fast-mode (compressive) MHD wave, launched by a pressure pulse associated with the flare \citep{1968SoPh....4...30U}.
This idea led to a successful theoretical explanation that combined the meter-wave type~II burst with the Moreton wave (see Section~\ref{sec:chromwave} below) and which did not involve a CME.
CME-associated type~II emission certainly does occur longer wavelengths, best observed via radio antennae in space (e.g. Bougeret et al. 1976).
\nocite{1976SSRv...19..511M,2008SSRv..136..487B}.

The Uchida theory has gradually become less probable for several reasons, though. 
First, the radio observations from space show the apparent low-frequency extensions of the metric type~II burst, even to the plasma frequency characteristic of one~AU.
It would be improbable for a flare-driven blast wave as such to survive as far as the Earth, since its energy must decay.
Second, as observations improved, it became clear that the powerful flares that make these type~II bursts and CMEs originated in regions of low plasma beta.
Thus the gas pressure pulse would not be relevant in comparison with the dynamics assocated with the field restructuring, and so the flare brightening itself would not show the existence of any energetically relevant pressure pulse.
Finally, it appeared in many cases that the drifting motions of the metric and long-wavelength type~II signatures could be identified with each other, and so via the Occam's Razor principle a desire arose to simplify the phenomenology and simply have all type~II bursts come from CME bow waves.
Nevertheless the metric type~II bursts have a clearly demonstrable association with an exciter at the 
onset of the impulsive phase, i.e. a strong flare association (Figure~\ref{fig:type_ii}).
This is convincing evidence of an association, but current explanations have indeed veered away from flares as direct causes of type~II bursts.
The change of paradigms has interesting consequences as discussed in Section~\ref{sec:synth}.

\subsubsection{Chromospheric waves}
\label{sec:chromwave}

\begin{figure}
\centering
\includegraphics[width=0.45\textwidth]{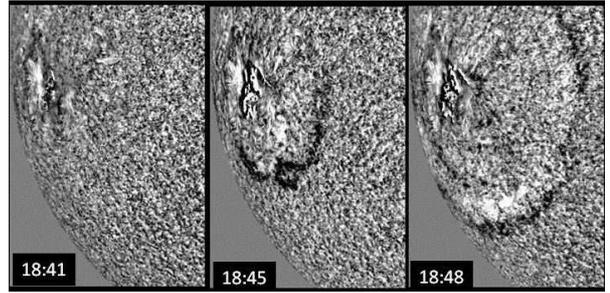}
\caption{One of the best examples of a Moreton wave, namely the ``solar tsunami'' of Balasubramaniam et al. (2010), from the flare SOL2006-12-06T18:40.
}
\label{fig:moreton}  
\end{figure}

Global wave signatures in the chromosphere made their appearance in 1959, with observations made possible by narrow-band H$\alpha$ filters and good seeing \citep{1960PASP...72..357M,1961ApJ...133..935A}.
These wave are now almost universally called Moreton waves, but had originally been noted by the developer of the filter, H.~E. Ramsey.
Figure~\ref{fig:moreton} shows an excellent recent example (SOL2006-12-06T18:40; Balasubramaniam et al. 2010).
\nocite{bala}

Theory was difficult with the Moreton waves at first, because it was clear that they were too fast ($\sim$10$^3$~km/s) for any known wave mode in the chromosphere, where they were observed to propagate.
This led to Uchida's theory of the fast-mode MHD shock wave \cite{1968SoPh....4...30U}.
In this picture a global wave radiated out into the corona from the flare site, but its energy refracted back down into the chromosphere as a result of the height dependence of the Alfv{\' e}n speed.
Thus the Moreton wave became the ``sweeping skirt'' of the true coronal wave.
In Uchida's view, this was a blast wave rather than a driven wave; indeed the refraction of the wave's energy could probably help stave off its decay, and many of the Moreton waves only appeared in limited angular sectors.
A particularly remarkable example of a Moreton wave, termed a ``solar tsunami,'' occurred with event SOL2006-12-06T18:40 (Balasumbramaniam et al., 2010).
 \nocite{bala}
This particular event, occurring near solar minimum, displayed an almost isotropic wave radiation and was not deflected by other active regions.

The chromospheric wave signatures also include the ``winking filament'' phenomenon, in which a remote filament bobs down and up as the Moreton wave front passes it \citep[e.g.][]{2004ApJ...608.1124O}.
The wink would result mainly from the Dop\-pler effect as the H$\alpha$ line shifts relative to the narrow filter bandpass.
\cite{2008ApJ...685..629G} estimate the energy in the filament motion to be 10$^{26-27}$~erg, even for a filament at a distance of $\sim$1~R$_\odot$ from the flare.
This suggests a total wave energy small compared with that of the flare itself, but not negligible.
We return to a discussion of wave energetics in Section~\ref{sec:ergs}

\subsubsection{X-ray waves}

\begin{figure}
\centering
\includegraphics[width=0.45\textwidth]{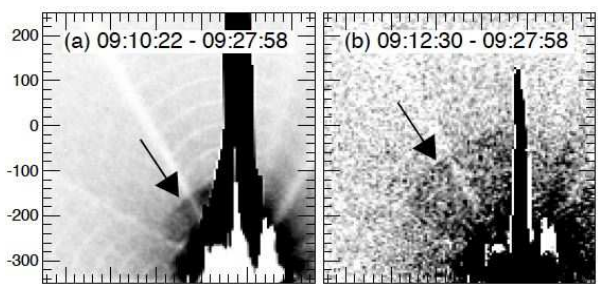}
\caption{A global coronal wave seen in soft X-rays with the Yohkoh/SXT telescope \citep[adapted from][]{2002A&A...383.1018K} for flare SOL1997-11-03T09:10.
Many image artifacts appear here because of the long exposure needed to capture the faint X-ray wave signatures marked with arrows; disregard the spokes, the light concentric rings, and the vertical spikes.
The flare is at lower right in the image; the left major tick mark on the X-axis is at $-$100$''$.
}
\label{fig:xray_wave}  
\end{figure}

Given Uchida's identification of the Moreton wave with the exciter of the type~II burst, it seemed logical that soft X-ray imaging observations of the corona would show the phenomenon directly; note that the high temperature of the corona makes this wavelength range (the soft X-rays and EUV) its natural emissions.
The Rankine-Hugoniot relations give values for the density and temperature jumps at a shock front; for a large Mach number the density jump approaches a factor of four asymptotically.
In spite of this clear expectation, though, plausible X-ray signatures of global coronal waves were not found by the Yohkoh/SXT instrument, launched in 1991, until the event SOL1996-05-08T08:08.
Analysis of this flare \citep{2003SoPh..212..121H} and others \citep{2002A&A...383.1018K,2002ApJ...572L.109N} helped to explain the lack of X-ray observations: the inferred Mach number was small; this coupled with the low image resolution and cadence of the SXT observations made detection and identification difficult.
Figure~\ref{fig:xray_wave} shows snapshots from the event observed by \cite{2002A&A...383.1018K}.

Other problems are the long-standing geometrical problems that apply to most wavelengths: the projection of a spherical shock front resembles a loop, and the the fact that the corona is optically thin means that the driver and its wave overlap in the image plane.
In addition the radial decrease in density means that X-ray brightness of the wave would drop rapidly with height, a problem both because of signal-to-noise ratio and also with respect to image dynamic range, which for soft X-rays can be many decades of intensity.

\subsubsection{Coronal shocks}

\begin{figure*}
\centering
\includegraphics[width=0.394\textwidth]{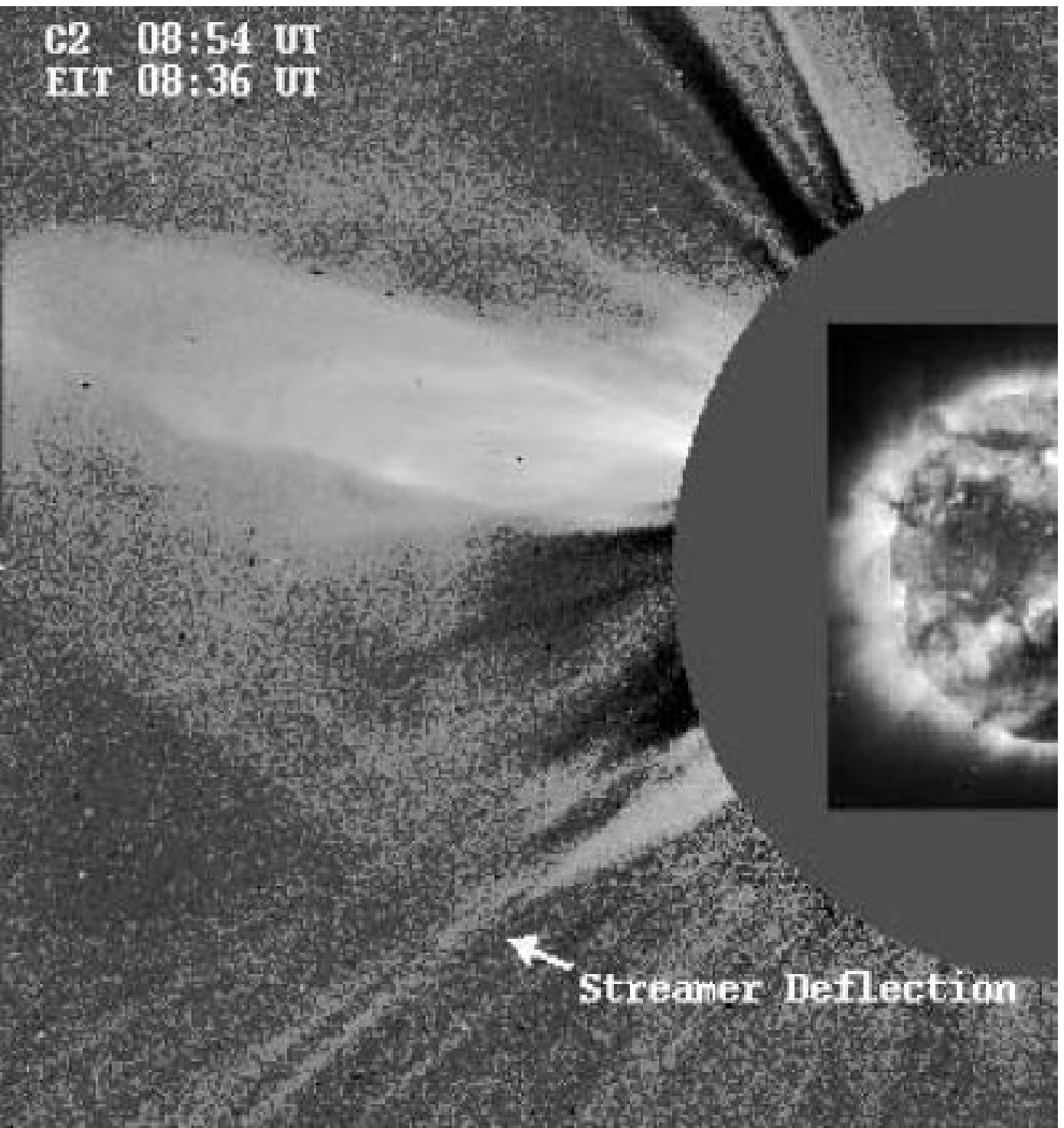}
\includegraphics[width=0.42\textwidth]{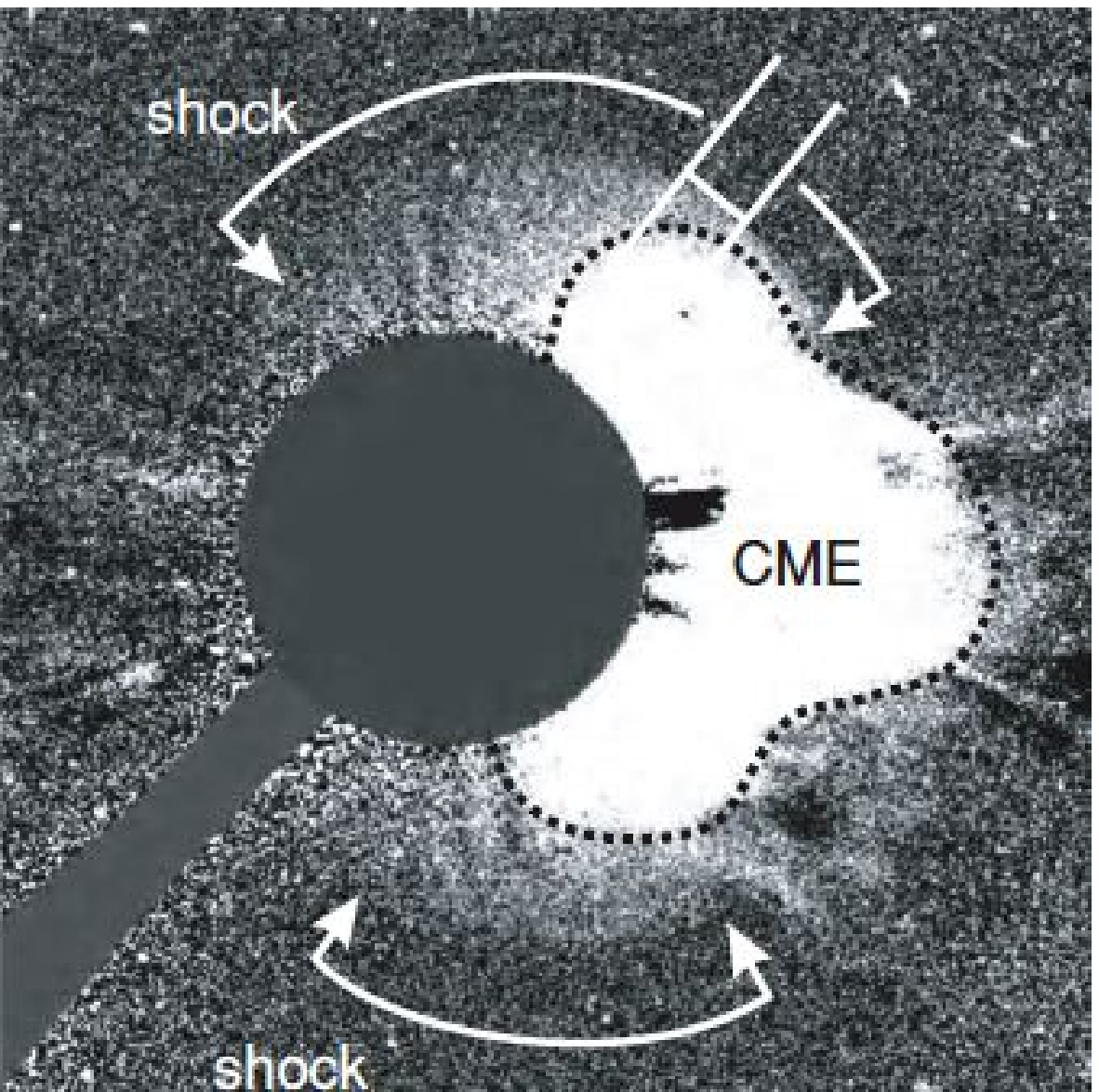}
\caption{Coronal shock waves identified in a SOHO/LASCO coronagraph images adapted from \cite{2003ApJ...598.1392V} (left; SOL1999-04-02T08:21), and from \cite{2009ApJ...693..267O} (right, SOL1997-11-06T11:55).
}
\label{fig:flank_wave}  
\end{figure*}     

The signatures of shocks are now known to be well defined in some coronagraph images
\citep{2003ApJ...598.1392V,2009ApJ...693..267O}, although this discovery took time and has perhaps produced surprises.
Contrary to the cartoon description of something like a bow shock, still suitable as a description of the ICME structure, the main coronal effects often appear to be in the flanks of the wave.
Some of the published examples have clear linear features trailing behind the CME, and these can be seen to affect ambient streamer structures that they interact with.
The shock signatures in the CME-driven wave flanks suggest that these parts of the structure may serve to link the type~II and Moreton waves, suggesting a replacement for the Uchida blast-wave theory.
If this could be established we would finally be able to link the processes that produce the coronal and interplanetary type~II signatures, which has been a perplexing problem for a long time.
We discuss this further in Section~\ref{sec:cmeless} below.
Figure~\ref{fig:flank_wave} shows two examples.
The left panel \citep{2003ApJ...598.1392V} illustrates the importance of the wave flank, while the event in the right panel \citep{2009ApJ...693..267O} has a disk-center origin that obscures this geometrically.

There have been many previous hints about the importance of non-radial motions in CME development in the low corona:
(i) ejecta seen in soft X-rays often have a distinctly non-radial component.
(ii) Type~II radio bursts also often have a non-radial component.
This of itself is not so conclusive, since we do not understand the special condition \citep{1999SoPh..187...89C} that allows the emission to occur in the first place.
(iii) CMEs often appear to expand drastically from a compact source in the lower atmosphere (Dere et al. 1997).
Adding to this there are recent observational and modeling results \cite[e.g.,][]{2009ApJ...702.1343T}
associating the Moreton wave with lateral expansion of the CME in the low corona.

\subsubsection{EIT waves}

\begin{figure}
\centering
\includegraphics[width=0.48\textwidth]{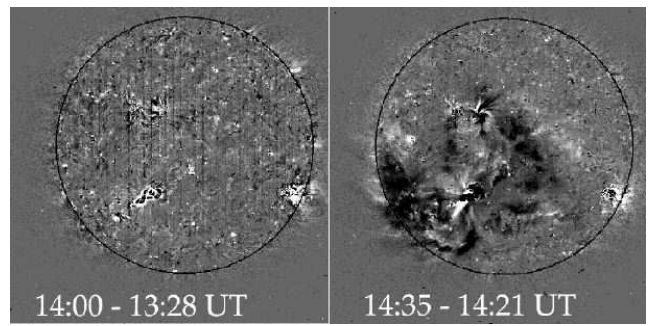}
\caption{Two frames from the SOHO/EIT observation of the global EIT wave of SOL1997-04-07T14:07 \citep{1999ApJ...517L.151T}.
These are running difference images as indicated.
}
\label{fig:eit_wave}  
\end{figure}

The ``EIT wave'' observed in one of the EUV wave bands (typically 171\AA~or 195\AA) was a striking observation that came with SOHO's EIT instrument 
(Moses et al. 1997; Thompson et al. 1999),
\nocite{1997SoPh..175..571M,1999ApJ...517L.151T}
although earlier EUV observations of course showed similar structures if not so well from the point of view of imaging \citep[e.g.,][]{1989ApJ...344..504N}.
Figure~\ref{fig:eit_wave} shows one of the early examples, that of SOL1997-04-07T14:03 \citep{1999ApJ...517L.151T}.
The initial thought was to identify the EIT waves with the shock front responsible for the Moreton wave and metric type~II burst.
This thought certainly seemed insufficient when \cite{2000ApJ...545..512D} pointed out that some EIT waves had velocities much smaller than any feasible Alfv{\' e}n speed in the corona.
We now understand the EIT signatures to result from more than one mechanism; \cite{2002ApJ...569.1009B} identify a small subset of them that one could associate with global shocks.
As with the X-ray observations of the corona, a shock signature would certainly be expected, and so why are so few observed?
Again there are the geometrical problems and signal-to-noise problems, as with the soft X-rays. 
Furthermore there is the passband issue for any line detection: do the ambient temperature and dynamics of the medium through which the wave passes allow the particular line to be excited?
Would a shock-associated temperature increase actually result in a brightening, or a dimming, because of the temperature dependence?

The great sensitivity and high resolution of EIT, and of succeeding EUV imagers such as those on  satellites such as TRACE, STEREO, and SDO, make the observation of these global waves quite routine.
There is no question that different effects are being lumped together under one ``EIT wave'' heading \citep[e.g.,][]{2009SSRv..149..325W}.
It is also clear, especially from movie representations, that much of the EIT wave effect is in fact a signature of the CME mass loss from the corona, i.e. an EUV dimming analogous to those seen in soft X-rays \cite[e.g.,][]{1998GeoRL..25.2481H}.
Thus the EIT  wave phenomena not only waves, but transient coronal holes and more diffuse brightenings associated with CME launches.

\subsubsection{Seismic waves}
\label{sec:seiswave}

A solar flare produces effects in the solar interior even though its main energy release appears to be in the solar atmosphere.
\cite{1972ApJ...176..833W} predicted these effects, writing ``...a major solar flare, which releases more than 10$^{32}$~ergs during its explosive phase, should deliver organized acoustical impulses to the solar interior containing about 10$^{28}$~ergs...'' and noted an uncertainty in his estimate of about a factor of 10.
\cite{1995ESASP.376b.341K} modeled this process, anticipating their observational discovery of the waves \citep{1998Natur.393..317K} at about the right magnitude.
It is interesting that (a) Wolff assumed minimal effects of ionization in the generation of the pressure pulse, 
while at the same time \cite{1972SoPh...24..414H} was assuming maximal ionization in order to explain white-light flares; and (b) that Wolff anticipated the possible magnetic effects as discussed below.

\begin{figure*}
\centering
\includegraphics[width=0.9\textwidth]{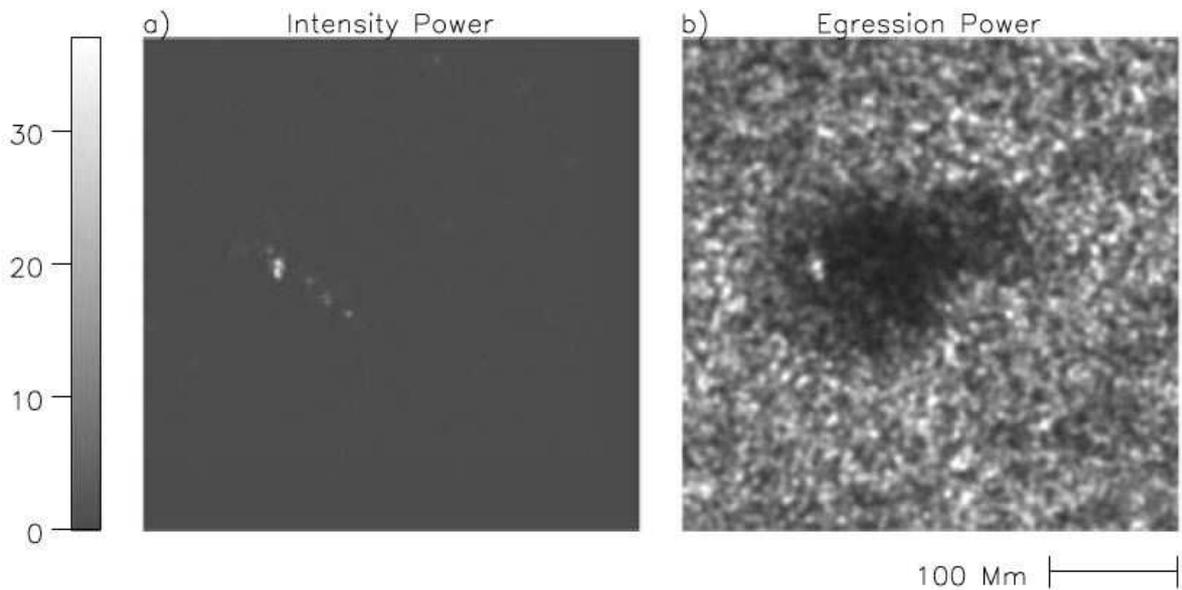}
\caption{\textit{Left:} white-light flare emission for the event SOL2003-10-29T20:49; \textit{right:} acoustic power sources (``egression power'').
The white light has been time-filtered in the same 5-7~mHz bandpass as in for the holographic reconstruction \citep{2008SoPh..251..627L}.
Note the easier visibility of the acoustic sources in the umbra, where the acoustic noise levels are lower.
}
\label{fig:mechanics}  
\end{figure*}     

The flare seismic waves consist of a set of ripples on the photosphere, observable via their Dop\-pler signature as they move radially away from the flare site at apparent velocities of about 40~km/s, often focused into sectors 
and sometimes revealing multiple radiant points \citep{2006SoPh..238....1K}.
The radiant points of the waves closely match the various signatures of the impulsive phase
\citep[Figure~\ref{fig:mechanics};][]{2008SoPh..251..627L}, and the optimum frequency bands for the holographic technique \citep{1990SoPh..126..101L,1999ApJ...513L.143D} appear to be 3-7~mHz, corresponding to time scales (2$\pi \omega$)$^{-1}$ of 10-50~s.

This ready agreement suggests that acoustic-wave excitation depends more the simple physics of conservation of energy and momentum than on the details of the way this happens (see Section~\ref{sec:synth}).
The momentum balance should be helpful diagnostically but is seldom discussed in this context,
except for the case of the evaporative flow and its effects on chromospheric line profiles \citep{1987Natur.326..165C}.
For comparison with the acoustic wave momentum, let us estimate the momentum present in surges, flares, and CMEs.
Table~\ref{tab:mom} does this by crudely scaling all properties to typical values observable in an X1-class flare.
Such a scaling is unsupported by correlation analysis, but is consistent with the estimates of 
\cite{2009ApJ...693..267O}, noting that most of their events originated behind the limb and are thus not at all suitable for scaling.
The duration of the impulse matters, because the detection sensitivity for helioseismic waves depends upon the temporal frequency (and resulting spatial structure) of the photospheric ripples.

\begin{table}
\caption{Vertical momentum components}
\label{tab:mom}       
\begin{tabular}{l r r r r}
\hline\noalign{\smallskip}
Phenomenon & Mass & Velocity & $\Delta$t & Momentum \\
                          & g & km/s & s & gm cm/s \\
\noalign{\smallskip}\hline\noalign{\smallskip}
Surge/jet$^a$& 2~$\times$~10$^{15}$& 500 & $\sim$300 &  --1.1~$\times$~10$^{23}$ \\
Flare energy$^b$& Small & 10$^4$ &100 & 10$^{23}$ \\
CME mass$^c$& 10$^{16}$ & 1000 & 100 & --10$^{24}$ \\
Evaporation$^{d}$ &2~$\times$~10$^{15}$ & 500 & 30 & --1~$\times$~10$^{23}$\\
Trapping$^{c}$ & 2~$\times$~10$^{15}$& 500 & 30 & $-$1~$\times$~10$^{23}$\\
Draining &2~$\times$~10$^{15}$ & 10 & $\sim$10$^4$ & 2~$\times$~10$^{21}$\\
\noalign{\smallskip}\hline
Seismic wave$^e$ & & 40 & 20-50 & 2.5~$\times$~10$^{22}$\\
\noalign{\smallskip}\hline
\end{tabular}
$^a$\cite{2009A&A...508.1443B}; $^b$Alfv{\' e}nic \citep{2008ApJ...675.1645F}; $^c$Rough estimate; $^d$\cite{1987Natur.326..165C}; $^e$10$^{29}$~erg
\end{table}

\begin{figure}
\centering
\includegraphics[width=0.35\textwidth]{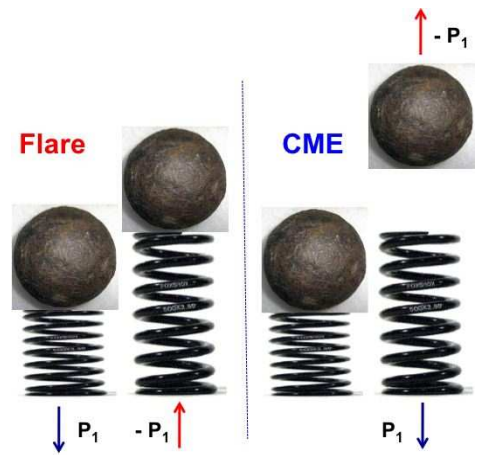}
\caption{Conservation of momentum, comparing an idealized flare and CME, inspired by \cite{2001AGUGM.125..143K}.
For a CME the mass never returns and the impulse has no compensation, but for a flare the evaporative motion is arrested by the closed field.
Note that illustrating momentum with a moving cannonball misrepresents the component present in the wave field.
}
\label{fig:mom_cons}  
\end{figure}

The momentum comparisons in Table~\ref{tab:mom} require some discussion.
The comparison with CMEs is a crucial one, but the acceleration parameters of CMEs are extremely ill-defined because of geometrical constraints (the coronagraph field of view does not extend down to the important region, especially for CMEs with sources not at the limb itself).
This component of the momentum may or may not couple well into the frequency range of the seismic observations; we do not understand the physics of the CME acceleration.
Note that an approach through height-vs.-time profiles probably does not have sufficient accuracy for the reasons cited above.
The same caveats would apply to surges, sprays, or flare-associated jets.
The momentum in the evaporative flow implies both signs, as noted by \cite{1987Natur.326..165C}.
Ideally a simple explosion in the chromosphere sends mass both up and down, the latter associated with the gas-dynamic shock \citep{1974AZh....51.1002K}.
The upward momentum in this flow is arrested by trapping in the closed magnetic field, which then must exert a vertical force in compensation (Figure~\ref{fig:mom_cons}).
Thus the flare ejecta on closed-field regions should ultimately produce both forward and reverse acoustic waves, separated by a time scale estimated from the loop heights and the evaporation flow speed.
For reasonable parameters, this might be one minute, but such a time scale competes with the duration of the impulsive phase.
If this latter results, as is probable, from a system of filamentary structures each following its own evolutionary path, the net momentum transfer would be reduced.
It therefore seems likely that the flare seismic waves be associated mainly with the process of eruption.

\subsubsection{CMEless flares}\label{sec:cmeless}

\begin{figure}
\centering
\includegraphics[width=0.35\textwidth]{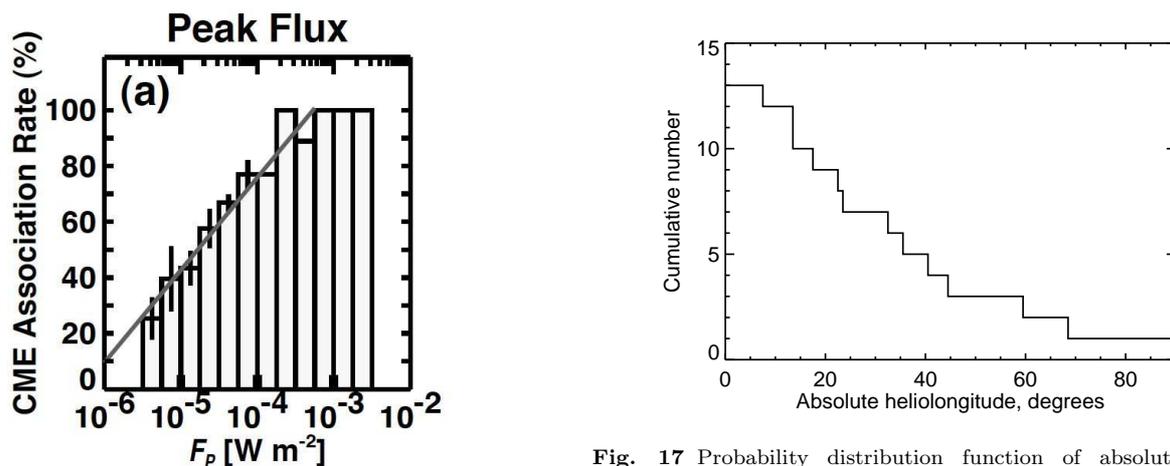}
\caption{Probability of CME occurrence as a function of GOES peak flux (adapted from Yashiro et al., 2006).
}
\label{fig:yashiro}       
\end{figure}

As the GOES class of a flare increases, so does its probability of association with a CME. 
The LASCO data demonstrate this compellingly \citep{2006ApJ...650L.143Y}, revealing that essentially all
major flares have CMEs (Figure~\ref{fig:yashiro}).
Above GOES~X2 the association reported thus far is 100\%, implying that a flare anywhere on the disk at this magnitude will have an accompanying CME.
It is therefore interesting to study the most powerful events that \textit{do not} have CMEs; is this the result of detection bias (e.g. the well-known ``plane of the sky'' bias resulting from the angular dependence of the Thomson cross-section), or is it due to a physical mechanism that we can identify?

\nocite{2006ApJ...650L.143Y}

\begin{figure}
\centering
\includegraphics[width=0.45\textwidth]{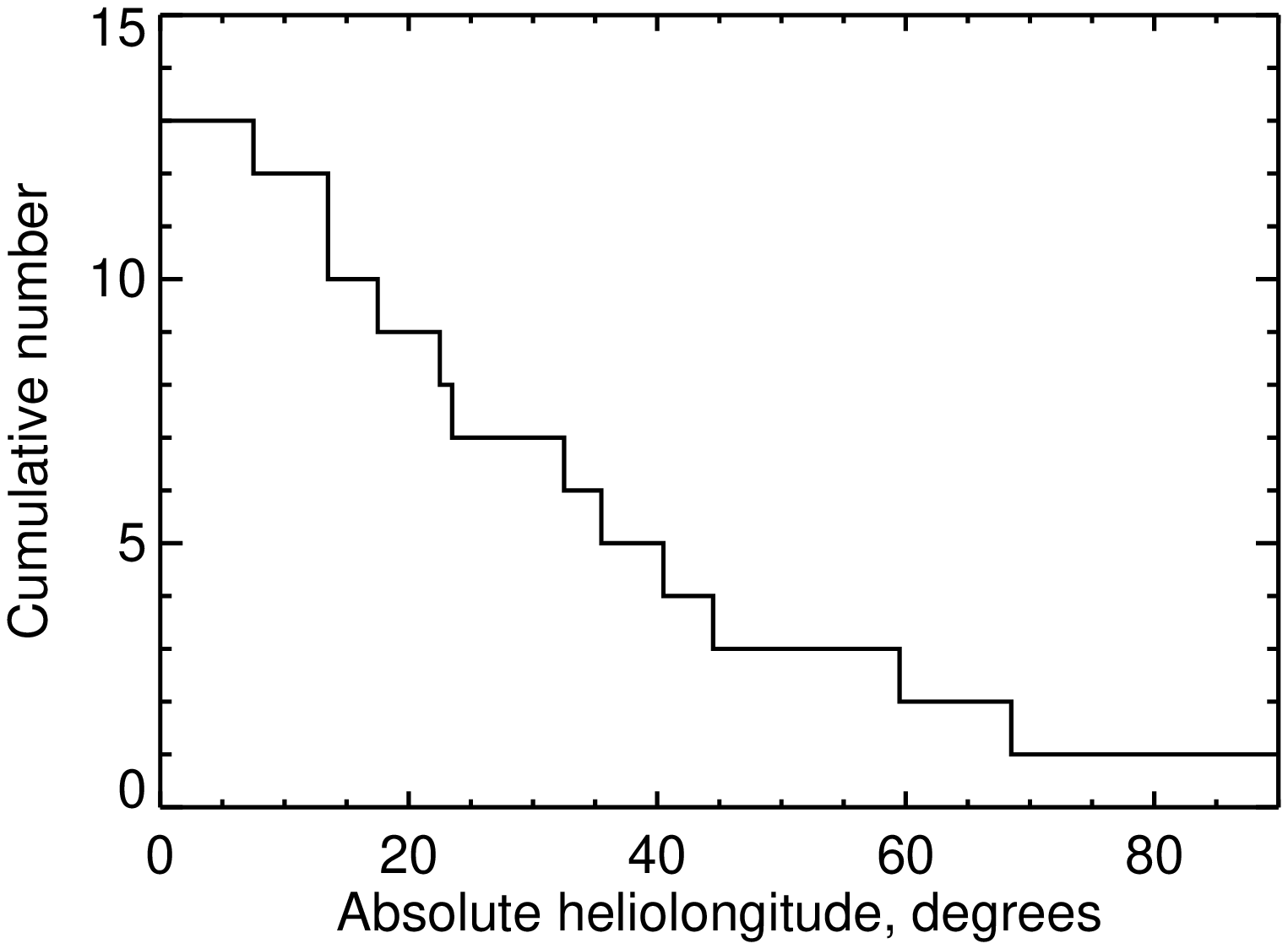}
\caption{Probability distribution function of absolute heliolongitudes of the union of the CMEless X-class flare lists of \cite{2007ApJ...665.1428W} and \cite{2009IAUS..257..283G} (13~events total, of which 3
originated outside $\pm$45$^\circ$ heliolongitude). 
}
\label{fig:cmeless}       
\end{figure}

There are X-class ``CMEless'' flares, but the list is short.
\cite{2007ApJ...665.1428W} and \cite{2009IAUS..257..283G} find a total of 13~events over the interval 2000-2005, with maximum GOES class~X1.7 and generally no trace of eruption or coronal ``extended flare'' (Section~\ref{sec:ext}) activity (generally including type~II bursts).
This list of CMEless flares, also describable as ``confined flares'' \citep[e.g.,][]{2001ApJ...552..833M} has a hint of Thomson-scattering angular bias, in that only 3/13 of the CMEless events occurred at heliolongitudes $>$45$^\circ$ (Y. Liu, personal communication 2010; see Figure~\ref{fig:cmeless}), where one might expect twice as many.
But there is no question that major CMEless flares do occur, and this has important consequences for our understanding of global wave origins.

\subsection{Wave synthesis: the Huygens Principle}\label{sec:synth}

We have seen that there are several kinds of global wave signatures now identifiable in the solar atmosphere and interior: the interplanetary shock, the metric type~II exciter, the Moreton wave, the direct coronal observation in white light, the X-ray signature, and the interior seismic wave.
Each of these signatures, except for the interplanetary shock, appears to have its origin in the lower atmosphere and to be both spatially and temporally consistent with excitation during the impulsive phase of the flare.
A single sudden restructuring of a part of the solar atmosphere might explain all of these features, but the physical details tend to be obscured by the flare itself and our understanding may require insights gotten from theory or modeling.
Clearly, since the impulsive phase contains a large fraction of the flare energy (see Section~\ref{sec:ergs}) our understanding of wave excitation will be closely linked to our understanding of the restructuring that produces the flare in the first place.

\begin{figure}
\includegraphics[width=0.45\textwidth]{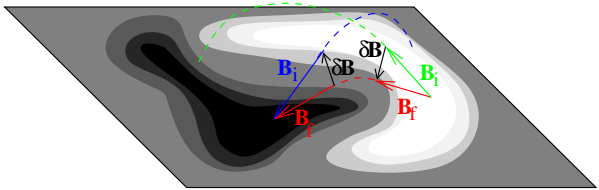}
\caption{Illustration of the implosion expected as a source of flare energy \citep{2008ASPC..383..221H}.
The sketch, courtesy of B.~J. Welsch, schematically shows magnetic regions of both polarities.
The red and green fields reconnect, and the retraction of the red field releases the energy and communicates
some of it into the solar interior as the source of a seismic wave.
Reconnection is not required for a restructuring of this kind, but it is likely to make the field more horizontal.
}
\label{fig:welsch}       
\end{figure}

How do we extract information about flare/CME physics from global  waves?
The clear answer is to apply the Huygens principle to the observations to seek information about the nature of the wave radiant source \citep{1999SoPh..190..467W,2002A&A...383.1018K,2003SoPh..212..121H,2009ApJ...702.1343T}.
It is limited by the resolution and sensitivity of the observations in several ways, but in principle Huygens reconstructions can give us key information about the shape, location, and activation time of the actual region of magnetic energy release.
Note that this region must have finite dimensions to be able to store sufficient energy, as described in the next Section.

\section{Energetics}
\label{sec:ergs}

\subsection{Overview}

\subsubsection{Spectral energy distribution}
\label{sec:sed}

A flare consists of the sudden release of magnetically stored energy.
``Follow the energy'' should be a good principle by which to learn how and why the flare develops.
In this section we describe the state of the art of energy storage, transport, and release in flares.
This task has received some closure recently because, at last, we have the first ``Sun as a star'' observations of total luminosity as seen in measurements of the total solar irradiance (TSI).
\cite{2006JGRA..11110S14W} observed an individual event (SOL2003-10-28T19:54, one of the ``Halloween flares'') using data from the SORCE spacecraft, and summed-epoch analysis of SOHO observations have shown this to be a systematic effect (Kretzschmar, 2010; Quesnel et al., 2010).
\nocite{2010NatPh...6..690K}
According to these results, the total flare luminosity is of order 100$\times$ the reported GOES soft X-ray luminosity, or L$_X$/L$_{tot} \sim$~0.01.
A major component of the luminosity is in the impulsive phase, consistent with the white light signatures already known but implying a significant extension into the near~UV (Figure~\ref{fig:over_spec}).
Note that \cite{1971SoPh...16..431T} estimated L$_X$/L$_{tot}$ at about 10\%, an order of magnitude above our current understanding.

\subsubsection{Energy storage}

\begin{figure}
\centering
\includegraphics[width=0.45\textwidth]{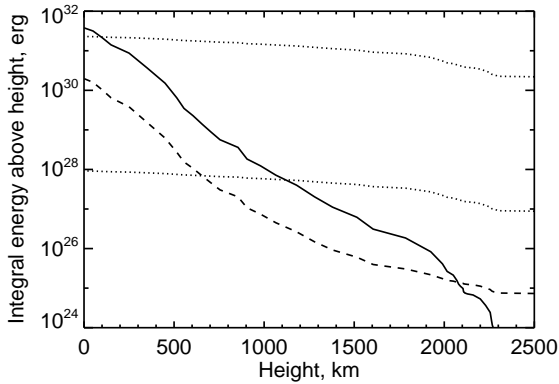}
\caption{The energy contained in the chromosphere, according to the VAL-C model
\citep{1981ApJS...45..635V}, evaluated for an area 10$^{18}$~cm$^2$: dashed line, thermal energy; solid line, latent heat of ionization; dotted lines, total magnetic energy for uniform fluxes of  30~and 1500~G.
}
\label{fig:valc_energy}       
\end{figure}

Although the main luminosity of a solar flare comes from the chromosphere, its volume does not appear to allow it to contain enough energy \citep[e.g.,][]{2003LNP...612...58F}.
Figure~\ref{fig:valc_energy} illustrates this.
One could push the estimate of magnetic energy by increasing the area above the 10$^{18}$~cm$^2$ assumed, increasing $|${\bf B}$|$ above the 1500~G assumed, and/or increasing the thickness of the layer above the 2500~km adopted for the VAL-C model \citep{1981ApJS...45..635V}.
Clearly, if not the chromosphere, then the low coronal volume could contain the canonical 10$^{33}$~erg needed for a major flare.
We discuss further intricacies of energy storage in Section~\ref{sec:b} below, but the general conclusion will be that we do not know now where flare energy can be stored in sufficient magnitude.

We take the opportunity here to discuss rarely mentioned alternative sites of flare energy storage.
Table~\ref{tab:glob} summarizes some of the other possibilities: gravitational potential energy, estimated from the Wilson depression and from filament levitation; energy flow in real time through the photosphere; escape of optically thin radiation such as Ly$\alpha$; and trapped particle radiation. 
This latter mechanism \citep{1969sfsr.conf..356E,2009ApJ...698L..86H} is particularly interesting,
since it appeals to a method for storing free energy in the magnetic field without current systems.
High-energy particles could store energy at some fraction of  the field energy $\int{(B^2/8\pi)dV}$, which we have estimated at 1\% of the energy content of a 10~G field in a volume of scale 0.1~R$_\odot$.
For each mechanism we estimate a feasible amount of energy (erg) and a time scale $\Delta$t (s).
The UV radiation estimate refers to the radiant energy trapped in the chromosphere by opacity, which we estimate as $\sigma (T_{phot}^4 - T_{eff}^4) / c$~erg/cm$^3$, where $\sigma$~is the Stefan-Boltzmann constant; $T_{phot}$~=~6420~K from \cite{1981ApJS...45..635V} and $T_{eff}$~=~5777~K.
Similar order-of-magnitude estimate have often been made previously for most of these mechanisms, with the same conclusion: each of them except for magnetic storage in the corona fails on either the energy scale or time scale criterion.
The energy for a flare or CME must reside in the solar atmosphere because of the low Alfv{\' e}n speed in the photosphere, and because the observations rule out the photospheric motions expected \citep[e.g.,][]{2010arXiv1003.1647S}.

\begin{table}
\caption{Energy sources for global effects}
\label{tab:glob}       
\begin{tabular}{l l r l}
\hline\noalign{\smallskip}
Site & Calculation & Energy & $\Delta$t \\
\noalign{\smallskip}\hline\noalign{\smallskip}
Coronal field & NLFFF & 1~$\times$~10$^{33}$ & Fast \\
Gravitational & Wilson depression  & 4~$\times$~10$^{27}$ & 60 s\\
Gravitational & Filament  & 3~$\times$~10$^{29}$ & 300 s\\
Subphotospheric & Wave energy  & Alfv{\' e}nic & 100~s \\
UV radiation &  $\Delta\mathcal{F}_\odot$/c & 3~$\times$~10$^{26}$  & Fast \\
Trapped radiation & 100~MeV  & 10$^{29}$ & 100~s \\
\noalign{\smallskip}\hline
\end{tabular}
\end{table}

\subsubsection{Energy transport}

Energy storage in the corona and release in the chromosphere, on time scales of a few seconds, means that we require a fast and efficient transport mechanism.
We also have the evidence from hard X-rays, $\gamma$-rays, and solar energetic particles that particle acceleration plays a major role in energy transport
The thick-target model invoking electrons \citep{1968ApJ...153L..59N,1971ApJ...164..151K,1971SoPh...18..489B,1972SoPh...24..414H} resulted naturally from these requirements.
Such a model works for the electron beams known from \textit{in situ} observations of interplanetary electron streams associated with type~III bursts.

In the case of the intense energy of the impulsive-phase energy transport, though, there have always been important theoretical questions about beam stability \citep[e.g.,][]{1978ApJ...221.1068C,1990A&A...234..496V}.
With modern data these beam-stability questions have become inescapable, and so energy transport by other means seems inevitable.
\cite{2008ApJ...675.1645F} propose transport by Alfv{\' e}n waves generated in the corona, for example by flows associated with magnetic reconnection.
This mechanism also may lead to electron acceleration by wave damping in the denser chromosphere, and so it would have many of the properties of the standard thick-target model.

\subsubsection{Role of CMEs}
\label{sec:role}

The role of CMEs in flare energetics is complicated.
Most flares, even including some at the GOES X-class level \citep[][see Section~\ref{sec:cmeless}]{2007ApJ...665.1428W} do not have associated CMEs.
Yet when they do, the estimated energy of the CME can rival that of the flare emission itself
\citep[e.g.,][]{2005JGRA..11011103E}.
We note that the CME energy usually is broken down into kinetic energy, enthalpy, gravitational potential energy, and magnetic energy \citep[e.g.,][]{2000ApJ...534..456V}.
The latter term dominates, especially since recent data suggest strongly that CMEs arise in coronal regions with low plasma~$\beta$, and yet the field cannot be directly observed.
Worse yet, since the restructuring of the field is generally thought to drive the CME, its magnetic energy must therefore be taken as negative, rather than positive.

\begin{figure}
\centering
\includegraphics[width=0.35\textwidth]{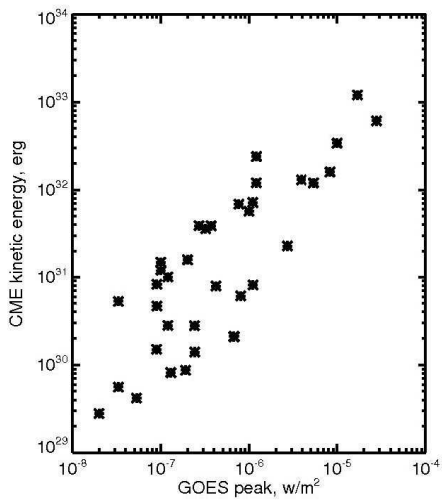}
\caption{Correlation of CME kinetic energy with GOES peak flux, based on the Halloween flare data given by \cite{2005JGRA..11009S15G}.
}
\label{fig:gopal}  
\end{figure}

\begin{figure}
\centering
\includegraphics[width=0.49\textwidth]{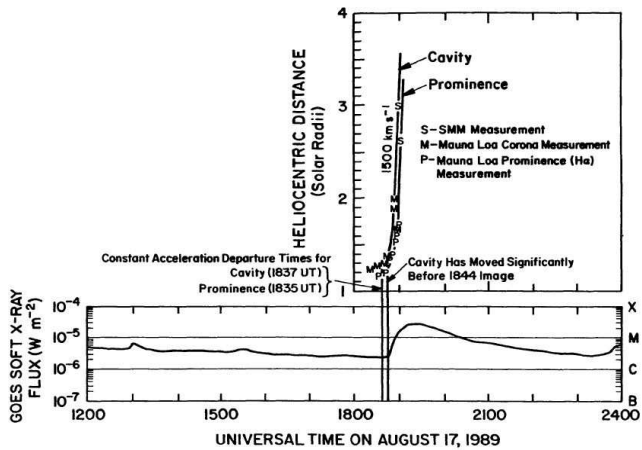}
\caption{A figure frequently shown to illustrate the early onset of a CME relative to its associated flare (from Sime \& Hundhausen 1992, as cited by Kahler, 1992).
\textit{Upper:} coronagraph height-vs-time plot for the CME SOL1989-08-17T20:07; \textit{lower:} GOES low-energy  time history showing its closely associated flare.
The decreasing background in the GOES data is from a preceding X-class flare.
}
\label{fig:sime}  
\end{figure}

When a CME occurs, its kinetic energy correlates strongly with the peak GOES flux \citep{2004JGRA..10903103B,2005JGRA..11009S15G}.
We illustrate this point in Figure~\ref{fig:gopal}, based on data compiled by \cite{2005JGRA..11009S15G}.
Timing arguments may give a different impression of the energetics (e.g. Kahler, 1992; Harrison, 1995), but we regard these as less direct.
Figure~\ref{fig:sime} (Sime \& Hundhausen 1992), frequently cited to illustrate this timing argument, makes it clear that the CME and its flare are virtually simultaneous, and indeed that the acceleration of the CME matches the impulsive phase of the flare as well as can be judged.
This confirms the conclusion drawn later, from better data, that there is no appreciable delay between the energetically significant phases of the CME and its associated flare (Dere et al. 1997, Zarro et al. 1999, Zhang et al. 2001).
\nocite{1992ARAA...30..113K}.

\nocite{1992ARAA...30..113K}

Nevertheless the CME energy is large; \cite{2007AIPC..932..277M} argue that even the SEP particles accelerated by the CME-driven interplanetary shock wave, in our current understanding, may contain as much as 10\% of the total event energy.
Further complication regarding this conclusion comes from the effect discovered by \cite{1995ApJ...453..973K} and now confirmed by \cite{2009ApJ...707.1588G}: flares with associated SEPs, normally associated with particle acceleration by the CME shock,  have a characteristic hard X-ray behavior pattern.
The spectrum gradually hardens with time, the ``soft-hard-harder'' (SHH) pattern.
Table~\ref{tab:kipl} shows the clear RHESSI association of SEPs with this SHH pattern.

\subsection{The fields}\label{sec:b}

The discussion above confirms what many authors have concluded, namely that the coronal magnetic field is the only plausible repository for flare energy prior to the event.
We discuss below how to understand this reservoir better.

\subsubsection{Static}\label{sec:static}

In practice one can measure the vector magnetic field in the photosphere via use of the linear and circular polarization of a magnetically sensitive Fraunhofer line.
This measurement is stable enough to permit the inference of vertical currents via Ampere's law, although there are several subtleties in this inference.
The extrapolation of the (static) photospheric field into the corona can be done at three levels of mathematical sophistication.
A \textit{potential field} has no body currents in the corona; very commonly one sees representations
of the ``potential-field source-surface'' (PFSS) field, which is a potential field in a concentric shell of the corona, nominally extended from the photosphere to 2.5~R$_\odot$; in this construction the field at this ``source surface'' of the solar wind then is required to be radial \citep{1969SoPh....9..131A,1969SoPh....6..442S}.
Physically this requires a fictitious current system to flow at and above the source surface itself.

More sophisticated models are either \textit{linear force-free field} models (LFFF) or, almost generally, \textit{non-linear force-free field} (NLFFF) models.
In the NLFFF models the force-free parameter $\alpha(x,y) = (\nabla\times B)/ B$ is considered to be a function of position (x, y) at the level of the photosphere. 
The magnitude of this variable can be estimated directly from the vector-magnetographic data by an Ampere's law integration in a horizontal plane, yielding the vertical current density $j_z (x,y)$.

Several groups now pursue the increasingly better observational material in attempts to obtain accurate extrapolations (DeRosa et al. 2009),
\nocite{2009ApJ...696.1780D}
but such work has many limitations.
The almost complete lack of observations of the coronal magnetic field would have to rank high on the
list of problems, but we could also note the incompleteness of the physics.
As an illustration of the latter, consider the simplest possible coronal magnetic structure, a loop anchored (``line-tied'') in the photosphere at both ends.
\nocite{2009ApJ...696.1780D}. 
Because of the force-free condition $\nabla \times B \cdot B = 0$, the parameter $\alpha(x,y)$ must be constant along a field line.
Thus the two ends of a field line within a magnetic loop must have the same value of $\alpha$ at either end.
This however is non-physical because the function $\alpha(x, y)$ must be determined by the solar interior via its dynamo physics and flows, and generally cannot match at two independently given points.
DeRosa et al. (2009) discuss this problem further.
To obtain a static MHD model of a coronal loop therefore requires the application of a physically incorrect boundary condition, and the MHD framework would not be appropriate for describing even this basic and well-observed coronal structure.
We note though that most of the work on photospheric field extrapolation does not go as far as the MHD approximation and relies on purely mathematical techniques.

\begin{figure*}
\centering
\includegraphics[width=0.52\textwidth]{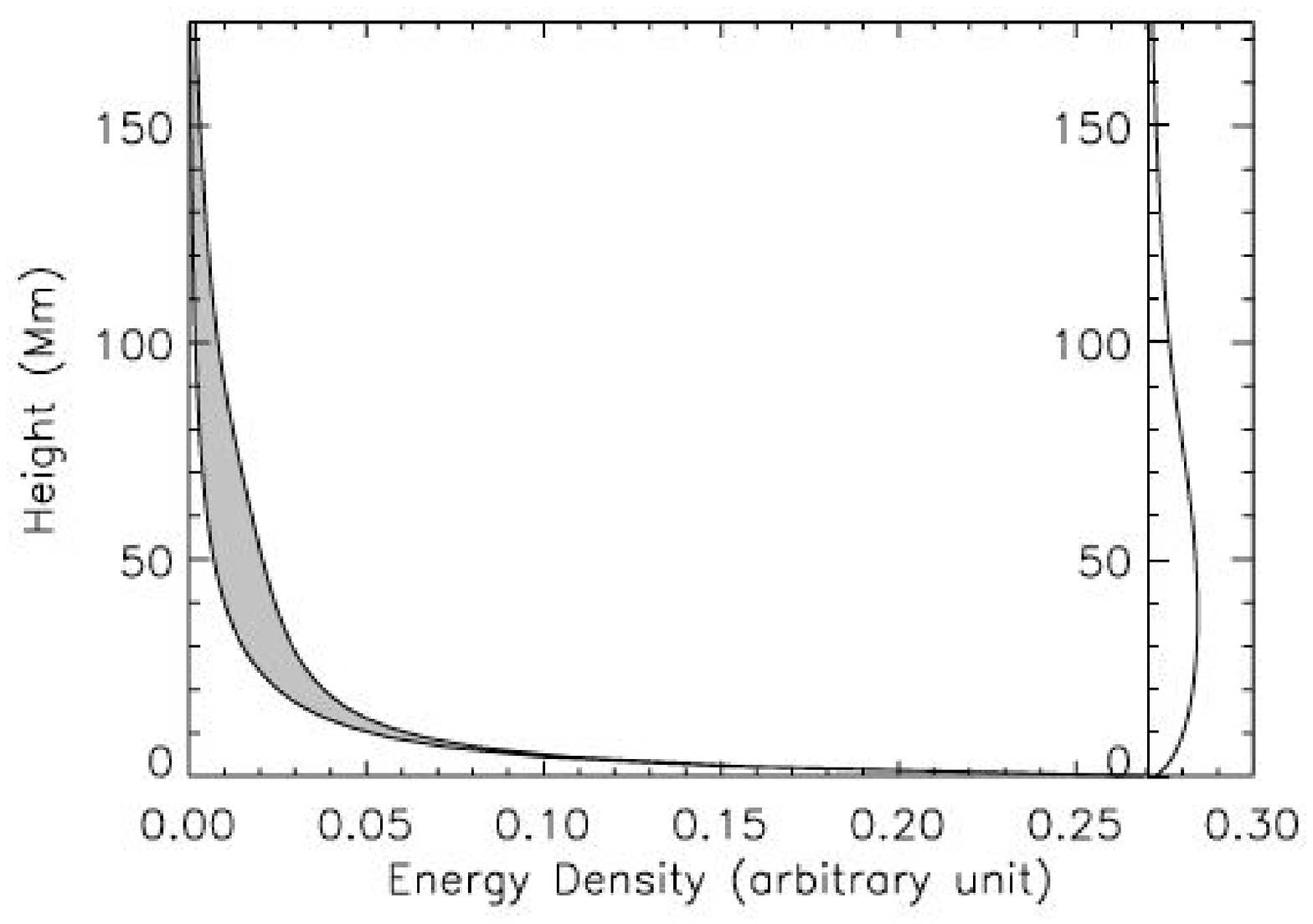}
\includegraphics[width=0.47\textwidth]{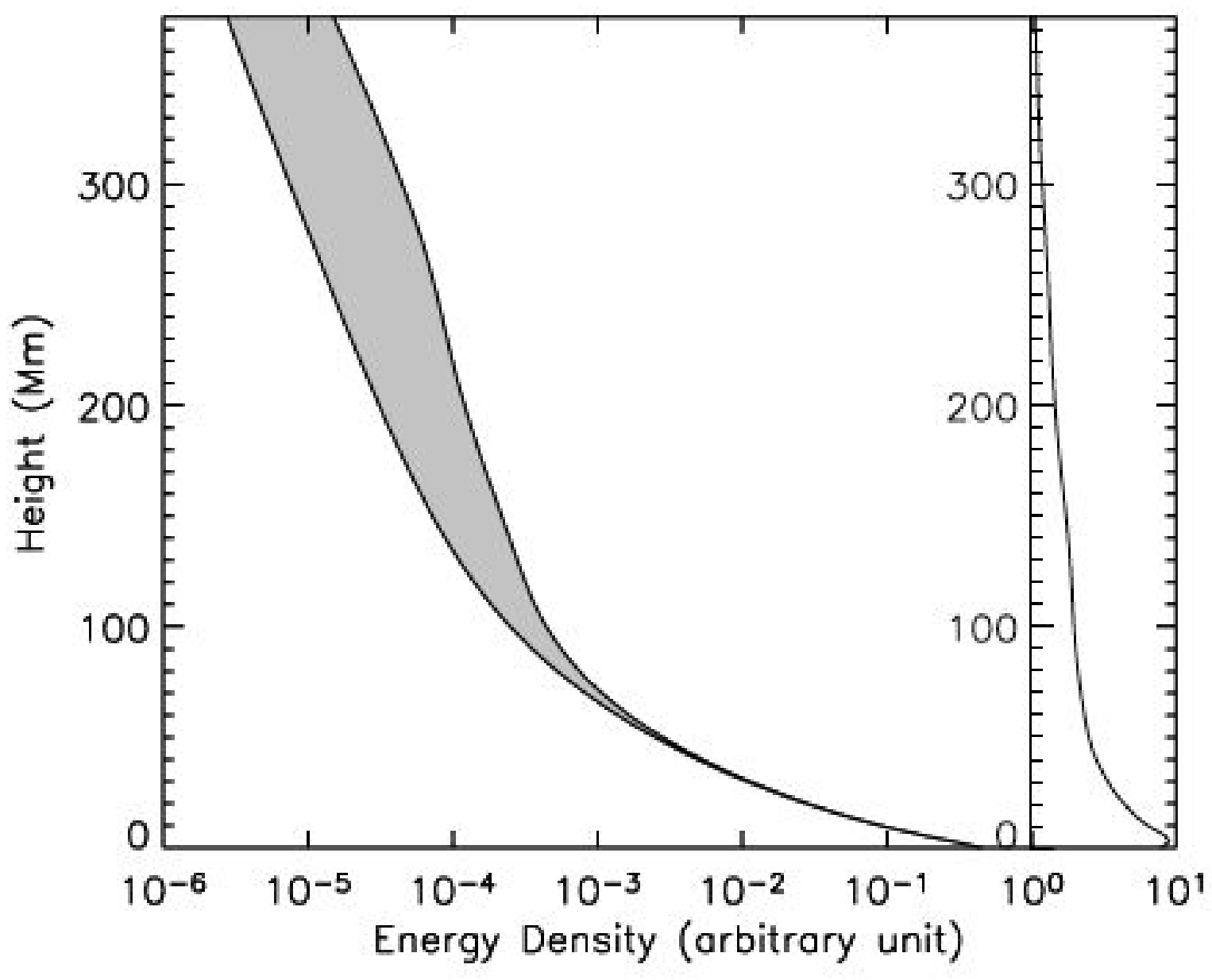}
\caption{The results of potential (lower curves) and non-linear force-free (upper curves) extrapolations from photospheric vector field measurements from \cite{2007A&A...468..701R} for two solar active regions (NOAA AR 8151 and 8210, respectively).
The inset plots show the variation of non-potential energy density with height
}
\label{fig:regnier}  
\end{figure*}

The problem described above appears to conflict with the broadly accepted idea that an LFFF state must describe a relaxed field, as enunciated by \cite{1958PNAS...44..489W} for a closed system.
The solar corona is not closed, though, because most of its magnetic flux evidently links through the
photosphere.
Thus from the argument given above it seems unlikely that a relaxed LFFF state can ever exist in the corona, except for field volumes physically detached from the photosphere. 
For a photospheric Alfv{\' e}n speed of 20~km/s and a spatial scale of R$_\odot$, this suggests a time scale greater than a day or so within which a relaxed equilibrium could never be achieved.
We discuss this problem further in the following section (\ref{sec:bvar}) and note that careful analysis with vector magnetogram data may in fact show no sign of relaxation to an LFFF state \citep{2002A&A...395..685B}.

Figure~\ref{fig:regnier} shows representative examples of NLFFF field extrapolation from the photosphere into the corona for two particular active regions \citep{2007A&A...468..701R}.
For the volume modeled in each region, the plots show the mean energy density vs height, and the fraction of ``non-potential energy density'' as an upper limit to what could be released by a flare.
The non-potential energy is concentrated near the base of the coronal volume, especially for AR~8210,
as one would expect.
It is noteworthy that \cite{2007A&A...468..701R} find non-potential excess energies considerably less than 10$^{32}$~erg for each region, even though AR~8210 did produce an X-class flare.
In general it appears that the coronal field extrapolations do not capture the flare energy sources well at the present time (DeRosa et al. 2009).
\nocite{2009ApJ...696.1780D}.
 
 \subsubsection{Dynamic}\label{sec:bvar}
 
\begin{figure*}
\centering
\includegraphics[width=.9\textwidth]{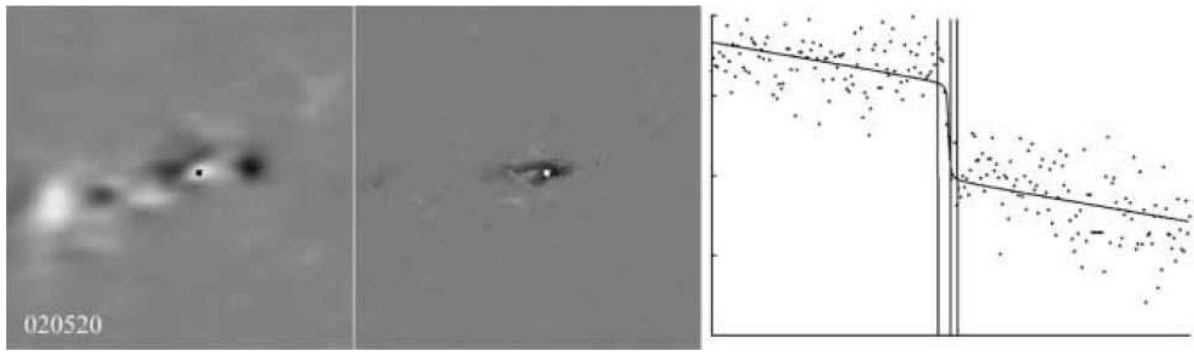}
\caption{One of the examples of stepwise changes in the photospheric  longitudinal magnetic field by \cite{2005ApJ...635..647S} as ``ubiquitous features of X-class flares.''
\textit{Left:} a GONG longitudinal magnetogram (10-m average prior to the event) for the flare SOL2002-05-20T15:29 (X2.1, S21~E65); \textit{middle:} a representation of the field change across the flare time; \textit{right:} the field variation in a representative pixel.
In the time-series plot the vertical lines denote the GOES start, peak (15:29~UT), and end times.
The total duration of the plot is four hours, and the vertical scale spans 400~G.
}
\label{fig:sudol-harvey}       
\end{figure*}

Flares do result in photospheric image changes associated with magnetic fields  \citep{1972SoPh...25..141R}.
The photospheric fields themselves also show striking flare effects when viewed via Zeeman splitting, rather than as an image feature such as umbra or facula \citep{1992SoPh..140...85W,2001ApJ...550L.105K,2005ApJ...635..647S} (see Figure~\ref{fig:sudol-harvey}).
The observed variations thus far have only been clearly recognized in the line-of-sight component of the field, partly because the sampling is much more efficient; the GONG magnetograms used by \cite{2005ApJ...635..647S} have a regular one-minute cadence.
This flare-associated variations appear as stepwise changes, usually superposed on a varying background. Because only the line-of-sight component is measured, the change can be either positive or negative, and cannot give any information about the energy contained in the field as such.

The stepwise variation of the field associated with a given flare can be determined across the image, as in the example in Figure~\ref{fig:sudol-harvey}. 
Note the gradual trend in the background in this plot, and the relatively large scatter from point to point.
The continuous stream of data at a regular cadence nevertheless allows a fit to the time series pixel by pixel, and the fit also determines an epoch for each pixel even in the presence of flare interference
in the magnetic signals \citep[e.g.,][]{2003ApJ...599..615Q}.
The epochs of the variations measured in this way by \cite{2005ApJ...635..647S} turn out to coincide as a rule with the impulsive phase of the flare (Section~\ref{sec:xray}), and the image configurations of the altered pixels resemble the flare ribbons.
These properties generally fit the expectation that a global-scale change of the field would drive the various forms of energy release observed in a flare.
The line-of-sight field variations are only qualitative in nature, but we can expect a much more exact understanding of the coronal field variations to come from systematic vector measurements now becoming available.

\subsection{Large-scale motions and a myth}\label{sec:large-scale}

A major flare and/or CME releases an energy equivalent to about one second's worth of solar luminosity.
This by consensus view must have come from stored magnetic energy (but see the pessimstic comment in Section~\ref{sec:static}). 
\cite{1972ApJ...176..833W} remarked ``...an enormous amount of magnetic field energy, comparable to the
total flare release of 10$^{32}$~ergs, seems to be annihilated during the flare. 
This should cause a subsequent relaxation of the entire field structure surrounding the flare, moving large masses of material in the process."
This was restated slightly more generally by \cite{2000ApJ...531L..75H} in the sense of a magnetic implosion, and Figure~\ref{fig:welsch} describes this process.
If these simplistic but physically reasonable views hold, then the Lorentz force thus invoked would be an excellent source of the global wave motions we described in Section~\ref{sec:waves}.

The problem with this view has always been that the bulk of flare mass motions have seemed \textit{explosive} in nature \citep[e.g.,][]{2001ApJ...552..833M}, with various things such as surges, sprays, filaments, and CMEs flying \textit{away} from the core of the flare.
The CME in particular contains a great deal of kinetic energy \citep[e.g.,][]{2000ApJ...534..456V}; \cite{2005JGRA..11011103E} show, for two major flares (the RHESSI flares of SOL2002-04-21T01:51 and SOL2002-07-23T00:35), that the CME kinetic energy is of the same order of magnitude as the radiated energy.
In some well-studied cases \citep[e.g.,][]{1998GeoRL..25.2469W,2009ApJ...701..283R} virtually no flare-like emissions can be detected at all.
Many events of this kind call to mind the flares in spotless active regions \citep[e.g.,][]{1970SoPh...13..401D,1986STP.....2..198H,1995JGR...100.3473H}.
For events of this kind one can imagine that the magnetic environment of the disturbance simply cannot confine the energy release, leading to a relatively free expansion (see Section~\ref{sec:cmeless}) and a rough equipartition between upward (CME) and downward (flare) effects, as observed.

Can a part of the coronal magnetic field simply expand unstably, with no compensating contractions elsewhere?
A CME gives every appearance of doing just that, and indeed the standard flare model assumes that this expansion of the field results in the formation of a large-scale current sheet from which flare energy can then be extracted.
Where is the energy for this process coming from?
A part of the answer to this problem may be hidden in one of the major ``myths'' of flare physics: the structural significance of the observable parts of a flare or CME.
Since the plasma beta is low \citep[e.g.,][]{2001SoPh..203...71G} in the cores of active regions, the 
flare and CME effects originating there have a predominantly electrodynamic character.
The visible mass mainly serves mainly as a marker, which in some circumstances we can use to help
define the velocity field.
But to the extent that the plasma beta is small, the energy and momentum transport reside in the
field, not in the gas component of the plasma.

\section{Conclusions}
\label{sec:concl}

In this review we have had two objectives: first, to review the phenomenology and physics of flares and associated CMEs, and second, to put these ideas in the context of global effects such as the large-scale
waves of several types.
For flares or the onsets of CMEs we can never directly observe the microphysics, as is possible in arguably related physical phenomena in space plasmas close enough to permit \textit{in situ} observations.
Thus in a sense we are always going to be stuck with the problem of inference from remote data.

The data and theoretical ideas available are generally consistent with the ideas that both flares and CMEs originate in magnetic energy storage concentrated in the low corona.
When flares and CMEs occur together, they correlate strongly and we find that the impulsive phase of
the flare coincides with the acceleration phase of the CME.
When they occur separately, which is seldom the case for powerful events, the distinction appears to be in the environment of the energy release, i.e. in the properties of unrelated coronal magnetic structure nearby.

The large-scale wave signatures all point back to the impulsive phase as nearly as we can tell.
This is consistent with the recent finding -- via flare effects on the total solar irradiance -- that the impulsive 
phase contains a large fraction of the flare luminosity.
This recent work on energetics suggests $L_X / L_{tot} \sim$~0.01 for at least the major flares.
This contrasts interestingly with results for stellar flares, for which \cite{2004A&ARv..12...71G} suggests values of 0.5-1 for this ratio.


\bigskip\noindent
{\bf Acknowledgements:}
Figure~\ref{fig:welsch} was kindly provided by Brian Welsch.
The author thanks NASA for support under contract NAS 5-98033 for RHESSI, and the University of Glasgow for hospitality while this paper was written.


\bibliographystyle{aps-nameyear}      
\bibliography{hu}   

\begin{thebibliography}{176}
\ifx \bisbn   \undefined \def \bisbn  #1{ISBN #1}\fi
\ifx \binits  \undefined \def \binits#1{#1} \fi
\ifx \bauthor  \undefined \def \bauthor#1{#1} \fi
\ifx \bjtitle  \undefined \def \bjtitle#1{\textrm{#1}}\fi
\ifx \batitle  \undefined \def \batitle#1{#1} \fi
\ifx \bctitle  \undefined \def \bctitle#1{#1} \fi
\ifx \bvolume  \undefined \def \bvolume#1{\textbf{#1}}\fi
\ifx \byear  \undefined \def \byear#1{#1} \fi
\ifx \bissue  \undefined \def \bissue#1{#1} \fi
\ifx \bfpage  \undefined \def \bfpage#1{#1} \fi
\ifx \blpage  \undefined \def \blpage #1{#1} \fi
\ifx \burl  \undefined \def \burl#1{#1} \fi
\ifx \doiurl  \undefined \def \doiurl#1{#1} \fi
\ifx \betal  \undefined \def \betal{et al.} \fi
\ifx \binstitute  \undefined \def \binstitute#1{#1} \fi
\ifx \beditor  \undefined \def \beditor#1{#1} \fi
\ifx \bpublisher  \undefined \def \bpublisher#1{#1} \fi
\ifx \bbtitle  \undefined \def \bbtitle#1{\textit{#1}} \fi
\ifx \bedition  \undefined \def \bedition#1{#1} \fi
\ifx \bseriesno  \undefined \def \bseriesno#1{#1} \fi
\ifx \blocation  \undefined \def \blocation#1{#1} \fi
\ifx \bsertitle  \undefined \def \bsertitle#1{#1} \fi
\ifx \bsnm \undefined \def \bsnm#1{#1} \fi
\ifx \bsuffix \undefined \def \bsuffix#1{#1} \fi
\ifx \bparticle \undefined \def \bparticle#1{#1} \fi
\ifx \barticle \undefined \def \barticle#1{#1} \fi
\ifx \botherref \undefined \def \botherref #1{#1} \fi
\ifx \url \undefined \def \url#1{#1} \fi
\ifx \bchapter \undefined \def \bchapter#1{#1} \fi
\ifx \bbook \undefined \def \bbook#1{#1} \fi
\ifx \bcomment \undefined \def \bcomment#1{#1} \fi
\ifx \oauthor \undefined \def \oauthor#1{#1} \fi
\ifx \citeauthoryear \undefined \def \citeauthoryear#1{#1} \fi
\ifx \texttildelow  \undefined \def \texttildelow{\symbol{126}} \fi
\def \endbibitem {}

\bibitem[\protect\citeauthoryear{{Altschuler} and
  {Newkirk}}{1969}]{1969SoPh....9..131A}
\begin{barticle}
\bauthor{\binits{M.D.} \bsnm{{Altschuler}}}, \bauthor{\binits{G.}
  \bsnm{{Newkirk}}},
\batitle{{Magnetic Fields and the Structure of the Solar Corona. I: Methods of
  Calculating Coronal Fields}}.
\bjtitle{\solphys}
\bvolume{9},
\bfpage{131}--\blpage{149}
(\byear{1969}).
doi:\doiurl{10.1007/BF00145734}
\end{barticle}
\endbibitem

\bibitem[\protect\citeauthoryear{{Aly}}{1984}]{1984ApJ...283..349A}
\begin{barticle}
\bauthor{\binits{J.J.} \bsnm{{Aly}}},
\batitle{{On some properties of force-free magnetic fields in infinite regions
  of space}}.
\bjtitle{\apj}
\bvolume{283},
\bfpage{349}--\blpage{362}
(\byear{1984}).
doi:\doiurl{10.1086/162313}
\end{barticle}
\endbibitem

\bibitem[\protect\citeauthoryear{{Asai} et~al.}{2004}]{2004ApJ...605L..77A}
\begin{barticle}
\bauthor{\binits{A.} \bsnm{{Asai}}}, \bauthor{\binits{T.} \bsnm{{Yokoyama}}},
  \bauthor{\binits{M.} \bsnm{{Shimojo}}}, \bauthor{\binits{K.}
  \bsnm{{Shibata}}},
\batitle{{Downflow Motions Associated with Impulsive Nonthermal Emissions
  Observed in the 2002 July 23 Solar Flare}}.
\bjtitle{\apjl}
\bvolume{605},
\bfpage{77}--\blpage{80}
(\byear{2004}).
doi:\doiurl{10.1086/420768}
\end{barticle}
\endbibitem

\bibitem[\protect\citeauthoryear{{Athay} and
  {Moreton}}{1961}]{1961ApJ...133..935A}
\begin{barticle}
\bauthor{\binits{R.G.} \bsnm{{Athay}}}, \bauthor{\binits{G.E.}
  \bsnm{{Moreton}}},
\batitle{{Impulsive Phenomena of the Solar Atmosphere. I. Some Optical Events
  Associated with Flares Showing Explosive Phase.}}
\bjtitle{\apj}
\bvolume{133},
\bfpage{935}
(\byear{1961}).
doi:\doiurl{10.1086/147098}
\end{barticle}
\endbibitem

\bibitem[\protect\citeauthoryear{{Bain} and
  {Fletcher}}{2009}]{2009A&A...508.1443B}
\begin{barticle}
\bauthor{\binits{H.M.} \bsnm{{Bain}}}, \bauthor{\binits{L.} \bsnm{{Fletcher}}},
\batitle{{Hard X-ray emission from a flare-related jet}}.
\bjtitle{\aap}
\bvolume{508},
\bfpage{1443}--\blpage{1452}
(\byear{2009}).
doi:\doiurl{10.1051/0004-6361/200911876}
\end{barticle}
\endbibitem

\bibitem[\protect\citeauthoryear{{Balasubramaniam} and {and 14
  co-authors}}{2010}]{bala}
\begin{barticle}
\bauthor{\binits{K.S.} \bsnm{{Balasubramaniam}}}, \bauthor{\bsnm{{and 14
  co-authors}}},
\batitle{{On the origin of the solar Moreton wave of 2006 December 6}}.
\bjtitle{\apj}
\bvolume{723},
\bfpage{587}--\blpage{601}
(\byear{2010})
\end{barticle}
\endbibitem

\bibitem[\protect\citeauthoryear{{Bat\-ta\-glia}
  et~al.}{2009}]{2009A&A...498..891B}
\begin{barticle}
\bauthor{\binits{M.} \bsnm{{Bat\-ta\-glia}}}, \bauthor{\binits{L.}
  \bsnm{{Fletcher}}}, \bauthor{\binits{A.O.} \bsnm{{Benz}}},
\batitle{{Observations of conduction driven evaporation in the early rise phase
  of solar flares}}.
\bjtitle{\aap}
\bvolume{498},
\bfpage{891}--\blpage{900}
(\byear{2009}).
doi:\doiurl{10.1051/0004-6361/200811196}
\end{barticle}
\endbibitem

\bibitem[\protect\citeauthoryear{{Biesecker}
  et~al.}{2002}]{2002ApJ...569.1009B}
\begin{barticle}
\bauthor{\binits{D.A.} \bsnm{{Biesecker}}}, \bauthor{\binits{D.C.}
  \bsnm{{Myers}}}, \bauthor{\binits{B.J.} \bsnm{{Thompson}}},
  \bauthor{\binits{D.M.} \bsnm{{Hammer}}}, \bauthor{\binits{A.}
  \bsnm{{Vourlidas}}},
\batitle{{Solar Phenomena Associated with ``EIT Waves''}}.
\bjtitle{\apj}
\bvolume{569},
\bfpage{1009}--\blpage{1015}
(\byear{2002}).
doi:\doiurl{10.1086/339402}
\end{barticle}
\endbibitem

\bibitem[\protect\citeauthoryear{{Bleybel} et~al.}{2002}]{2002A&A...395..685B}
\begin{barticle}
\bauthor{\binits{A.} \bsnm{{Bleybel}}}, \bauthor{\binits{T.} \bsnm{{Amari}}},
  \bauthor{\binits{L.} \bsnm{{van Driel-Gesztelyi}}}, \bauthor{\binits{K.D.}
  \bsnm{{Leka}}},
\batitle{{Global budget for an eruptive active region . I. Equilibrium
  reconstruction approach}}.
\bjtitle{\aap}
\bvolume{395},
\bfpage{685}--\blpage{695}
(\byear{2002}).
doi:\doiurl{10.1051/0004-6361:20021332}
\end{barticle}
\endbibitem

\bibitem[\protect\citeauthoryear{{Bougeret} and {and 42
  co-authors}}{2008}]{2008SSRv..136..487B}
\begin{barticle}
\bauthor{\binits{J.L.} \bsnm{{Bougeret}}}, \bauthor{\bsnm{{and 42
  co-authors}}},
\batitle{{S/WAVES: The Radio and Plasma Wave Investigation on the STEREO
  Mission}}.
\bjtitle{Space Science Reviews}
\bvolume{136},
\bfpage{487}--\blpage{528}
(\byear{2008}).
doi:\doiurl{10.1007/s11214-007-9298-8}
\end{barticle}
\endbibitem

\bibitem[\protect\citeauthoryear{{Brown}}{1971}]{1971SoPh...18..489B}
\begin{barticle}
\bauthor{\binits{J.C.} \bsnm{{Brown}}},
\batitle{{The Deduction of Energy Spectra of Non-Thermal Electrons in Flares
  from the Observed Dynamic Spectra of Hard X-Ray Bursts}}.
\bjtitle{\solphys}
\bvolume{18},
\bfpage{489}--\blpage{502}
(\byear{1971}).
doi:\doiurl{10.1007/BF00149070}
\end{barticle}
\endbibitem

\bibitem[\protect\citeauthoryear{{Bruzek}}{1964}]{1964ApJ...140..746B}
\begin{barticle}
\bauthor{\binits{A.} \bsnm{{Bruzek}}},
\batitle{{On the Association Between Loop Prominences and Flares.}}
\bjtitle{\apj}
\bvolume{140},
\bfpage{746}
(\byear{1964}).
doi:\doiurl{10.1086/147969}
\end{barticle}
\endbibitem

\bibitem[\protect\citeauthoryear{{Burkepile}
  et~al.}{2004}]{2004JGRA..10903103B}
\begin{barticle}
\bauthor{\binits{J.T.} \bsnm{{Burkepile}}}, \bauthor{\binits{A.J.}
  \bsnm{{Hundhausen}}}, \bauthor{\binits{A.L.} \bsnm{{Stanger}}},
  \bauthor{\binits{O.C.} \bsnm{{St.~Cyr}}}, \bauthor{\binits{J.A.}
  \bsnm{{Seiden}}},
\batitle{{Role of projection effects on solar coronal mass ejection properties:
  1. A study of CMEs associated with limb activity}}.
\bjtitle{Journal of Geophysical Research (Space Physics)}
\bvolume{109},
\bfpage{3103}
(\byear{2004}).
doi:\doiurl{10.1029/2003JA010149}
\end{barticle}
\endbibitem

\bibitem[\protect\citeauthoryear{{Canfield} et~al.}{1987}]{1987Natur.326..165C}
\begin{barticle}
\bauthor{\binits{R.C.} \bsnm{{Canfield}}}, \bauthor{\binits{T.R.}
  \bsnm{{Metcalf}}}, \bauthor{\binits{K.T.} \bsnm{{Strong}}},
  \bauthor{\binits{D.M.} \bsnm{{Zarro}}},
\batitle{{A novel observational test of momentum balance in a solar flare}}.
\bjtitle{\nat}
\bvolume{326},
\bfpage{165}
(\byear{1987}).
doi:\doiurl{10.1038/326165a0}
\end{barticle}
\endbibitem

\bibitem[\protect\citeauthoryear{{Cargill}}{1994}]{1994ApJ...422..381C}
\begin{barticle}
\bauthor{\binits{P.J.} \bsnm{{Cargill}}},
\batitle{{Some implications of the nanoflare concept}}.
\bjtitle{\apj}
\bvolume{422},
\bfpage{381}--\blpage{393}
(\byear{1994}).
doi:\doiurl{10.1086/173733}
\end{barticle}
\endbibitem

\bibitem[\protect\citeauthoryear{{Cargill} and
  {Priest}}{1983}]{1983ApJ...266..383C}
\begin{barticle}
\bauthor{\binits{P.J.} \bsnm{{Cargill}}}, \bauthor{\binits{E.R.}
  \bsnm{{Priest}}},
\batitle{{The heating of postflare loops}}.
\bjtitle{\apj}
\bvolume{266},
\bfpage{383}--\blpage{389}
(\byear{1983}).
doi:\doiurl{10.1086/160786}
\end{barticle}
\endbibitem

\bibitem[\protect\citeauthoryear{{Carmichael}}{1964}]{1964NASSP..50..451C}
\begin{barticle}
\bauthor{\binits{H.} \bsnm{{Carmichael}}},
\batitle{{A Process for Flares}}.
\bjtitle{NASA Special Publication}
\bvolume{50},
\bfpage{451}
(\byear{1964})
\end{barticle}
\endbibitem

\bibitem[\protect\citeauthoryear{{Carrington}}{1859}]{1859MNRAs..20...13C}
\begin{barticle}
\bauthor{\binits{R.C.} \bsnm{{Carrington}}},
\batitle{{Description of a Singular Appearance seen in the Sun on September 1,
  1859}}.
\bjtitle{\mnras}
\bvolume{20},
\bfpage{13}--\blpage{16}
(\byear{1859})
\end{barticle}
\endbibitem

\bibitem[\protect\citeauthoryear{{Chapman} and
  {Bartels}}{1940}]{chapman-bartels}
\begin{botherref}
\oauthor{\binits{S.} \bsnm{{Chapman}}}, \oauthor{\binits{J.} \bsnm{{Bartels}}},
\textit{{Geomagnetism}}
1940
\end{botherref}
\endbibitem

\bibitem[\protect\citeauthoryear{{Chifor} et~al.}{2007}]{2007A&A...472..967C}
\begin{barticle}
\bauthor{\binits{C.} \bsnm{{Chifor}}}, \bauthor{\binits{D.} \bsnm{{Tripathi}}},
  \bauthor{\binits{H.E.} \bsnm{{Mason}}}, \bauthor{\binits{B.R.}
  \bsnm{{Dennis}}},
\batitle{{X-ray precursors to flares and filament eruptions}}.
\bjtitle{\aap}
\bvolume{472},
\bfpage{967}--\blpage{979}
(\byear{2007}).
doi:\doiurl{10.1051/0004-6361:20077771}
\end{barticle}
\endbibitem

\bibitem[\protect\citeauthoryear{{Chupp} et~al.}{1987}]{1987ApJ...318..913C}
\begin{barticle}
\bauthor{\binits{E.L.} \bsnm{{Chupp}}}, \bauthor{\binits{H.}
  \bsnm{{Debrunner}}}, \bauthor{\binits{E.} \bsnm{{Flueckiger}}},
  \bauthor{\binits{D.J.} \bsnm{{Forrest}}}, \bauthor{\binits{F.}
  \bsnm{{Golliez}}}, \bauthor{\binits{G.} \bsnm{{Kanbach}}},
  \bauthor{\binits{W.T.} \bsnm{{Vestrand}}}, \bauthor{\binits{J.}
  \bsnm{{Cooper}}}, \bauthor{\binits{G.} \bsnm{{Share}}},
\batitle{{Solar neutron emissivity during the large flare on 1982 June 3}}.
\bjtitle{\apj}
\bvolume{318},
\bfpage{913}--\blpage{925}
(\byear{1987}).
doi:\doiurl{10.1086/165423}
\end{barticle}
\endbibitem

\bibitem[\protect\citeauthoryear{{Cliver} et~al.}{1994}]{1994ApJ...426..767C}
\begin{barticle}
\bauthor{\binits{E.W.} \bsnm{{Cliver}}}, \bauthor{\binits{N.B.}
  \bsnm{{Crosby}}}, \bauthor{\binits{B.R.} \bsnm{{Dennis}}},
\batitle{{Are solar gamma-ray-line flares different from other large flares?}}
\bjtitle{\apj}
\bvolume{426},
\bfpage{767}--\blpage{773}
(\byear{1994}).
doi:\doiurl{10.1086/174113}
\end{barticle}
\endbibitem

\bibitem[\protect\citeauthoryear{{Cliver} et~al.}{1999}]{1999SoPh..187...89C}
\begin{barticle}
\bauthor{\binits{E.W.} \bsnm{{Cliver}}}, \bauthor{\binits{D.F.} \bsnm{{Webb}}},
  \bauthor{\binits{R.A.} \bsnm{{Howard}}},
\batitle{{On the origin of solar metric type II bursts}}.
\bjtitle{\solphys}
\bvolume{187},
\bfpage{89}--\blpage{114}
(\byear{1999})
\end{barticle}
\endbibitem

\bibitem[\protect\citeauthoryear{{Cliver} et~al.}{1986}]{1986ApJ...305..920C}
\begin{barticle}
\bauthor{\binits{E.W.} \bsnm{{Cliver}}}, \bauthor{\binits{B.R.}
  \bsnm{{Dennis}}}, \bauthor{\binits{A.L.} \bsnm{{Kiplinger}}},
  \bauthor{\binits{S.R.} \bsnm{{Kane}}}, \bauthor{\binits{D.F.}
  \bsnm{{Neidig}}}, \bauthor{\binits{N.R.} \bsnm{{Sheeley}} \bsuffix{Jr.}},
  \bauthor{\binits{M.J.} \bsnm{{Koomen}}},
\batitle{{Solar gradual hard X-ray bursts and associated phenomena}}.
\bjtitle{\apj}
\bvolume{305},
\bfpage{920}--\blpage{935}
(\byear{1986}).
doi:\doiurl{10.1086/164306}
\end{barticle}
\endbibitem

\bibitem[\protect\citeauthoryear{{Colgate}}{1978}]{1978ApJ...221.1068C}
\begin{barticle}
\bauthor{\binits{S.A.} \bsnm{{Colgate}}},
\batitle{{A phenomenological model of solar flares}}.
\bjtitle{\apj}
\bvolume{221},
\bfpage{1068}--\blpage{1083}
(\byear{1978}).
doi:\doiurl{10.1086/156111}
\end{barticle}
\endbibitem

\bibitem[\protect\citeauthoryear{{Cox}}{2000}]{2000asqu.book.....C}
\begin{botherref}
\oauthor{\binits{A.N.} \bsnm{{Cox}}},
\textit{{Allen's astrophysical quantities}}
2000
\end{botherref}
\endbibitem

\bibitem[\protect\citeauthoryear{{De Rosa} and {and 18
  co-authors}}{2009}]{2009ApJ...696.1780D}
\begin{barticle}
\bauthor{\binits{M.L.} \bsnm{{De Rosa}}}, \bauthor{\bsnm{{and 18 co-authors}}},
\batitle{{A Critical Assessment of Nonlinear Force-Free Field Modeling of the
  Solar Corona for Active Region 10953}}.
\bjtitle{\apj}
\bvolume{696},
\bfpage{1780}--\blpage{1791}
(\byear{2009}).
doi:\doiurl{10.1088/0004-637X/696/2/1780}
\end{barticle}
\endbibitem

\bibitem[\protect\citeauthoryear{{Delann{\'e}e}}{2000}]{2000ApJ...545..512D}
\begin{barticle}
\bauthor{\binits{C.} \bsnm{{Delann{\'e}e}}},
\batitle{{Another View of the EIT Wave Phenomenon}}.
\bjtitle{\apj}
\bvolume{545},
\bfpage{512}--\blpage{523}
(\byear{2000}).
doi:\doiurl{10.1086/317777}
\end{barticle}
\endbibitem

\bibitem[\protect\citeauthoryear{{Dennis} and
  {Zarro}}{1993}]{1993SoPh..146..177D}
\begin{barticle}
\bauthor{\binits{B.R.} \bsnm{{Dennis}}}, \bauthor{\binits{D.M.}
  \bsnm{{Zarro}}},
\batitle{{The Neupert effect - What can it tell us about the impulsive and
  gradual phases of solar flares?}}
\bjtitle{\solphys}
\bvolume{146},
\bfpage{177}--\blpage{190}
(\byear{1993}).
doi:\doiurl{10.1007/BF00662178}
\end{barticle}
\endbibitem

\bibitem[\protect\citeauthoryear{{Dere} and {and 33
  co-authors}}{1997}]{1997SoPh..175..601D}
\begin{barticle}
\bauthor{\binits{K.P.} \bsnm{{Dere}}}, \bauthor{\bsnm{{and 33 co-authors}}},
\batitle{{EIT and LASCO Observations of the Initiation of a Coronal Mass
  Ejection}}.
\bjtitle{\solphys}
\bvolume{175},
\bfpage{601}--\blpage{612}
(\byear{1997}).
doi:\doiurl{10.1023/A:1004907307376}
\end{barticle}
\endbibitem

\bibitem[\protect\citeauthoryear{{Dodson} and
  {Hedeman}}{1970}]{1970SoPh...13..401D}
\begin{barticle}
\bauthor{\binits{H.W.} \bsnm{{Dodson}}}, \bauthor{\binits{E.R.}
  \bsnm{{Hedeman}}},
\batitle{{Major H{$\alpha$} Flares in Centers of Activity with very Small or no
  Spots}}.
\bjtitle{\solphys}
\bvolume{13},
\bfpage{401}--\blpage{419}
(\byear{1970}).
doi:\doiurl{10.1007/BF00153560}
\end{barticle}
\endbibitem

\bibitem[\protect\citeauthoryear{{Donea} et~al.}{1999}]{1999ApJ...513L.143D}
\begin{barticle}
\bauthor{\binits{A.} \bsnm{{Donea}}}, \bauthor{\binits{D.C.} \bsnm{{Braun}}},
  \bauthor{\binits{C.} \bsnm{{Lindsey}}},
\batitle{{Seismic Images of a Solar Flare}}.
\bjtitle{\apjl}
\bvolume{513},
\bfpage{143}--\blpage{146}
(\byear{1999}).
doi:\doiurl{10.1086/311915}
\end{barticle}
\endbibitem

\bibitem[\protect\citeauthoryear{{Elliot}}{1969}]{1969sfsr.conf..356E}
\begin{botherref}
\oauthor{\binits{H.} \bsnm{{Elliot}}},
{A possible mechanism for solar flares},
in \textit{Solar Flares and Space Research},
ed. by C. {de Jager}, Z. {Svestka},
1969,
p. 356
\end{botherref}
\endbibitem

\bibitem[\protect\citeauthoryear{{Emslie} et~al.}{2005}]{2005JGRA..11011103E}
\begin{barticle}
\bauthor{\binits{A.G.} \bsnm{{Emslie}}}, \bauthor{\binits{B.R.}
  \bsnm{{Dennis}}}, \bauthor{\binits{G.D.} \bsnm{{Holman}}},
  \bauthor{\binits{H.S.} \bsnm{{Hudson}}},
\batitle{{Refinements to flare energy estimates: A followup to ``Energy
  partition in two solar flare/CME events'' by A. G. Emslie et al.}}
\bjtitle{Journal of Geophysical Research (Space Physics)}
\bvolume{110},
\bfpage{11103}
(\byear{2005}).
doi:\doiurl{10.1029/2005JA011305}
\end{barticle}
\endbibitem

\bibitem[\protect\citeauthoryear{{Erickson} et~al.}{1977}]{1977SoPh...54...57E}
\begin{barticle}
\bauthor{\binits{W.C.} \bsnm{{Erickson}}}, \bauthor{\binits{M.R.}
  \bsnm{{Kundu}}}, \bauthor{\binits{M.J.} \bsnm{{Mahoney}}},
  \bauthor{\binits{T.E.} \bsnm{{Gergely}}},
\batitle{{Determination of the decameter wavelength spectrum of the quiet
  sun}}.
\bjtitle{\solphys}
\bvolume{54},
\bfpage{57}--\blpage{63}
(\byear{1977}).
doi:\doiurl{10.1007/BF00146425}
\end{barticle}
\endbibitem

\bibitem[\protect\citeauthoryear{{F{\'a}rn{\'{\i}}k} and
  {Savy}}{1998}]{1998SoPh..183..339F}
\begin{barticle}
\bauthor{\binits{F.} \bsnm{{F{\'a}rn{\'{\i}}k}}}, \bauthor{\binits{S.K.}
  \bsnm{{Savy}}},
\batitle{{Soft X-Ray Pre-Flare Emission Studied in Yohkoh-SXT Images}}.
\bjtitle{\solphys}
\bvolume{183},
\bfpage{339}--\blpage{357}
(\byear{1998})
\end{barticle}
\endbibitem

\bibitem[\protect\citeauthoryear{{Fisher} and
  {Hawley}}{1990}]{1990ApJ...357..243F}
\begin{barticle}
\bauthor{\binits{G.H.} \bsnm{{Fisher}}}, \bauthor{\binits{S.L.}
  \bsnm{{Hawley}}},
\batitle{{An equation for the evolution of solar and stellar flare loops}}.
\bjtitle{\apj}
\bvolume{357},
\bfpage{243}--\blpage{258}
(\byear{1990}).
doi:\doiurl{10.1086/168911}
\end{barticle}
\endbibitem

\bibitem[\protect\citeauthoryear{{Fletcher} and
  {Hudson}}{2008}]{2008ApJ...675.1645F}
\begin{barticle}
\bauthor{\binits{L.} \bsnm{{Fletcher}}}, \bauthor{\binits{H.S.}
  \bsnm{{Hudson}}},
\batitle{{Impulsive Phase Flare Energy Transport by Large-Scale Alfv{\'e}n
  Waves and the Electron Acceleration Problem}}.
\bjtitle{\apj}
\bvolume{675},
\bfpage{1645}--\blpage{1655}
(\byear{2008}).
doi:\doiurl{10.1086/527044}
\end{barticle}
\endbibitem

\bibitem[\protect\citeauthoryear{{Fletcher} and
  {Warren}}{2003}]{2003LNP...612...58F}
\begin{botherref}
\oauthor{\binits{L.} \bsnm{{Fletcher}}}, \oauthor{\binits{H.P.}
  \bsnm{{Warren}}},
{The Energy Release Process in Solar Flares; Constraints from TRACE
  Observations},
in \textit{Energy Conversion and Particle Acceleration in the Solar Corona},
ed. by {L.~Klein}.
Lecture Notes in Physics, Berlin Springer Verlag,
vol. 612,
2003,
pp. 58--79
\end{botherref}
\endbibitem

\bibitem[\protect\citeauthoryear{{Forbes} and
  {Acton}}{1996}]{1996ApJ...459..330F}
\begin{barticle}
\bauthor{\binits{T.G.} \bsnm{{Forbes}}}, \bauthor{\binits{L.W.}
  \bsnm{{Acton}}},
\batitle{{Reconnection and Field Line Shrinkage in Solar Flares}}.
\bjtitle{\apj}
\bvolume{459},
\bfpage{330}
(\byear{1996}).
doi:\doiurl{10.1086/176896}
\end{barticle}
\endbibitem

\bibitem[\protect\citeauthoryear{{Forbes} et~al.}{2006}]{2006SSRv..123..251F}
\begin{barticle}
\bauthor{\binits{T.G.} \bsnm{{Forbes}}}, \bauthor{\binits{J.A.}
  \bsnm{{Linker}}}, \bauthor{\binits{J.} \bsnm{{Chen}}}, \bauthor{\binits{C.}
  \bsnm{{Cid}}}, \bauthor{\binits{J.} \bsnm{{K{\'o}ta}}},
  \bauthor{\binits{M.A.} \bsnm{{Lee}}}, \bauthor{\binits{G.} \bsnm{{Mann}}},
  \bauthor{\binits{Z.} \bsnm{{Miki{\'c}}}}, \bauthor{\binits{M.S.}
  \bsnm{{Potgieter}}}, \bauthor{\binits{J.M.} \bsnm{{Schmidt}}},
  \bauthor{\binits{G.L.} \bsnm{{Siscoe}}}, \bauthor{\binits{R.}
  \bsnm{{Vainio}}}, \bauthor{\binits{S.K.} \bsnm{{Antiochos}}},
  \bauthor{\binits{P.} \bsnm{{Riley}}},
\batitle{{CME Theory and Models}}.
\bjtitle{Space Science Reviews}
\bvolume{123},
\bfpage{251}--\blpage{302}
(\byear{2006}).
doi:\doiurl{10.1007/s11214-006-9019-8}
\end{barticle}
\endbibitem

\bibitem[\protect\citeauthoryear{{Frost} and
  {Dennis}}{1971}]{1971ApJ...165..655F}
\begin{barticle}
\bauthor{\binits{K.J.} \bsnm{{Frost}}}, \bauthor{\binits{B.R.}
  \bsnm{{Dennis}}},
\batitle{{Evidence from Hard X-Rays for Two-Stage Particle Acceleration in a
  Solar Flare}}.
\bjtitle{\apj}
\bvolume{165},
\bfpage{655}
(\byear{1971}).
doi:\doiurl{10.1086/150932}
\end{barticle}
\endbibitem

\bibitem[\protect\citeauthoryear{{Gallagher}
  et~al.}{2002}]{2002SoPh..210..341G}
\begin{barticle}
\bauthor{\binits{P.T.} \bsnm{{Gallagher}}}, \bauthor{\binits{B.R.}
  \bsnm{{Dennis}}}, \bauthor{\binits{S.} \bsnm{{Krucker}}},
  \bauthor{\binits{R.A.} \bsnm{{Schwartz}}}, \bauthor{\binits{A.K.}
  \bsnm{{Tolbert}}},
\batitle{{Rhessi and Trace Observations of the 21 April 2002 x1.5 Flare}}.
\bjtitle{\solphys}
\bvolume{210},
\bfpage{341}--\blpage{356}
(\byear{2002}).
doi:\doiurl{10.1023/A:1022422019779}
\end{barticle}
\endbibitem

\bibitem[\protect\citeauthoryear{{Gary}}{2001}]{2001SoPh..203...71G}
\begin{barticle}
\bauthor{\binits{G.A.} \bsnm{{Gary}}},
\batitle{{Plasma Beta above a Solar Active Region: Rethinking the Paradigm}}.
\bjtitle{\solphys}
\bvolume{203},
\bfpage{71}--\blpage{86}
(\byear{2001})
\end{barticle}
\endbibitem

\bibitem[\protect\citeauthoryear{{Gilbert} et~al.}{2008}]{2008ApJ...685..629G}
\begin{barticle}
\bauthor{\binits{H.R.} \bsnm{{Gilbert}}}, \bauthor{\binits{A.G.}
  \bsnm{{Daou}}}, \bauthor{\binits{D.} \bsnm{{Young}}}, \bauthor{\binits{D.}
  \bsnm{{Tripathi}}}, \bauthor{\binits{D.} \bsnm{{Alexander}}},
\batitle{{The Filament-Moreton Wave Interaction of 2006 December 6}}.
\bjtitle{\apj}
\bvolume{685},
\bfpage{629}--\blpage{645}
(\byear{2008}).
doi:\doiurl{10.1086/590545}
\end{barticle}
\endbibitem

\bibitem[\protect\citeauthoryear{{Gopalswamy}
  et~al.}{2009}]{2009IAUS..257..283G}
\begin{botherref}
\oauthor{\binits{N.} \bsnm{{Gopalswamy}}}, \oauthor{\binits{S.}
  \bsnm{{Akiyama}}}, \oauthor{\binits{S.} \bsnm{{Yashiro}}},
{Major solar flares without coronal mass ejections},
in \textit{IAU Symposium},
ed. by {N.~Gopalswamy \& D.~F.~Webb}.
IAU Symposium,
vol. 257,
2009,
pp. 283--286.
doi:\doiurl{10.1017/S174392130902941X}
\end{botherref}
\endbibitem

\bibitem[\protect\citeauthoryear{{Gopalswamy}
  et~al.}{2005}]{2005JGRA..11009S15G}
\begin{barticle}
\bauthor{\binits{N.} \bsnm{{Gopalswamy}}}, \bauthor{\binits{S.}
  \bsnm{{Yashiro}}}, \bauthor{\binits{Y.} \bsnm{{Liu}}}, \bauthor{\binits{G.}
  \bsnm{{Michalek}}}, \bauthor{\binits{A.} \bsnm{{Vourlidas}}},
  \bauthor{\binits{M.L.} \bsnm{{Kaiser}}}, \bauthor{\binits{R.A.}
  \bsnm{{Howard}}},
\batitle{{Coronal mass ejections and other extreme characteristics of the 2003
  October-November solar eruptions}}.
\bjtitle{Journal of Geophysical Research (Space Physics)}
\bvolume{110},
\bfpage{9}
(\byear{2005}).
doi:\doiurl{10.1029/2004JA010958}
\end{barticle}
\endbibitem

\bibitem[\protect\citeauthoryear{{Grayson} et~al.}{2009}]{2009ApJ...707.1588G}
\begin{barticle}
\bauthor{\binits{J.A.} \bsnm{{Grayson}}}, \bauthor{\binits{S.}
  \bsnm{{Krucker}}}, \bauthor{\binits{R.P.} \bsnm{{Lin}}},
\batitle{{A Statistical Study of Spectral Hardening in Solar Flares and Related
  Solar Energetic Particle Events}}.
\bjtitle{\apj}
\bvolume{707},
\bfpage{1588}--\blpage{1594}
(\byear{2009}).
doi:\doiurl{10.1088/0004-637X/707/2/1588}
\end{barticle}
\endbibitem

\bibitem[\protect\citeauthoryear{{Grigis} and
  {Benz}}{2004}]{2004A&A...426.1093G}
\begin{barticle}
\bauthor{\binits{P.C.} \bsnm{{Grigis}}}, \bauthor{\binits{A.O.} \bsnm{{Benz}}},
\batitle{{The spectral evolution of impulsive solar X-ray flares}}.
\bjtitle{\aap}
\bvolume{426},
\bfpage{1093}--\blpage{1101}
(\byear{2004}).
doi:\doiurl{10.1051/0004-6361:20041367}
\end{barticle}
\endbibitem

\bibitem[\protect\citeauthoryear{{G{\"u}del}}{2004}]{2004A&ARv..12...71G}
\begin{barticle}
\bauthor{\binits{M.} \bsnm{{G{\"u}del}}},
\batitle{{X-ray astronomy of stellar coronae}}.
\bjtitle{\aapr}
\bvolume{12},
\bfpage{71}--\blpage{237}
(\byear{2004}).
doi:\doiurl{10.1007/s00159-004-0023-2}
\end{barticle}
\endbibitem

\bibitem[\protect\citeauthoryear{{Hanaoka} and {and 19
  co-authors}}{1994}]{1994PASJ...46..205H}
\begin{barticle}
\bauthor{\binits{Y.} \bsnm{{Hanaoka}}}, \bauthor{\bsnm{{and 19 co-authors}}},
\batitle{{Simultaneous observations of a prominence eruption followed by a
  coronal arcade formation in radio, soft X-rays, and H(alpha)}}.
\bjtitle{\pasj}
\bvolume{46},
\bfpage{205}--\blpage{216}
(\byear{1994})
\end{barticle}
\endbibitem

\bibitem[\protect\citeauthoryear{{Hannah} et~al.}{2008}]{2008ApJ...677..704H}
\begin{barticle}
\bauthor{\binits{I.G.} \bsnm{{Hannah}}}, \bauthor{\binits{S.}
  \bsnm{{Christe}}}, \bauthor{\binits{S.} \bsnm{{Krucker}}},
  \bauthor{\binits{G.J.} \bsnm{{Hurford}}}, \bauthor{\binits{H.S.}
  \bsnm{{Hudson}}}, \bauthor{\binits{R.P.} \bsnm{{Lin}}},
\batitle{{RHESSI Microflare Statistics. II. X-Ray Imaging, Spectroscopy, and
  Energy Distributions}}.
\bjtitle{\apj}
\bvolume{677},
\bfpage{704}--\blpage{718}
(\byear{2008}).
doi:\doiurl{10.1086/529012}
\end{barticle}
\endbibitem

\bibitem[\protect\citeauthoryear{{Harrison}}{1995}]{1995A&A...304..585H}
\begin{barticle}
\bauthor{\binits{R.A.} \bsnm{{Harrison}}},
\batitle{{The nature of solar flares associated with coronal mass ejection.}}
\bjtitle{\aap}
\bvolume{304},
\bfpage{585}
(\byear{1995})
\end{barticle}
\endbibitem

\bibitem[\protect\citeauthoryear{{Harvey} et~al.}{1986}]{1986STP.....2..198H}
\begin{barticle}
\bauthor{\binits{K.L.} \bsnm{{Harvey}}}, \bauthor{\binits{N.R.}
  \bsnm{{Sheeley}}}, \bauthor{\binits{J.W.} \bsnm{{Harvey}}},
\batitle{{He I 10830 {\AA} Observations of Two-Ribbon Flare-Like Events
  Associated with Filament Disappearances}}.
\bjtitle{Solar-Terrestrial Predictions, Volume 2nd, Meudon, France, 18-22 June,
  1984.~P.A.~Simon et al, p.~198-203}
\bvolume{2},
\bfpage{198}--\blpage{203}
(\byear{1986})
\end{barticle}
\endbibitem

\bibitem[\protect\citeauthoryear{{Hirayama}}{1974}]{1974SoPh...34..323H}
\begin{barticle}
\bauthor{\binits{T.} \bsnm{{Hirayama}}},
\batitle{{Theoretical Model of Flares and Prominences. I: Evaporating Flare
  Model}}.
\bjtitle{\solphys}
\bvolume{34},
\bfpage{323}--\blpage{338}
(\byear{1974}).
doi:\doiurl{10.1007/BF00153671}
\end{barticle}
\endbibitem

\bibitem[\protect\citeauthoryear{{Hodgson}}{1859}]{1859MNRAS..20...15H}
\begin{barticle}
\bauthor{\binits{R.} \bsnm{{Hodgson}}},
\batitle{{On a curious Appearance seen in the Sun}}.
\bjtitle{\mnras}
\bvolume{20},
\bfpage{15}--\blpage{16}
(\byear{1859})
\end{barticle}
\endbibitem

\bibitem[\protect\citeauthoryear{{Hudson} et~al.}{2010}]{2010arXiv1001.1005H}
\begin{botherref}
\oauthor{\binits{H.} \bsnm{{Hudson}}}, \oauthor{\binits{L.} \bsnm{{Fletcher}}},
  \oauthor{\binits{S.} \bsnm{{Krucker}}},
{The white-light continuum in the impulsive phase of a solar flare}.
\textrm{ArXiv e-prints}
(2010)
\end{botherref}
\endbibitem

\bibitem[\protect\citeauthoryear{{Hudson}}{1972}]{1972SoPh...24..414H}
\begin{barticle}
\bauthor{\binits{H.S.} \bsnm{{Hudson}}},
\batitle{{Thick-Target Processes and White-Light Flares}}.
\bjtitle{\solphys}
\bvolume{24},
\bfpage{414}--\blpage{428}
(\byear{1972}).
doi:\doiurl{10.1007/BF00153384}
\end{barticle}
\endbibitem

\bibitem[\protect\citeauthoryear{{Hudson}}{1978}]{1978ApJ...224..235H}
\begin{barticle}
\bauthor{\binits{H.S.} \bsnm{{Hudson}}},
\batitle{{A purely coronal hard X-ray event}}.
\bjtitle{\apj}
\bvolume{224},
\bfpage{235}--\blpage{240}
(\byear{1978}).
doi:\doiurl{10.1086/156370}
\end{barticle}
\endbibitem

\bibitem[\protect\citeauthoryear{{Hudson}}{1988}]{1988ARA&A..26..473H}
\begin{barticle}
\bauthor{\binits{H.S.} \bsnm{{Hudson}}},
\batitle{{Observed variability of the solar luminosity}}.
\bjtitle{\araa}
\bvolume{26},
\bfpage{473}--\blpage{507}
(\byear{1988}).
doi:\doiurl{10.1146/annurev.aa.26.090188.002353}
\end{barticle}
\endbibitem

\bibitem[\protect\citeauthoryear{{Hudson}}{1991}]{1991SoPh..133..357H}
\begin{barticle}
\bauthor{\binits{H.S.} \bsnm{{Hudson}}},
\batitle{{Solar flares, microflares, nanoflares, and coronal heating}}.
\bjtitle{\solphys}
\bvolume{133},
\bfpage{357}--\blpage{369}
(\byear{1991}).
doi:\doiurl{10.1007/BF00149894}
\end{barticle}
\endbibitem

\bibitem[\protect\citeauthoryear{{Hudson}}{2000}]{2000ApJ...531L..75H}
\begin{barticle}
\bauthor{\binits{H.S.} \bsnm{{Hudson}}},
\batitle{{Implosions in Coronal Transients}}.
\bjtitle{\apjl}
\bvolume{531},
\bfpage{75}--\blpage{77}
(\byear{2000}).
doi:\doiurl{10.1086/312516}
\end{barticle}
\endbibitem

\bibitem[\protect\citeauthoryear{{Hudson} and
  {Cliver}}{2001}]{2001JGR...10625199H}
\begin{barticle}
\bauthor{\binits{H.S.} \bsnm{{Hudson}}}, \bauthor{\binits{E.W.}
  \bsnm{{Cliver}}},
\batitle{{Observing coronal mass ejections without coronagraphs}}.
\bjtitle{\jgr}
\bvolume{106},
\bfpage{25199}--\blpage{25214}
(\byear{2001}).
doi:\doiurl{10.1029/2000JA904026}
\end{barticle}
\endbibitem

\bibitem[\protect\citeauthoryear{{Hudson} et~al.}{1996}]{1996ApJ...470..629H}
\begin{barticle}
\bauthor{\binits{H.S.} \bsnm{{Hudson}}}, \bauthor{\binits{L.W.}
  \bsnm{{Acton}}}, \bauthor{\binits{S.L.} \bsnm{{Freeland}}},
\batitle{{A Long-Duration Solar Flare with Mass Ejection and Global
  Consequences}}.
\bjtitle{\apj}
\bvolume{470},
\bfpage{629}
(\byear{1996}).
doi:\doiurl{10.1086/177894}
\end{barticle}
\endbibitem

\bibitem[\protect\citeauthoryear{{Hudson} et~al.}{2008}]{2008ASPC..383..221H}
\begin{botherref}
\oauthor{\binits{H.S.} \bsnm{{Hudson}}}, \oauthor{\binits{G.H.}
  \bsnm{{Fisher}}}, \oauthor{\binits{B.T.} \bsnm{{Welsch}}},
{Flare Energy and Magnetic Field Variations},
in \textit{Subsurface and Atmospheric Influences on Solar Activity},
ed. by {R.~Howe, R.~W.~Komm, K.~S.~Balasubramaniam, \& G.~J.~D.~Petrie }.
Astronomical Society of the Pacific Conference Series,
vol. 383,
2008,
p. 221
\end{botherref}
\endbibitem

\bibitem[\protect\citeauthoryear{{Hudson} et~al.}{1995}]{1995JGR...100.3473H}
\begin{barticle}
\bauthor{\binits{H.S.} \bsnm{{Hudson}}}, \bauthor{\binits{B.} \bsnm{{Haisch}}},
  \bauthor{\binits{K.T.} \bsnm{{Strong}}},
\batitle{{Comment on 'The solar flare myth' by J. T. Gosling}}.
\bjtitle{\jgr}
\bvolume{100},
\bfpage{3473}--\blpage{3477}
(\byear{1995}).
doi:\doiurl{10.1029/94JA02710}
\end{barticle}
\endbibitem

\bibitem[\protect\citeauthoryear{{Hudson} et~al.}{2006}]{2006SoPh..234...79H}
\begin{barticle}
\bauthor{\binits{H.S.} \bsnm{{Hudson}}}, \bauthor{\binits{C.J.}
  \bsnm{{Wolfson}}}, \bauthor{\binits{T.R.} \bsnm{{Metcalf}}},
\batitle{{White-Light Flares: A TRACE/RHESSI Overview}}.
\bjtitle{\solphys}
\bvolume{234},
\bfpage{79}--\blpage{93}
(\byear{2006}).
doi:\doiurl{10.1007/s11207-006-0056-y}
\end{barticle}
\endbibitem

\bibitem[\protect\citeauthoryear{{Hudson} et~al.}{1992}]{1992PASJ...44L..77H}
\begin{barticle}
\bauthor{\binits{H.S.} \bsnm{{Hudson}}}, \bauthor{\binits{L.W.}
  \bsnm{{Acton}}}, \bauthor{\binits{T.} \bsnm{{Hirayama}}},
  \bauthor{\binits{Y.} \bsnm{{Uchida}}},
\batitle{{White-light flares observed by YOHKOH}}.
\bjtitle{\pasj}
\bvolume{44},
\bfpage{77}--\blpage{81}
(\byear{1992})
\end{barticle}
\endbibitem

\bibitem[\protect\citeauthoryear{{Hudson} et~al.}{1994}]{1994ApJ...422L..25H}
\begin{barticle}
\bauthor{\binits{H.S.} \bsnm{{Hudson}}}, \bauthor{\binits{K.T.}
  \bsnm{{Strong}}}, \bauthor{\binits{B.R.} \bsnm{{Dennis}}},
  \bauthor{\binits{D.} \bsnm{{Zarro}}}, \bauthor{\binits{M.} \bsnm{{Inda}}},
  \bauthor{\binits{T.} \bsnm{{Kosugi}}}, \bauthor{\binits{T.} \bsnm{{Sakao}}},
\batitle{{Impulsive behavior in solar soft X-radiation}}.
\bjtitle{\apjl}
\bvolume{422},
\bfpage{25}--\blpage{27}
(\byear{1994}).
doi:\doiurl{10.1086/187203}
\end{barticle}
\endbibitem

\bibitem[\protect\citeauthoryear{{Hudson} et~al.}{1998}]{1998GeoRL..25.2481H}
\begin{barticle}
\bauthor{\binits{H.S.} \bsnm{{Hudson}}}, \bauthor{\binits{J.R.}
  \bsnm{{Lemen}}}, \bauthor{\binits{O.C.} \bsnm{{St.~Cyr}}},
  \bauthor{\binits{A.C.} \bsnm{{Sterling}}}, \bauthor{\binits{D.F.}
  \bsnm{{Webb}}},
\batitle{{X-ray coronal changes during halo CMEs}}.
\bjtitle{\grl}
\bvolume{25},
\bfpage{2481}--\blpage{2484}
(\byear{1998}).
doi:\doiurl{10.1029/98GL01303}
\end{barticle}
\endbibitem

\bibitem[\protect\citeauthoryear{{Hudson} et~al.}{2001}]{2001ApJ...561L.211H}
\begin{barticle}
\bauthor{\binits{H.S.} \bsnm{{Hudson}}}, \bauthor{\binits{T.} \bsnm{{Kosugi}}},
  \bauthor{\binits{N.V.} \bsnm{{Nitta}}}, \bauthor{\binits{M.}
  \bsnm{{Shimojo}}},
\batitle{{Hard X-Radiation from a Fast Coronal Ejection}}.
\bjtitle{\apjl}
\bvolume{561},
\bfpage{211}--\blpage{214}
(\byear{2001}).
doi:\doiurl{10.1086/324760}
\end{barticle}
\endbibitem

\bibitem[\protect\citeauthoryear{{Hudson} et~al.}{2003}]{2003SoPh..212..121H}
\begin{barticle}
\bauthor{\binits{H.S.} \bsnm{{Hudson}}}, \bauthor{\binits{J.I.} \bsnm{{Khan}}},
  \bauthor{\binits{J.R.} \bsnm{{Lemen}}}, \bauthor{\binits{N.V.}
  \bsnm{{Nitta}}}, \bauthor{\binits{Y.} \bsnm{{Uchida}}},
\batitle{{Soft X-ray observation of a large-scale coronal wave and its
  exciter}}.
\bjtitle{\solphys}
\bvolume{212},
\bfpage{121}--\blpage{149}
(\byear{2003}).
doi:\doiurl{10.1023/A:1022904125479}
\end{barticle}
\endbibitem

\bibitem[\protect\citeauthoryear{{Hudson} et~al.}{2009}]{2009ApJ...698L..86H}
\begin{barticle}
\bauthor{\binits{H.S.} \bsnm{{Hudson}}}, \bauthor{\binits{A.L.}
  \bsnm{{MacKinnon}}}, \bauthor{\binits{M.L.} \bsnm{{De Rosa}}},
  \bauthor{\binits{S.F.N.} \bsnm{{Frewen}}},
\batitle{{Coronal Radiation Belts}}.
\bjtitle{\apjl}
\bvolume{698},
\bfpage{86}--\blpage{89}
(\byear{2009}).
doi:\doiurl{10.1088/0004-637X/698/2/L86}
\end{barticle}
\endbibitem

\bibitem[\protect\citeauthoryear{{Hurford} et~al.}{2003}]{2003ApJ...595L..77H}
\begin{barticle}
\bauthor{\binits{G.J.} \bsnm{{Hurford}}}, \bauthor{\binits{R.A.}
  \bsnm{{Schwartz}}}, \bauthor{\binits{S.} \bsnm{{Krucker}}},
  \bauthor{\binits{R.P.} \bsnm{{Lin}}}, \bauthor{\binits{D.M.} \bsnm{{Smith}}},
  \bauthor{\binits{N.} \bsnm{{Vilmer}}},
\batitle{{First Gamma-Ray Images of a Solar Flare}}.
\bjtitle{\apjl}
\bvolume{595},
\bfpage{77}--\blpage{80}
(\byear{2003}).
doi:\doiurl{10.1086/378179}
\end{barticle}
\endbibitem

\bibitem[\protect\citeauthoryear{{Jess} et~al.}{2008}]{2008ApJ...688L.119J}
\begin{barticle}
\bauthor{\binits{D.B.} \bsnm{{Jess}}}, \bauthor{\binits{M.}
  \bsnm{{Mathioudakis}}}, \bauthor{\binits{P.J.} \bsnm{{Crockett}}},
  \bauthor{\binits{F.P.} \bsnm{{Keenan}}},
\batitle{{Do All Flares Have White-Light Emission?}}
\bjtitle{\apjl}
\bvolume{688},
\bfpage{119}--\blpage{122}
(\byear{2008}).
doi:\doiurl{10.1086/595588}
\end{barticle}
\endbibitem

\bibitem[\protect\citeauthoryear{{Ji} et~al.}{2007}]{2007ApJ...660..893J}
\begin{barticle}
\bauthor{\binits{H.} \bsnm{{Ji}}}, \bauthor{\binits{G.} \bsnm{{Huang}}},
  \bauthor{\binits{H.} \bsnm{{Wang}}},
\batitle{{The Relaxation of Sheared Magnetic Fields: A Contracting Process}}.
\bjtitle{\apj}
\bvolume{660},
\bfpage{893}--\blpage{900}
(\byear{2007}).
doi:\doiurl{10.1086/513017}
\end{barticle}
\endbibitem

\bibitem[\protect\citeauthoryear{{Kahler}}{1982}]{1992ARAA...30..113K}
\begin{barticle}
\bauthor{\binits{S.W.} \bsnm{{Kahler}}},
\batitle{{Solar flares and coronal mass ejections}}.
\bjtitle{\araa}
\bvolume{30},
\bfpage{113}--\blpage{141}
(\byear{1982})
\end{barticle}
\endbibitem

\bibitem[\protect\citeauthoryear{{Kahler}}{1982}]{1982JGR....87.3439K}
\begin{barticle}
\bauthor{\binits{S.W.} \bsnm{{Kahler}}},
\batitle{{The role of the big flare syndrome in correlations of solar energetic
  proton fluxes and associated microwave burst parameters}}.
\bjtitle{\jgr}
\bvolume{87},
\bfpage{3439}--\blpage{3448}
(\byear{1982}).
doi:\doiurl{10.1029/JA087iA05p03439}
\end{barticle}
\endbibitem

\bibitem[\protect\citeauthoryear{{Kanbach} et~al.}{1993}]{1993A&AS...97..349K}
\begin{barticle}
\bauthor{\binits{G.} \bsnm{{Kanbach}}}, \bauthor{\binits{D.L.}
  \bsnm{{Bertsch}}}, \bauthor{\binits{C.E.} \bsnm{{Fichtel}}},
  \bauthor{\binits{R.C.} \bsnm{{Hartman}}}, \bauthor{\binits{S.D.}
  \bsnm{{Hunter}}}, \bauthor{\binits{D.A.} \bsnm{{Kniffen}}},
  \bauthor{\binits{P.W.} \bsnm{{Kwok}}}, \bauthor{\binits{Y.C.} \bsnm{{Lin}}},
  \bauthor{\binits{J.R.} \bsnm{{Mattox}}}, \bauthor{\binits{H.A.}
  \bsnm{{Mayer-Hasselwander}}},
\batitle{{Detection of a long-duration solar gamma-ray flare on June 11, 1991
  with EGRET on COMPTON-GRO}}.
\bjtitle{\aaps}
\bvolume{97},
\bfpage{349}--\blpage{353}
(\byear{1993})
\end{barticle}
\endbibitem

\bibitem[\protect\citeauthoryear{{Kane} and
  {Anderson}}{1970}]{1970ApJ...162.1003K}
\begin{barticle}
\bauthor{\binits{S.R.} \bsnm{{Kane}}}, \bauthor{\binits{K.A.}
  \bsnm{{Anderson}}},
\batitle{{Spectral Characteristics of Impulsive Solar-Flare X-Rays $>$10 keV}}.
\bjtitle{\apj}
\bvolume{162},
\bfpage{1003}
(\byear{1970}).
doi:\doiurl{10.1086/150732}
\end{barticle}
\endbibitem

\bibitem[\protect\citeauthoryear{{Kane} and
  {Donnelly}}{1971}]{1971ApJ...164..151K}
\begin{barticle}
\bauthor{\binits{S.R.} \bsnm{{Kane}}}, \bauthor{\binits{R.F.}
  \bsnm{{Donnelly}}},
\batitle{{Impulsive Hard X-Ray and Ultraviolet Emission during Solar Flares}}.
\bjtitle{\apj}
\bvolume{164},
\bfpage{151}
(\byear{1971}).
doi:\doiurl{10.1086/150826}
\end{barticle}
\endbibitem

\bibitem[\protect\citeauthoryear{{Kaufmann} et~al.}{2002}]{2002ApJ...574.1059K}
\begin{barticle}
\bauthor{\binits{P.} \bsnm{{Kaufmann}}}, \bauthor{\binits{J.} \bsnm{{Raulin}}},
  \bauthor{\binits{A.M.} \bsnm{{Melo}}}, \bauthor{\binits{E.}
  \bsnm{{Correia}}}, \bauthor{\binits{J.E.R.} \bsnm{{Costa}}},
  \bauthor{\binits{C.G.G.} \bsnm{{de Castro}}}, \bauthor{\binits{A.V.R.}
  \bsnm{{Silva}}}, \bauthor{\binits{M.} \bsnm{{Yoshimori}}},
  \bauthor{\binits{H.S.} \bsnm{{Hudson}}}, \bauthor{\binits{W.Q.}
  \bsnm{{Gan}}}, \bauthor{\binits{D.E.} \bsnm{{Gary}}}, \bauthor{\binits{P.T.}
  \bsnm{{Gallagher}}}, \bauthor{\binits{H.} \bsnm{{Levato}}},
  \bauthor{\binits{A.} \bsnm{{Marun}}}, \bauthor{\binits{M.} \bsnm{{Rovira}}},
\batitle{{Solar Submillimeter and Gamma-Ray Burst Emission}}.
\bjtitle{\apj}
\bvolume{574},
\bfpage{1059}--\blpage{1065}
(\byear{2002}).
doi:\doiurl{10.1086/341061}
\end{barticle}
\endbibitem

\bibitem[\protect\citeauthoryear{{Khan} and
  {Aurass}}{2002}]{2002A&A...383.1018K}
\begin{barticle}
\bauthor{\binits{J.I.} \bsnm{{Khan}}}, \bauthor{\binits{H.} \bsnm{{Aurass}}},
\batitle{{X-ray observations of a large-scale solar coronal shock wave}}.
\bjtitle{\aap}
\bvolume{383},
\bfpage{1018}--\blpage{1031}
(\byear{2002}).
doi:\doiurl{10.1051/0004-6361:20011707}
\end{barticle}
\endbibitem

\bibitem[\protect\citeauthoryear{{Kiplinger}}{1995}]{1995ApJ...453..973K}
\begin{barticle}
\bauthor{\binits{A.L.} \bsnm{{Kiplinger}}},
\batitle{{Comparative Studies of Hard X-Ray Spectral Evolution in Solar Flares
  with High-Energy Proton Events Observed at Earth}}.
\bjtitle{\apj}
\bvolume{453},
\bfpage{973}
(\byear{1995}).
doi:\doiurl{10.1086/176457}
\end{barticle}
\endbibitem

\bibitem[\protect\citeauthoryear{{Klimchuk}}{2001}]{2001AGUGM.125..143K}
\begin{barticle}
\bauthor{\binits{J.A.} \bsnm{{Klimchuk}}},
\batitle{{Theory of Coronal Mass Ejections}}.
\bjtitle{Space Weather (Geophysical Monograph 125), ed.~P.~Song, H.~Singer,
  G.~Siscoe (Washington: Am.~Geophys.~Un.), 143 (2001)}
\bvolume{125},
\bfpage{143}
(\byear{2001})
\end{barticle}
\endbibitem

\bibitem[\protect\citeauthoryear{{Kosovichev}}{2006}]{2006SoPh..238....1K}
\begin{barticle}
\bauthor{\binits{A.G.} \bsnm{{Kosovichev}}},
\batitle{{Properties of Flares-Generated Seismic Waves on the Sun}}.
\bjtitle{\solphys}
\bvolume{238},
\bfpage{1}--\blpage{11}
(\byear{2006}).
doi:\doiurl{10.1007/s11207-006-0190-6}
\end{barticle}
\endbibitem

\bibitem[\protect\citeauthoryear{{Kosovichev} and
  {Zharkova}}{1995}]{1995ESASP.376b.341K}
\begin{botherref}
\oauthor{\binits{A.G.} \bsnm{{Kosovichev}}}, \oauthor{\binits{V.V.}
  \bsnm{{Zharkova}}},
{Seismic Response to Solar Flares: Theoretical Predictions},
in \textit{Helioseismology}.
ESA Special Publication,
vol. 376,
1995,
p. 341
\end{botherref}
\endbibitem

\bibitem[\protect\citeauthoryear{{Kosovichev} and
  {Zharkova}}{1998}]{1998Natur.393..317K}
\begin{barticle}
\bauthor{\binits{A.G.} \bsnm{{Kosovichev}}}, \bauthor{\binits{V.V.}
  \bsnm{{Zharkova}}},
\batitle{{X-ray flare sparks quake inside Sun}}.
\bjtitle{\nat}
\bvolume{393},
\bfpage{317}--\blpage{318}
(\byear{1998}).
doi:\doiurl{10.1038/30629}
\end{barticle}
\endbibitem

\bibitem[\protect\citeauthoryear{{Kosovichev} and
  {Zharkova}}{2001}]{2001ApJ...550L.105K}
\begin{barticle}
\bauthor{\binits{A.G.} \bsnm{{Kosovichev}}}, \bauthor{\binits{V.V.}
  \bsnm{{Zharkova}}},
\batitle{{Magnetic Energy Release and Transients in the Solar Flare of 2000
  July 14}}.
\bjtitle{\apjl}
\bvolume{550},
\bfpage{105}--\blpage{108}
(\byear{2001}).
doi:\doiurl{10.1086/319484}
\end{barticle}
\endbibitem

\bibitem[\protect\citeauthoryear{{Kostiuk} and
  {Pikel'ner}}{1974}]{1974AZh....51.1002K}
\begin{barticle}
\bauthor{\binits{N.D.} \bsnm{{Kostiuk}}}, \bauthor{\binits{S.B.}
  \bsnm{{Pikel'ner}}},
\batitle{{Gas dynamics of a flare region heated by a flux of high-velocity
  electrons}}.
\bjtitle{\azh}
\bvolume{51},
\bfpage{1002}--\blpage{1016}
(\byear{1974})
\end{barticle}
\endbibitem

\bibitem[\protect\citeauthoryear{{Kretzschmar}
  et~al.}{2010}]{2010NatPh...6..690K}
\begin{barticle}
\bauthor{\binits{M.} \bsnm{{Kretzschmar}}}, \bauthor{\binits{T.D.} \bsnm{{de
  Wit}}}, \bauthor{\binits{W.} \bsnm{{Schmutz}}}, \bauthor{\binits{S.}
  \bsnm{{Mekaoui}}}, \bauthor{\binits{J.} \bsnm{{Hochedez}}},
  \bauthor{\binits{S.} \bsnm{{Dewitte}}},
\batitle{{The effect of flares on total solar irradiance}}.
\bjtitle{Nature Physics}
\bvolume{6},
\bfpage{690}--\blpage{692}
(\byear{2010}).
doi:\doiurl{10.1038/nphys1741}
\end{barticle}
\endbibitem

\bibitem[\protect\citeauthoryear{{Krucker} et~al.}{2007}]{2007ApJ...669L..49K}
\begin{barticle}
\bauthor{\binits{S.} \bsnm{{Krucker}}}, \bauthor{\binits{S.M.} \bsnm{{White}}},
  \bauthor{\binits{R.P.} \bsnm{{Lin}}},
\batitle{{Solar Flare Hard X-Ray Emission from the High Corona}}.
\bjtitle{\apjl}
\bvolume{669},
\bfpage{49}--\blpage{52}
(\byear{2007}).
doi:\doiurl{10.1086/523759}
\end{barticle}
\endbibitem

\bibitem[\protect\citeauthoryear{{Krucker} et~al.}{2008}]{2008A&ARv..16..155K}
\begin{barticle}
\bauthor{\binits{S.} \bsnm{{Krucker}}}, \bauthor{\binits{M.}
  \bsnm{{Battaglia}}}, \bauthor{\binits{P.J.} \bsnm{{Cargill}}},
  \bauthor{\binits{L.} \bsnm{{Fletcher}}}, \bauthor{\binits{H.S.}
  \bsnm{{Hudson}}}, \bauthor{\binits{A.L.} \bsnm{{MacKinnon}}},
  \bauthor{\binits{S.} \bsnm{{Masuda}}}, \bauthor{\binits{L.} \bsnm{{Sui}}},
  \bauthor{\binits{M.} \bsnm{{Tomczak}}}, \bauthor{\binits{A.L.}
  \bsnm{{Veronig}}}, \bauthor{\binits{L.} \bsnm{{Vlahos}}},
  \bauthor{\binits{S.M.} \bsnm{{White}}},
\batitle{{Hard X-ray emission from the solar corona}}.
\bjtitle{\aapr}
\bvolume{16},
\bfpage{155}--\blpage{208}
(\byear{2008}).
doi:\doiurl{10.1007/s00159-008-0014-9}
\end{barticle}
\endbibitem

\bibitem[\protect\citeauthoryear{{Li} and {Zhang}}{2009}]{2009ApJ...690..347L}
\begin{barticle}
\bauthor{\binits{L.} \bsnm{{Li}}}, \bauthor{\binits{J.} \bsnm{{Zhang}}},
\batitle{{On the Brightening Propagation of Post-Flare Loops Observed by
  TRACE}}.
\bjtitle{\apj}
\bvolume{690},
\bfpage{347}--\blpage{357}
(\byear{2009}).
doi:\doiurl{10.1088/0004-637X/690/1/347}
\end{barticle}
\endbibitem

\bibitem[\protect\citeauthoryear{{Lin} and
  {Hudson}}{1976}]{1976SoPh...50..153L}
\begin{barticle}
\bauthor{\binits{R.P.} \bsnm{{Lin}}}, \bauthor{\binits{H.S.} \bsnm{{Hudson}}},
\batitle{{Non-thermal processes in large solar flares}}.
\bjtitle{\solphys}
\bvolume{50},
\bfpage{153}--\blpage{178}
(\byear{1976}).
doi:\doiurl{10.1007/ BF00206199}
\end{barticle}
\endbibitem

\bibitem[\protect\citeauthoryear{{Lin} et~al.}{2003}]{2003ApJ...595L..69L}
\begin{barticle}
\bauthor{\binits{R.P.} \bsnm{{Lin}}}, \bauthor{\binits{S.} \bsnm{{Krucker}}},
  \bauthor{\binits{G.J.} \bsnm{{Hurford}}}, \bauthor{\binits{D.M.}
  \bsnm{{Smith}}}, \bauthor{\binits{H.S.} \bsnm{{Hudson}}},
  \bauthor{\binits{G.D.} \bsnm{{Holman}}}, \bauthor{\binits{R.A.}
  \bsnm{{Schwartz}}}, \bauthor{\binits{B.R.} \bsnm{{Dennis}}},
  \bauthor{\binits{G.H.} \bsnm{{Share}}}, \bauthor{\binits{R.J.}
  \bsnm{{Murphy}}}, \bauthor{\binits{A.G.} \bsnm{{Emslie}}},
  \bauthor{\binits{C.} \bsnm{{Johns-Krull}}}, \bauthor{\binits{N.}
  \bsnm{{Vilmer}}},
\batitle{{RHESSI Observations of Particle Acceleration and Energy Release in an
  Intense Solar Gamma-Ray Line Flare}}.
\bjtitle{\apjl}
\bvolume{595},
\bfpage{69}--\blpage{76}
(\byear{2003}).
doi:\doiurl{10.1086/378932}
\end{barticle}
\endbibitem

\bibitem[\protect\citeauthoryear{{Lindsey} and
  {Braun}}{1990}]{1990SoPh..126..101L}
\begin{barticle}
\bauthor{\binits{C.} \bsnm{{Lindsey}}}, \bauthor{\binits{D.C.} \bsnm{{Braun}}},
\batitle{{Helioseismic imaging of sunspots at their antipodes}}.
\bjtitle{\solphys}
\bvolume{126},
\bfpage{101}--\blpage{115}
(\byear{1990}).
doi:\doiurl{10.1007/BF00158301}
\end{barticle}
\endbibitem

\bibitem[\protect\citeauthoryear{{Lindsey} and
  {Donea}}{2008}]{2008SoPh..251..627L}
\begin{barticle}
\bauthor{\binits{C.} \bsnm{{Lindsey}}}, \bauthor{\binits{A.} \bsnm{{Donea}}},
\batitle{{Mechanics of Seismic Emission from Solar Flares}}.
\bjtitle{\solphys}
\bvolume{251},
\bfpage{627}--\blpage{639}
(\byear{2008}).
doi:\doiurl{10.1007/s11207-008-9140-9}
\end{barticle}
\endbibitem

\bibitem[\protect\citeauthoryear{{Liu} et~al.}{2009}]{2009ApJ...703..757L}
\begin{barticle}
\bauthor{\binits{C.} \bsnm{{Liu}}}, \bauthor{\binits{J.} \bsnm{{Lee}}},
  \bauthor{\binits{M.} \bsnm{{Karlick{\'y}}}}, \bauthor{\binits{D.}
  \bsnm{{Prasad Choudhary}}}, \bauthor{\binits{N.} \bsnm{{Deng}}},
  \bauthor{\binits{H.} \bsnm{{Wang}}},
\batitle{{Successive Solar Flares and Coronal Mass Ejections on 2005 September
  13 from NOAA AR 10808}}.
\bjtitle{\apj}
\bvolume{703},
\bfpage{757}--\blpage{768}
(\byear{2009}).
doi:\doiurl{10.1088/0004-637X/703/1/757}
\end{barticle}
\endbibitem

\bibitem[\protect\citeauthoryear{{Liu}}{2009}]{ruiliu}
\begin{barticle}
\bauthor{\binits{R.} \bsnm{{Liu}}},
\batitle{{On the Brightening Propagation of Post-Flare Loops Observed by
  TRACE}}.
\bjtitle{\apj}
\bvolume{690},
\bfpage{347}--\blpage{357}
(\byear{2009}).
doi:\doiurl{10.1088/0004-637X/690/1/347}
\end{barticle}
\endbibitem

\bibitem[\protect\citeauthoryear{{MacCombie} and
  {Rust}}{1979}]{1979SoPh...61...69M}
\begin{barticle}
\bauthor{\binits{W.J.} \bsnm{{MacCombie}}}, \bauthor{\binits{D.M.}
  \bsnm{{Rust}}},
\batitle{{Physical parameters in long-decay coronal enhancements}}.
\bjtitle{\solphys}
\bvolume{61},
\bfpage{69}--\blpage{88}
(\byear{1979}).
doi:\doiurl{10.1007/BF00155447}
\end{barticle}
\endbibitem

\bibitem[\protect\citeauthoryear{{Malitson} et~al.}{1976}]{1976SSRv...19..511M}
\begin{barticle}
\bauthor{\binits{H.H.} \bsnm{{Malitson}}}, \bauthor{\binits{J.}
  \bsnm{{Fainberg}}}, \bauthor{\binits{R.G.} \bsnm{{Stone}}},
\batitle{{Hectometric and kilometric solar radio emission observed from
  satellites in August 1972}}.
\bjtitle{Space Science Reviews}
\bvolume{19},
\bfpage{511}--\blpage{531}
(\byear{1976}).
doi:\doiurl{10.1007/BF00210640}
\end{barticle}
\endbibitem

\bibitem[\protect\citeauthoryear{{Matthews} et~al.}{2003}]{2003A&A...409.1107M}
\begin{barticle}
\bauthor{\binits{S.A.} \bsnm{{Matthews}}}, \bauthor{\binits{L.} \bsnm{{van
  Driel-Gesztelyi}}}, \bauthor{\binits{H.S.} \bsnm{{Hudson}}},
  \bauthor{\binits{N.V.} \bsnm{{Nitta}}},
\batitle{{A catalogue of white-light flares observed by Yohkoh}}.
\bjtitle{\aap}
\bvolume{409},
\bfpage{1107}--\blpage{1125}
(\byear{2003}).
doi:\doiurl{10.1051/0004-6361:20031187}
\end{barticle}
\endbibitem

\bibitem[\protect\citeauthoryear{{McKenzie} and
  {Hudson}}{1999}]{1999ApJ...519L..93M}
\begin{barticle}
\bauthor{\binits{D.E.} \bsnm{{McKenzie}}}, \bauthor{\binits{H.S.}
  \bsnm{{Hudson}}},
\batitle{{X-Ray Observations of Motions and Structure above a Solar Flare
  Arcade}}.
\bjtitle{\apjl}
\bvolume{519},
\bfpage{93}--\blpage{96}
(\byear{1999}).
doi:\doiurl{10.1086/312110}
\end{barticle}
\endbibitem

\bibitem[\protect\citeauthoryear{{McTiernan}
  et~al.}{1993}]{1993ApJ...416L..91M}
\begin{barticle}
\bauthor{\binits{J.M.} \bsnm{{McTiernan}}}, \bauthor{\binits{S.R.}
  \bsnm{{Kane}}}, \bauthor{\binits{J.M.} \bsnm{{Loran}}},
  \bauthor{\binits{J.R.} \bsnm{{Lemen}}}, \bauthor{\binits{L.W.}
  \bsnm{{Acton}}}, \bauthor{\binits{H.} \bsnm{{Hara}}}, \bauthor{\binits{S.}
  \bsnm{{Tsuneta}}}, \bauthor{\binits{T.} \bsnm{{Kosugi}}},
\batitle{{Temperature and Density Structure of the 1991 November 2 Flare
  Observed by the YOHKOH Soft X-Ray Telescope and Hard X-Ray Telescope}}.
\bjtitle{\apjl}
\bvolume{416},
\bfpage{91}
(\byear{1993}).
doi:\doiurl{10.1086/187078}
\end{barticle}
\endbibitem

\bibitem[\protect\citeauthoryear{{Metcalf} et~al.}{2003}]{2003ApJ...595..483M}
\begin{barticle}
\bauthor{\binits{T.R.} \bsnm{{Metcalf}}}, \bauthor{\binits{D.}
  \bsnm{{Alexander}}}, \bauthor{\binits{H.S.} \bsnm{{Hudson}}},
  \bauthor{\binits{D.W.} \bsnm{{Longcope}}},
\batitle{{TRACE and Yohkoh Observations of a White-Light Flare}}.
\bjtitle{\apj}
\bvolume{595},
\bfpage{483}--\blpage{492}
(\byear{2003}).
doi:\doiurl{10.1086/377217}
\end{barticle}
\endbibitem

\bibitem[\protect\citeauthoryear{{Mewaldt} et~al.}{2007}]{2007AIPC..932..277M}
\begin{botherref}
\oauthor{\binits{R.A.} \bsnm{{Mewaldt}}}, \oauthor{\binits{C.M.S.}
  \bsnm{{Cohen}}}, \oauthor{\binits{D.K.} \bsnm{{Haggerty}}},
  \oauthor{\binits{G.M.} \bsnm{{Mason}}}, \oauthor{\binits{M.L.}
  \bsnm{{Looper}}}, \oauthor{\binits{T.T.} \bsnm{{von Rosenvinge}}},
  \oauthor{\binits{M.E.} \bsnm{{Wiedenbeck}}},
{Radiation risks from large solar energetic particle events},
in \textit{American Institute of Physics Conference Series}.
American Institute of Physics Conference Series,
vol. 932,
2007,
pp. 277--282.
doi:\doiurl{10.1063/1.2778975}
\end{botherref}
\endbibitem

\bibitem[\protect\citeauthoryear{{Mewaldt} et~al.}{2008}]{2008AIPC.1039..111M}
\begin{botherref}
\oauthor{\binits{R.A.} \bsnm{{Mewaldt}}}, \oauthor{\binits{C.M.S.}
  \bsnm{{Cohen}}}, \oauthor{\binits{J.} \bsnm{{Giacalone}}},
  \oauthor{\binits{G.M.} \bsnm{{Mason}}}, \oauthor{\binits{E.E.}
  \bsnm{{Chollet}}}, \oauthor{\binits{M.I.} \bsnm{{Desai}}},
  \oauthor{\binits{D.K.} \bsnm{{Haggerty}}}, \oauthor{\binits{M.D.}
  \bsnm{{Looper}}}, \oauthor{\binits{R.S.} \bsnm{{Selesnick}}},
  \oauthor{\binits{A.} \bsnm{{Vourlidas}}},
{How Efficient are Coronal Mass Ejections at Accelerating Solar Energetic
  Particles?},
in \textit{American Institute of Physics Conference Series},
ed. by {G.~Li, Q.~Hu, O.~Verkhoglyadova, G.~P.~Zank, R.~P.~Lin, \& J.~Luhmann
  }.
American Institute of Physics Conference Series,
vol. 1039,
2008,
pp. 111--117.
doi:\doiurl{10.1063/1.2982431}
\end{botherref}
\endbibitem

\bibitem[\protect\citeauthoryear{{Mittal} and
  {Narain}}{2010}]{2010JASTP..72..643M}
\begin{barticle}
\bauthor{\binits{N.} \bsnm{{Mittal}}}, \bauthor{\binits{U.} \bsnm{{Narain}}},
\batitle{{Initiation of CMEs: A review}}.
\bjtitle{Journal of Atmospheric and Solar-Terrestrial Physics}
\bvolume{72},
\bfpage{643}--\blpage{652}
(\byear{2010}).
doi:\doiurl{10.1016/j.jastp.2010.03.011}
\end{barticle}
\endbibitem

\bibitem[\protect\citeauthoryear{{Moore} et~al.}{2001}]{2001ApJ...552..833M}
\begin{barticle}
\bauthor{\binits{R.L.} \bsnm{{Moore}}}, \bauthor{\binits{A.C.}
  \bsnm{{Sterling}}}, \bauthor{\binits{H.S.} \bsnm{{Hudson}}},
  \bauthor{\binits{J.R.} \bsnm{{Lemen}}},
\batitle{{Onset of the Magnetic Explosion in Solar Flares and Coronal Mass
  Ejections}}.
\bjtitle{\apj}
\bvolume{552},
\bfpage{833}--\blpage{848}
(\byear{2001}).
doi:\doiurl{10.1086/320559}
\end{barticle}
\endbibitem

\bibitem[\protect\citeauthoryear{{Moore} et~al.}{1980}]{1980sfsl.work..341M}
\begin{botherref}
\oauthor{\binits{R.} \bsnm{{Moore}}}, \oauthor{\binits{D.L.}
  \bsnm{{McKenzie}}}, \oauthor{\binits{Z.} \bsnm{{Svestka}}},
  \oauthor{\binits{K.G.} \bsnm{{Widing}}}, \oauthor{\binits{K.P.}
  \bsnm{{Dere}}}, \oauthor{\binits{S.K.} \bsnm{{Antiochos}}},
  \oauthor{\binits{H.W.} \bsnm{{Dodson-Prince}}}, \oauthor{\binits{E.}
  \bsnm{{Hiei}}}, \oauthor{\binits{K.R.} \bsnm{{Krall}}},
  \oauthor{\binits{A.S.} \bsnm{{Krieger}}},
{The thermal X-ray flare plasma},
in \textit{Skylab Solar Workshop II},
ed. by {P.~A.~Sturrock},
1980,
pp. 341--409
\end{botherref}
\endbibitem

\bibitem[\protect\citeauthoryear{{Moreton} and
  {Ramsey}}{1960}]{1960PASP...72..357M}
\begin{barticle}
\bauthor{\binits{G.E.} \bsnm{{Moreton}}}, \bauthor{\binits{H.E.}
  \bsnm{{Ramsey}}},
\batitle{{Recent Observations of Dynamical Phenomena Associated with Solar
  Flares}}.
\bjtitle{\pasp}
\bvolume{72},
\bfpage{357}
(\byear{1960}).
doi:\doiurl{10.1086/127549}
\end{barticle}
\endbibitem

\bibitem[\protect\citeauthoryear{{Moses} and {and 34
  co-authors}}{1997}]{1997SoPh..175..571M}
\begin{barticle}
\bauthor{\binits{D.} \bsnm{{Moses}}}, \bauthor{\bsnm{{and 34 co-authors}}},
\batitle{{EIT Observations of the Extreme Ultraviolet Sun}}.
\bjtitle{\solphys}
\bvolume{175},
\bfpage{571}--\blpage{599}
(\byear{1997}).
doi:\doiurl{10.1023/A:1004902913117}
\end{barticle}
\endbibitem

\bibitem[\protect\citeauthoryear{{Narukage} et~al.}{2002}]{2002ApJ...572L.109N}
\begin{barticle}
\bauthor{\binits{N.} \bsnm{{Narukage}}}, \bauthor{\binits{H.S.}
  \bsnm{{Hudson}}}, \bauthor{\binits{T.} \bsnm{{Morimoto}}},
  \bauthor{\binits{S.} \bsnm{{Akiyama}}}, \bauthor{\binits{R.} \bsnm{{Kitai}}},
  \bauthor{\binits{H.} \bsnm{{Kurokawa}}}, \bauthor{\binits{K.}
  \bsnm{{Shibata}}},
\batitle{{Simultaneous Observation of a Moreton Wave on 1997 November 3 in
  H{$\alpha$} and Soft X-Rays}}.
\bjtitle{\apjl}
\bvolume{572},
\bfpage{109}--\blpage{112}
(\byear{2002}).
doi:\doiurl{10.1086/341599}
\end{barticle}
\endbibitem

\bibitem[\protect\citeauthoryear{{Neidig} and
  {Cliver}}{1983}]{1983STIN...8424521N}
\begin{barticle}
\bauthor{\binits{D.F.} \bsnm{{Neidig}}}, \bauthor{\binits{E.W.}
  \bsnm{{Cliver}}},
\batitle{{A catalog of solar white-light flares, including their statistical
  properties and associated emissions, 1859 - 1982}}.
\bjtitle{NASA STI/Recon Technical Report N}
\bvolume{84},
\bfpage{24521}
(\byear{1983})
\end{barticle}
\endbibitem

\bibitem[\protect\citeauthoryear{{Neupert}}{1968}]{1968ApJ...153L..59N}
\begin{barticle}
\bauthor{\binits{W.M.} \bsnm{{Neupert}}},
\batitle{{Comparison of Solar X-Ray Line Emission with Microwave Emission
  during Flares}}.
\bjtitle{ApJ}
\bvolume{153},
\bfpage{59}
(\byear{1968})
\end{barticle}
\endbibitem

\bibitem[\protect\citeauthoryear{{Neupert}}{1989}]{1989ApJ...344..504N}
\begin{barticle}
\bauthor{\binits{W.M.} \bsnm{{Neupert}}},
\batitle{{Transient coronal extreme ultraviolet emission before and during the
  impulsive phase of a solar flare}}.
\bjtitle{\apj}
\bvolume{344},
\bfpage{504}--\blpage{512}
(\byear{1989}).
doi:\doiurl{10.1086/167819}
\end{barticle}
\endbibitem

\bibitem[\protect\citeauthoryear{{Neupert} et~al.}{1967}]{1967ApJ...149L..79N}
\begin{barticle}
\bauthor{\binits{W.M.} \bsnm{{Neupert}}}, \bauthor{\binits{W.} \bsnm{{Gates}}},
  \bauthor{\binits{M.} \bsnm{{Swartz}}}, \bauthor{\binits{R.} \bsnm{{Young}}},
\batitle{{Observation of the Solar Flare X-Ray Emission-Line Spectrum of Iron
  from 1.3 to 20 {\AA}}}.
\bjtitle{\apjl}
\bvolume{149},
\bfpage{79}
(\byear{1967}).
doi:\doiurl{10.1086/180061}
\end{barticle}
\endbibitem

\bibitem[\protect\citeauthoryear{{Nitta} and
  {Hudson}}{2001}]{2001GeoRL..28.3801N}
\begin{barticle}
\bauthor{\binits{N.V.} \bsnm{{Nitta}}}, \bauthor{\binits{H.S.}
  \bsnm{{Hudson}}},
\batitle{{Recurrent flare/CME events from an emerging flux region}}.
\bjtitle{\grl}
\bvolume{28},
\bfpage{3801}--\blpage{3804}
(\byear{2001}).
doi:\doiurl{10.1029/2001GL013261}
\end{barticle}
\endbibitem

\bibitem[\protect\citeauthoryear{{Okamoto} et~al.}{2004}]{2004ApJ...608.1124O}
\begin{barticle}
\bauthor{\binits{T.J.} \bsnm{{Okamoto}}}, \bauthor{\binits{H.} \bsnm{{Nakai}}},
  \bauthor{\binits{A.} \bsnm{{Keiyama}}}, \bauthor{\binits{N.}
  \bsnm{{Narukage}}}, \bauthor{\binits{S.} \bsnm{{UeNo}}}, \bauthor{\binits{R.}
  \bsnm{{Kitai}}}, \bauthor{\binits{H.} \bsnm{{Kurokawa}}},
  \bauthor{\binits{K.} \bsnm{{Shibata}}},
\batitle{{Filament Oscillations and Moreton Waves Associated with EIT Waves}}.
\bjtitle{\apj}
\bvolume{608},
\bfpage{1124}--\blpage{1132}
(\byear{2004}).
doi:\doiurl{10.1086/420838}
\end{barticle}
\endbibitem

\bibitem[\protect\citeauthoryear{{Ontiveros} and
  {Vourlidas}}{2009}]{2009ApJ...693..267O}
\begin{barticle}
\bauthor{\binits{V.} \bsnm{{Ontiveros}}}, \bauthor{\binits{A.}
  \bsnm{{Vourlidas}}},
\batitle{{Quantitative Measurements of Coronal Mass Ejection-Driven Shocks from
  LASCO Observations}}.
\bjtitle{\apj}
\bvolume{693},
\bfpage{267}--\blpage{275}
(\byear{2009}).
doi:\doiurl{10.1088/0004-637X/693/1/267}
\end{barticle}
\endbibitem

\bibitem[\protect\citeauthoryear{{Orlando} et~al.}{2009}]{2009arXiv0912.3775O}
\begin{botherref}
\oauthor{\binits{E.} \bsnm{{Orlando}}}, \oauthor{\binits{N.}
  \bsnm{{Giglietto}}}, \oauthor{\bsnm{{for the Fermi Large Area Telescope
  Collaboration}}},
{Fermi-LAT Observation of Quiet Solar Emission}.
\textrm{ArXiv e-prints}
(2009)
\end{botherref}
\endbibitem

\bibitem[\protect\citeauthoryear{{Parker}}{1988}]{1988ApJ...330..474P}
\begin{barticle}
\bauthor{\binits{E.N.} \bsnm{{Parker}}},
\batitle{{Nanoflares and the solar X-ray corona}}.
\bjtitle{\apj}
\bvolume{330},
\bfpage{474}--\blpage{479}
(\byear{1988}).
doi:\doiurl{10.1086/166485}
\end{barticle}
\endbibitem

\bibitem[\protect\citeauthoryear{{Parks} and
  {Winckler}}{1971}]{1971SoPh...16..186P}
\begin{barticle}
\bauthor{\binits{G.K.} \bsnm{{Parks}}}, \bauthor{\binits{J.R.}
  \bsnm{{Winckler}}},
\batitle{{The relation of energetic solar X-rays (h$\nu >$60 keV) and high
  frequency microwaves deduced from the periodic bursts of August 8, 1968
  flare}}.
\bjtitle{\solphys}
\bvolume{16},
\bfpage{186}--\blpage{197}
(\byear{1971}).
doi:\doiurl{10.1007/BF00154511}
\end{barticle}
\endbibitem

\bibitem[\protect\citeauthoryear{{Pick} and
  {Vilmer}}{2008}]{2008A&ARv..16....1P}
\begin{barticle}
\bauthor{\binits{M.} \bsnm{{Pick}}}, \bauthor{\binits{N.} \bsnm{{Vilmer}}},
\batitle{{Sixty-five years of solar radioastronomy: flares, coronal mass
  ejections and Sun Earth connection}}.
\bjtitle{\aapr}
\bvolume{16},
\bfpage{1}--\blpage{153}
(\byear{2008}).
doi:\doiurl{10.1007/s00159-008-0013-x}
\end{barticle}
\endbibitem

\bibitem[\protect\citeauthoryear{{Qiu} and {Gary}}{2003}]{2003ApJ...599..615Q}
\begin{barticle}
\bauthor{\binits{J.} \bsnm{{Qiu}}}, \bauthor{\binits{D.E.} \bsnm{{Gary}}},
\batitle{{Flare-related Magnetic Anomaly with a Sign Reversal}}.
\bjtitle{\apj}
\bvolume{599},
\bfpage{615}--\blpage{625}
(\byear{2003}).
doi:\doiurl{10.1086/379146}
\end{barticle}
\endbibitem

\bibitem[\protect\citeauthoryear{{Qiu} et~al.}{2004}]{2004ApJ...603..335Q}
\begin{barticle}
\bauthor{\binits{J.} \bsnm{{Qiu}}}, \bauthor{\binits{J.} \bsnm{{Lee}}},
  \bauthor{\binits{D.E.} \bsnm{{Gary}}},
\batitle{{Impulsive and Gradual Nonthermal Emissions in an X-Class Flare}}.
\bjtitle{\apj}
\bvolume{603},
\bfpage{335}--\blpage{347}
(\byear{2004}).
doi:\doiurl{10.1086/381353}
\end{barticle}
\endbibitem

\bibitem[\protect\citeauthoryear{{Quesnel} et~al.}{2010}]{2010arXiv1003.4194Q}
\begin{botherref}
\oauthor{\binits{A.} \bsnm{{Quesnel}}}, \oauthor{\binits{B.R.}
  \bsnm{{Dennis}}}, \oauthor{\binits{B.} \bsnm{{Fleck}}}, \oauthor{\binits{C.}
  \bsnm{{Froelich}}}, \oauthor{\binits{H.S.} \bsnm{{Hudson}}},
{The Signature of Flares in VIRGO Total Solar Irradiance Measurements}.
\textrm{ArXiv e-prints}
(2010)
\end{botherref}
\endbibitem

\bibitem[\protect\citeauthoryear{{Ramaty} et~al.}{1995}]{1995ApJ...455L.193R}
\begin{barticle}
\bauthor{\binits{R.} \bsnm{{Ramaty}}}, \bauthor{\binits{N.}
  \bsnm{{Mandzhavidze}}}, \bauthor{\binits{B.} \bsnm{{Kozlovsky}}},
  \bauthor{\binits{R.J.} \bsnm{{Murphy}}},
\batitle{{Solar Atmopheric Abundances and Energy Content in Flare Accelerated
  Ions from Gamma-Ray Spectroscopy}}.
\bjtitle{\apjl}
\bvolume{455},
\bfpage{193}
(\byear{1995}).
doi:\doiurl{10.1086/309841}
\end{barticle}
\endbibitem

\bibitem[\protect\citeauthoryear{{R{\'e}gnier} and
  {Priest}}{2007}]{2007A&A...468..701R}
\begin{barticle}
\bauthor{\binits{S.} \bsnm{{R{\'e}gnier}}}, \bauthor{\binits{E.R.}
  \bsnm{{Priest}}},
\batitle{{Nonlinear force-free models for the solar corona. I. Two active
  regions with very different structure}}.
\bjtitle{\aap}
\bvolume{468},
\bfpage{701}--\blpage{709}
(\byear{2007}).
doi:\doiurl{10.1051/0004-6361:20077318}
\end{barticle}
\endbibitem

\bibitem[\protect\citeauthoryear{{Riddle}}{1970}]{1970SoPh...13..448R}
\begin{barticle}
\bauthor{\binits{A.C.} \bsnm{{Riddle}}},
\batitle{{80 MHz Observations of a Moving Type IV Solar Burst, March 1, 1969}}.
\bjtitle{\solphys}
\bvolume{13},
\bfpage{448}--\blpage{457}
(\byear{1970}).
doi:\doiurl{10.1007/BF00153563}
\end{barticle}
\endbibitem

\bibitem[\protect\citeauthoryear{{Robbrecht}
  et~al.}{2009}]{2009ApJ...701..283R}
\begin{barticle}
\bauthor{\binits{E.} \bsnm{{Robbrecht}}}, \bauthor{\binits{S.}
  \bsnm{{Patsourakos}}}, \bauthor{\binits{A.} \bsnm{{Vourlidas}}},
\batitle{{No Trace Left Behind: STEREO Observation of a Coronal Mass Ejection
  Without Low Coronal Signatures}}.
\bjtitle{\apj}
\bvolume{701},
\bfpage{283}--\blpage{291}
(\byear{2009}).
doi:\doiurl{10.1088/0004-637X/701/1/283}
\end{barticle}
\endbibitem

\bibitem[\protect\citeauthoryear{{Rust}}{1972}]{1972SoPh...25..141R}
\begin{barticle}
\bauthor{\binits{D.M.} \bsnm{{Rust}}},
\batitle{{Flares and Changing Magnetic Fields}}.
\bjtitle{\solphys}
\bvolume{25},
\bfpage{141}--\blpage{157}
(\byear{1972}).
doi:\doiurl{10.1007/BF00155753}
\end{barticle}
\endbibitem

\bibitem[\protect\citeauthoryear{{Rust} and
  {Hildner}}{1976}]{1976SoPh...48..381R}
\begin{barticle}
\bauthor{\binits{D.M.} \bsnm{{Rust}}}, \bauthor{\binits{E.} \bsnm{{Hildner}}},
\batitle{{Expansion of an X-ray coronal arch into the outer corona}}.
\bjtitle{\solphys}
\bvolume{48},
\bfpage{381}--\blpage{387}
(\byear{1976}).
doi:\doiurl{10.1007/BF00152003}
\end{barticle}
\endbibitem

\bibitem[\protect\citeauthoryear{{Schatten} et~al.}{1969}]{1969SoPh....6..442S}
\begin{barticle}
\bauthor{\binits{K.H.} \bsnm{{Schatten}}}, \bauthor{\binits{J.M.}
  \bsnm{{Wilcox}}}, \bauthor{\binits{N.F.} \bsnm{{Ness}}},
\batitle{{A model of interplanetary and coronal magnetic fields}}.
\bjtitle{\solphys}
\bvolume{6},
\bfpage{442}--\blpage{455}
(\byear{1969}).
doi:\doiurl{10.1007/BF00146478}
\end{barticle}
\endbibitem

\bibitem[\protect\citeauthoryear{{Schmit} et~al.}{2009}]{2009ApJ...700L..96S}
\begin{barticle}
\bauthor{\binits{D.J.} \bsnm{{Schmit}}}, \bauthor{\binits{S.E.}
  \bsnm{{Gibson}}}, \bauthor{\binits{S.} \bsnm{{Tomczyk}}},
  \bauthor{\binits{K.K.} \bsnm{{Reeves}}}, \bauthor{\binits{A.C.}
  \bsnm{{Sterling}}}, \bauthor{\binits{D.H.} \bsnm{{Brooks}}},
  \bauthor{\binits{D.R.} \bsnm{{Williams}}}, \bauthor{\binits{D.}
  \bsnm{{Tripathi}}},
\batitle{{Large-Scale Flows in Prominence Cavities}}.
\bjtitle{\apjl}
\bvolume{700},
\bfpage{96}--\blpage{98}
(\byear{2009}).
doi:\doiurl{10.1088/0004-637X/700/2/L96}
\end{barticle}
\endbibitem

\bibitem[\protect\citeauthoryear{{Schuck}}{2010}]{2010arXiv1003.1647S}
\begin{botherref}
\oauthor{\binits{P.W.} \bsnm{{Schuck}}},
{The photospheric energy and helicity budgets of the flux-injection
  hypothesis}.
\textrm{ArXiv e-prints}
(2010)
\end{botherref}
\endbibitem

\bibitem[\protect\citeauthoryear{{Serio} et~al.}{1991}]{1991A&A...241..197S}
\begin{barticle}
\bauthor{\binits{S.} \bsnm{{Serio}}}, \bauthor{\binits{F.} \bsnm{{Reale}}},
  \bauthor{\binits{J.} \bsnm{{Jakimiec}}}, \bauthor{\binits{B.}
  \bsnm{{Sylwester}}}, \bauthor{\binits{J.} \bsnm{{Sylwester}}},
\batitle{{Dynamics of flaring loops. I - Thermodynamic decay scaling laws}}.
\bjtitle{\aap}
\bvolume{241},
\bfpage{197}--\blpage{202}
(\byear{1991})
\end{barticle}
\endbibitem

\bibitem[\protect\citeauthoryear{{Sheeley} et~al.}{2004}]{2004ApJ...616.1224S}
\begin{barticle}
\bauthor{\binits{N.R.} \bsnm{{Sheeley}} \bsuffix{Jr.}}, \bauthor{\binits{H.P.}
  \bsnm{{Warren}}}, \bauthor{\binits{Y.} \bsnm{{Wang}}},
\batitle{{The Origin of Postflare Loops}}.
\bjtitle{\apj}
\bvolume{616},
\bfpage{1224}--\blpage{1231}
(\byear{2004}).
doi:\doiurl{10.1086/425126}
\end{barticle}
\endbibitem

\bibitem[\protect\citeauthoryear{{Shih} et~al.}{2009}]{2009ApJ...698L.152S}
\begin{barticle}
\bauthor{\binits{A.Y.} \bsnm{{Shih}}}, \bauthor{\binits{R.P.} \bsnm{{Lin}}},
  \bauthor{\binits{D.M.} \bsnm{{Smith}}},
\batitle{{RHESSI Observations of the Proportional Acceleration of Relativistic
  $>$0.3 MeV Electrons and $>$30 MeV Protons in Solar Flares}}.
\bjtitle{\apjl}
\bvolume{698},
\bfpage{152}--\blpage{157}
(\byear{2009}).
doi:\doiurl{10.1088/0004-637X/698/2/L152}
\end{barticle}
\endbibitem

\bibitem[\protect\citeauthoryear{{Shimizu}}{1995}]{1995PASJ...47..251S}
\begin{barticle}
\bauthor{\binits{T.} \bsnm{{Shimizu}}},
\batitle{{Energetics and Occurrence Rate of Active-Region Transient
  Brightenings and Implications for the Heating of the Active-Region Corona}}.
\bjtitle{\pasj}
\bvolume{47},
\bfpage{251}--\blpage{263}
(\byear{1995})
\end{barticle}
\endbibitem

\bibitem[\protect\citeauthoryear{{Shimojo} et~al.}{1996}]{1996PASJ...48..123S}
\begin{barticle}
\bauthor{\binits{M.} \bsnm{{Shimojo}}}, \bauthor{\binits{S.}
  \bsnm{{Hashimoto}}}, \bauthor{\binits{K.} \bsnm{{Shibata}}},
  \bauthor{\binits{T.} \bsnm{{Hirayama}}}, \bauthor{\binits{H.S.}
  \bsnm{{Hudson}}}, \bauthor{\binits{L.W.} \bsnm{{Acton}}},
\batitle{{Statistical Study of Solar X-Ray Jets Observed with the YOHKOH Sof t
  X-Ray Telescope}}.
\bjtitle{\pasj}
\bvolume{48},
\bfpage{123}--\blpage{136}
(\byear{1996})
\end{barticle}
\endbibitem

\bibitem[\protect\citeauthoryear{{Sturrock}}{1980}]{1980sfsl.work.....S}
\begin{botherref}
\oauthor{\binits{P.A.} \bsnm{{Sturrock}}} (ed.),
{Solar flares: A monograph from SKYLAB Solar Workshop II},
in \textit{Skylab Solar Workshop II}
1980
\end{botherref}
\endbibitem

\bibitem[\protect\citeauthoryear{{Sturrock}}{1991}]{1991ApJ...380..655S}
\begin{barticle}
\bauthor{\binits{P.A.} \bsnm{{Sturrock}}},
\batitle{{Maximum energy of semi-infinite magnetic field configurations}}.
\bjtitle{\apj}
\bvolume{380},
\bfpage{655}--\blpage{659}
(\byear{1991}).
doi:\doiurl{10.1086/170620}
\end{barticle}
\endbibitem

\bibitem[\protect\citeauthoryear{{Sudol} and
  {Harvey}}{2005}]{2005ApJ...635..647S}
\begin{barticle}
\bauthor{\binits{J.J.} \bsnm{{Sudol}}}, \bauthor{\binits{J.W.}
  \bsnm{{Harvey}}},
\batitle{{Longitudinal Magnetic Field Changes Accompanying Solar Flares}}.
\bjtitle{\apj}
\bvolume{635},
\bfpage{647}--\blpage{658}
(\byear{2005}).
doi:\doiurl{10.1086/497361}
\end{barticle}
\endbibitem

\bibitem[\protect\citeauthoryear{{Sui} et~al.}{2004}]{2004ApJ...612..546S}
\begin{barticle}
\bauthor{\binits{L.} \bsnm{{Sui}}}, \bauthor{\binits{G.D.} \bsnm{{Holman}}},
  \bauthor{\binits{B.R.} \bsnm{{Dennis}}},
\batitle{{Evidence for Magnetic Reconnection in Three Homologous Solar Flares
  Observed by RHESSI}}.
\bjtitle{\apj}
\bvolume{612},
\bfpage{546}--\blpage{556}
(\byear{2004}).
doi:\doiurl{10.1086/422515}
\end{barticle}
\endbibitem

\bibitem[\protect\citeauthoryear{{Svestka} and
  {Cliver}}{1992}]{1992LNP...399....1S}
\begin{botherref}
\oauthor{\binits{Z.} \bsnm{{Svestka}}}, \oauthor{\binits{E.W.}
  \bsnm{{Cliver}}},
{History and Basic Characteristics of Eruptive Flares},
in \textit{IAU Colloq. 133: Eruptive Solar Flares},
ed. by {Z.~Svestka, B.~V.~Jackson, \& M.~E.~Machado}.
Lecture Notes in Physics, Berlin Springer Verlag,
vol. 399,
1992,
p. 1.
doi:\doiurl{10.1007/3-540-55246-4\_70}
\end{botherref}
\endbibitem

\bibitem[\protect\citeauthoryear{{Svestka} et~al.}{1987}]{1987SoPh..108..237S}
\begin{barticle}
\bauthor{\binits{Z.F.} \bsnm{{Svestka}}}, \bauthor{\binits{J.M.}
  \bsnm{{Fontenla}}}, \bauthor{\binits{M.E.} \bsnm{{Machado}}},
  \bauthor{\binits{S.F.} \bsnm{{Martin}}}, \bauthor{\binits{D.F.}
  \bsnm{{Neidig}}},
\batitle{{Multi-thermal observations of newly formed loops in a dynamic
  flare}}.
\bjtitle{\solphys}
\bvolume{108},
\bfpage{237}--\blpage{250}
(\byear{1987}).
doi:\doiurl{10.1007/BF00214164}
\end{barticle}
\endbibitem

\bibitem[\protect\citeauthoryear{{Temmer} et~al.}{2009}]{2009ApJ...702.1343T}
\begin{barticle}
\bauthor{\binits{M.} \bsnm{{Temmer}}}, \bauthor{\binits{B.} \bsnm{{Vr{\v
  s}nak}}}, \bauthor{\binits{T.} \bsnm{{{\v Z}ic}}}, \bauthor{\binits{A.M.}
  \bsnm{{Veronig}}},
\batitle{{Analytic Modeling of the Moreton Wave Kinematics}}.
\bjtitle{\apj}
\bvolume{702},
\bfpage{1343}--\blpage{1352}
(\byear{2009}).
doi:\doiurl{10.1088/0004-637X/702/2/1343}
\end{barticle}
\endbibitem

\bibitem[\protect\citeauthoryear{{Thomas} and
  {Teske}}{1971}]{1971SoPh...16..431T}
\begin{barticle}
\bauthor{\binits{R.J.} \bsnm{{Thomas}}}, \bauthor{\binits{R.G.}
  \bsnm{{Teske}}},
\batitle{{Solar Soft X-Rays and Solar Activity. II: Soft X-Ray Emission during
  Solar Flares}}.
\bjtitle{\solphys}
\bvolume{16},
\bfpage{431}--\blpage{453}
(\byear{1971}).
doi:\doiurl{10.1007/BF00162486}
\end{barticle}
\endbibitem

\bibitem[\protect\citeauthoryear{{Thompson} et~al.}{1999}]{1999ApJ...517L.151T}
\begin{barticle}
\bauthor{\binits{B.J.} \bsnm{{Thompson}}}, \bauthor{\binits{J.B.}
  \bsnm{{Gurman}}}, \bauthor{\binits{W.M.} \bsnm{{Neupert}}},
  \bauthor{\binits{J.S.} \bsnm{{Newmark}}}, \bauthor{\binits{J.}
  \bsnm{{Delaboudini{\`e}re}}}, \bauthor{\binits{O.C.} \bsnm{{St.~Cyr}}},
  \bauthor{\binits{S.} \bsnm{{Stezelberger}}}, \bauthor{\binits{K.P.}
  \bsnm{{Dere}}}, \bauthor{\binits{R.A.} \bsnm{{Howard}}},
  \bauthor{\binits{D.J.} \bsnm{{Michels}}},
\batitle{{SOHO/EIT Observations of the 1997 April 7 Coronal Transient: Possible
  Evidence of Coronal Moreton Waves}}.
\bjtitle{\apjl}
\bvolume{517},
\bfpage{151}--\blpage{154}
(\byear{1999}).
doi:\doiurl{10.1086/312030}
\end{barticle}
\endbibitem

\bibitem[\protect\citeauthoryear{{Tripathi} et~al.}{2004}]{2004A&A...422..337T}
\begin{barticle}
\bauthor{\binits{D.} \bsnm{{Tripathi}}}, \bauthor{\binits{V.}
  \bsnm{{Bothmer}}}, \bauthor{\binits{H.} \bsnm{{Cremades}}},
\batitle{{The basic characteristics of EUV post-eruptive arcades and their role
  as tracers of coronal mass ejection source regions}}.
\bjtitle{\aap}
\bvolume{422},
\bfpage{337}--\blpage{349}
(\byear{2004}).
doi:\doiurl{10.1051/0004-6361:20035815}
\end{barticle}
\endbibitem

\bibitem[\protect\citeauthoryear{{Tsuneta} et~al.}{1991}]{1991SoPh..136...37T}
\begin{barticle}
\bauthor{\binits{S.} \bsnm{{Tsuneta}}}, \bauthor{\binits{L.} \bsnm{{Acton}}},
  \bauthor{\binits{M.} \bsnm{{Bruner}}}, \bauthor{\binits{J.} \bsnm{{Lemen}}},
  \bauthor{\binits{W.} \bsnm{{Brown}}}, \bauthor{\binits{R.}
  \bsnm{{Caravalho}}}, \bauthor{\binits{R.} \bsnm{{Catura}}},
  \bauthor{\binits{S.} \bsnm{{Freeland}}}, \bauthor{\binits{B.}
  \bsnm{{Jurcevich}}}, \bauthor{\binits{J.} \bsnm{{Owens}}},
\batitle{{The soft X-ray telescope for the SOLAR-A mission}}.
\bjtitle{\solphys}
\bvolume{136},
\bfpage{37}--\blpage{67}
(\byear{1991}).
doi:\doiurl{10.1007/BF00151694}
\end{barticle}
\endbibitem

\bibitem[\protect\citeauthoryear{{Uchida}}{1968}]{1968SoPh....4...30U}
\begin{barticle}
\bauthor{\binits{Y.} \bsnm{{Uchida}}},
\batitle{{Propagation of Hydromagnetic Disturbances in the Solar Corona and
  Moreton's Wave Phenomenon}}.
\bjtitle{\solphys}
\bvolume{4},
\bfpage{30}--\blpage{44}
(\byear{1968}).
doi:\doiurl{10.1007/BF00146996}
\end{barticle}
\endbibitem

\bibitem[\protect\citeauthoryear{{{\v S}vestka}
  et~al.}{1998}]{1998SoPh..182..179S}
\begin{barticle}
\bauthor{\binits{Z.} \bsnm{{{\v S}vestka}}}, \bauthor{\binits{F.}
  \bsnm{{F{\'a}rn{\'{\i}}k}}}, \bauthor{\binits{H.S.} \bsnm{{Hudson}}},
  \bauthor{\binits{P.} \bsnm{{Hick}}},
\batitle{{Large-Scale Active Coronal Phenomena in Yohkoh SXT Images IV. Solar
  Wind Streams from Flaring Active Regions}}.
\bjtitle{\solphys}
\bvolume{182},
\bfpage{179}--\blpage{193}
(\byear{1998}).
doi:\doiurl{10.1023/A:1005033717284}
\end{barticle}
\endbibitem

\bibitem[\protect\citeauthoryear{{van den Oord}}{1990}]{1990A&A...234..496V}
\begin{barticle}
\bauthor{\binits{G.H.J.} \bsnm{{van den Oord}}},
\batitle{{The electrodynamics of beam/return current systems in the solar
  corona}}.
\bjtitle{\aap}
\bvolume{234},
\bfpage{496}--\blpage{518}
(\byear{1990})
\end{barticle}
\endbibitem

\bibitem[\protect\citeauthoryear{{Vernazza} et~al.}{1981}]{1981ApJS...45..635V}
\begin{barticle}
\bauthor{\binits{J.E.} \bsnm{{Vernazza}}}, \bauthor{\binits{E.H.}
  \bsnm{{Avrett}}}, \bauthor{\binits{R.} \bsnm{{Loeser}}},
\batitle{{Structure of the solar chromosphere. III - Models of the EUV
  brightness components of the quiet-sun}}.
\bjtitle{\apjs}
\bvolume{45},
\bfpage{635}--\blpage{725}
(\byear{1981}).
doi:\doiurl{10.1086/190731}
\end{barticle}
\endbibitem

\bibitem[\protect\citeauthoryear{{Veronig} et~al.}{2006}]{2006A&A...446..675V}
\begin{barticle}
\bauthor{\binits{A.M.} \bsnm{{Veronig}}}, \bauthor{\binits{M.}
  \bsnm{{Karlick{\'y}}}}, \bauthor{\binits{B.} \bsnm{{Vr{\v s}nak}}},
  \bauthor{\binits{M.} \bsnm{{Temmer}}}, \bauthor{\binits{J.}
  \bsnm{{Magdaleni{\'c}}}}, \bauthor{\binits{B.R.} \bsnm{{Dennis}}},
  \bauthor{\binits{W.} \bsnm{{Otruba}}}, \bauthor{\binits{W.}
  \bsnm{{P{\"o}tzi}}},
\batitle{{X-ray sources and magnetic reconnection in the X3.9 flare of 2003
  November 3}}.
\bjtitle{\aap}
\bvolume{446},
\bfpage{675}--\blpage{690}
(\byear{2006}).
doi:\doiurl{10.1051/0004-6361:20053112}
\end{barticle}
\endbibitem

\bibitem[\protect\citeauthoryear{{Veronig} et~al.}{2002}]{2002A&A...392..699V}
\begin{barticle}
\bauthor{\binits{A.} \bsnm{{Veronig}}}, \bauthor{\binits{B.} \bsnm{{Vr{\v
  s}nak}}}, \bauthor{\binits{B.R.} \bsnm{{Dennis}}}, \bauthor{\binits{M.}
  \bsnm{{Temmer}}}, \bauthor{\binits{A.} \bsnm{{Hanslmeier}}},
  \bauthor{\binits{J.} \bsnm{{Magdaleni{\'c}}}},
\batitle{{Investigation of the Neupert effect in solar flares. I. Statistical
  properties and the evaporation model}}.
\bjtitle{\aap}
\bvolume{392},
\bfpage{699}--\blpage{712}
(\byear{2002}).
doi:\doiurl{10.1051/0004-6361:20020947}
\end{barticle}
\endbibitem

\bibitem[\protect\citeauthoryear{{Vourlidas}
  et~al.}{2000}]{2000ApJ...534..456V}
\begin{barticle}
\bauthor{\binits{A.} \bsnm{{Vourlidas}}}, \bauthor{\binits{P.}
  \bsnm{{Subramanian}}}, \bauthor{\binits{K.P.} \bsnm{{Dere}}},
  \bauthor{\binits{R.A.} \bsnm{{Howard}}},
\batitle{{Large-Angle Spectrometric Coronagraph Measurements of the Energetics
  of Coronal Mass Ejections}}.
\bjtitle{\apj}
\bvolume{534},
\bfpage{456}--\blpage{467}
(\byear{2000}).
doi:\doiurl{10.1086/308747}
\end{barticle}
\endbibitem

\bibitem[\protect\citeauthoryear{{Vourlidas}
  et~al.}{2003}]{2003ApJ...598.1392V}
\begin{barticle}
\bauthor{\binits{A.} \bsnm{{Vourlidas}}}, \bauthor{\binits{S.T.} \bsnm{{Wu}}},
  \bauthor{\binits{A.H.} \bsnm{{Wang}}}, \bauthor{\binits{P.}
  \bsnm{{Subramanian}}}, \bauthor{\binits{R.A.} \bsnm{{Howard}}},
\batitle{{Direct Detection of a Coronal Mass Ejection-Associated Shock in Large
  Angle and Spectrometric Coronagraph Experiment White-Light Images}}.
\bjtitle{\apj}
\bvolume{598},
\bfpage{1392}--\blpage{1402}
(\byear{2003}).
doi:\doiurl{10.1086/379098}
\end{barticle}
\endbibitem

\bibitem[\protect\citeauthoryear{{Vr{\v s}nak}}{2005}]{2005EOSTr..86..112V}
\begin{barticle}
\bauthor{\binits{B.} \bsnm{{Vr{\v s}nak}}},
\batitle{{Terminology of Large-Scale Waves in the Solar Atmosphere}}.
\bjtitle{EOS Transactions}
\bvolume{86},
\bfpage{112}--\blpage{113}
(\byear{2005}).
doi:\doiurl{10.1029/2005EO110004}
\end{barticle}
\endbibitem

\bibitem[\protect\citeauthoryear{{Wang}}{1992}]{1992SoPh..140...85W}
\begin{barticle}
\bauthor{\binits{H.} \bsnm{{Wang}}},
\batitle{{Evolution of vector magnetic fields and the August 27 1990 X-3
  flare}}.
\bjtitle{\solphys}
\bvolume{140},
\bfpage{85}--\blpage{98}
(\byear{1992}).
doi:\doiurl{10.1007/BF00148431}
\end{barticle}
\endbibitem

\bibitem[\protect\citeauthoryear{{Wang}}{2009}]{2009RAA.....9..127W}
\begin{barticle}
\bauthor{\binits{H.} \bsnm{{Wang}}},
\batitle{{Study of white-light flares observed by Hinode}}.
\bjtitle{Research in Astronomy and Astrophysics}
\bvolume{9},
\bfpage{127}--\blpage{132}
(\byear{2009}).
doi:\doiurl{10.1088/1674-4527/9/2/001}
\end{barticle}
\endbibitem

\bibitem[\protect\citeauthoryear{{Wang} and
  {Zhang}}{2007}]{2007ApJ...665.1428W}
\begin{barticle}
\bauthor{\binits{Y.} \bsnm{{Wang}}}, \bauthor{\binits{J.} \bsnm{{Zhang}}},
\batitle{{A Comparative Study between Eruptive X-Class Flares Associated with
  Coronal Mass Ejections and Confined X-Class Flares}}.
\bjtitle{\apj}
\bvolume{665},
\bfpage{1428}--\blpage{1438}
(\byear{2007}).
doi:\doiurl{10.1086/519765}
\end{barticle}
\endbibitem

\bibitem[\protect\citeauthoryear{{Webb} et~al.}{1998}]{1998GeoRL..25.2469W}
\begin{barticle}
\bauthor{\binits{D.F.} \bsnm{{Webb}}}, \bauthor{\binits{E.W.} \bsnm{{Cliver}}},
  \bauthor{\binits{N.} \bsnm{{Gopalswamy}}}, \bauthor{\binits{H.S.}
  \bsnm{{Hudson}}}, \bauthor{\binits{O.C.} \bsnm{{St.~Cyr}}},
\batitle{{The solar origin of the January 1997 coronal mass ejection, magnetic
  cloud and geomagnetic storm}}.
\bjtitle{\grl}
\bvolume{25},
\bfpage{2469}--\blpage{2472}
(\byear{1998}).
doi:\doiurl{10.1029/98GL00493}
\end{barticle}
\endbibitem

\bibitem[\protect\citeauthoryear{{Wild} et~al.}{1963}]{1963ARA&A...1..291W}
\begin{barticle}
\bauthor{\binits{J.P.} \bsnm{{Wild}}}, \bauthor{\binits{S.F.} \bsnm{{Smerd}}},
  \bauthor{\binits{A.A.} \bsnm{{Weiss}}},
\batitle{{Solar Bursts}}.
\bjtitle{\araa}
\bvolume{1},
\bfpage{291}
(\byear{1963}).
doi:\doiurl{10.1146/annurev.aa.01.090163.001451}
\end{barticle}
\endbibitem

\bibitem[\protect\citeauthoryear{{Wills-Davey} and
  {Attrill}}{2009}]{2009SSRv..149..325W}
\begin{barticle}
\bauthor{\binits{M.J.} \bsnm{{Wills-Davey}}}, \bauthor{\binits{G.D.R.}
  \bsnm{{Attrill}}},
\batitle{{EIT Waves: A Changing Understanding over a Solar Cycle}}.
\bjtitle{Space Science Reviews}
\bvolume{149},
\bfpage{325}--\blpage{353}
(\byear{2009}).
doi:\doiurl{10.1007/s11214-009-9612-8}
\end{barticle}
\endbibitem

\bibitem[\protect\citeauthoryear{{Wills-Davey} and
  {Thompson}}{1999}]{1999SoPh..190..467W}
\begin{barticle}
\bauthor{\binits{M.J.} \bsnm{{Wills-Davey}}}, \bauthor{\binits{B.J.}
  \bsnm{{Thompson}}},
\batitle{{Observations of a Propagating Disturbance in TRACE}}.
\bjtitle{\solphys}
\bvolume{190},
\bfpage{467}--\blpage{483}
(\byear{1999}).
doi:\doiurl{10.1023/A:1005201500675}
\end{barticle}
\endbibitem

\bibitem[\protect\citeauthoryear{{Wolff}}{1972}]{1972ApJ...176..833W}
\begin{barticle}
\bauthor{\binits{C.L.} \bsnm{{Wolff}}},
\batitle{{Free Oscillations of the Sun and Their Possible Stimulation by Solar
  Flares}}.
\bjtitle{\apj}
\bvolume{176},
\bfpage{833}
(\byear{1972}).
doi:\doiurl{10.1086/151680}
\end{barticle}
\endbibitem

\bibitem[\protect\citeauthoryear{{Woltjer}}{1958}]{1958PNAS...44..489W}
\begin{barticle}
\bauthor{\binits{L.} \bsnm{{Woltjer}}},
\batitle{{A Theorem on Force-Free Magnetic Fields}}.
\bjtitle{Proceedings of the National Academy of Science}
\bvolume{44},
\bfpage{489}--\blpage{491}
(\byear{1958})
\end{barticle}
\endbibitem

\bibitem[\protect\citeauthoryear{{Woods} et~al.}{2006}]{2006JGRA..11110S14W}
\begin{barticle}
\bauthor{\binits{T.N.} \bsnm{{Woods}}}, \bauthor{\binits{G.} \bsnm{{Kopp}}},
  \bauthor{\binits{P.C.} \bsnm{{Chamberlin}}},
\batitle{{Contributions of the solar ultraviolet irradiance to the total solar
  irradiance during large flares}}.
\bjtitle{Journal of Geophysical Research (Space Physics)}
\bvolume{111},
\bfpage{10}
(\byear{2006}).
doi:\doiurl{10.1029/2005JA011507}
\end{barticle}
\endbibitem

\bibitem[\protect\citeauthoryear{{Yang} et~al.}{2009}]{2009ApJ...693..132Y}
\begin{barticle}
\bauthor{\binits{Y.} \bsnm{{Yang}}}, \bauthor{\binits{C.Z.} \bsnm{{Cheng}}},
  \bauthor{\binits{S.} \bsnm{{Krucker}}}, \bauthor{\binits{R.P.} \bsnm{{Lin}}},
  \bauthor{\binits{W.H.} \bsnm{{Ip}}},
\batitle{{A Statistical Study of Hard X-Ray Footpoint Motions in Large Solar
  Flares}}.
\bjtitle{\apj}
\bvolume{693},
\bfpage{132}--\blpage{139}
(\byear{2009}).
doi:\doiurl{10.1088/0004-637X/693/1/132}
\end{barticle}
\endbibitem

\bibitem[\protect\citeauthoryear{{Yashiro} et~al.}{2006}]{2006ApJ...650L.143Y}
\begin{barticle}
\bauthor{\binits{S.} \bsnm{{Yashiro}}}, \bauthor{\binits{S.} \bsnm{{Akiyama}}},
  \bauthor{\binits{N.} \bsnm{{Gopalswamy}}}, \bauthor{\binits{R.A.}
  \bsnm{{Howard}}},
\batitle{{Different Power-Law Indices in the Frequency Distributions of Flares
  with and without Coronal Mass Ejections}}.
\bjtitle{\apjl}
\bvolume{650},
\bfpage{143}--\blpage{146}
(\byear{2006}).
doi:\doiurl{10.1086/508876}
\end{barticle}
\endbibitem

\bibitem[\protect\citeauthoryear{{Zarro} et~al.}{1999}]{1999ApJ...520L.139Z}
\begin{barticle}
\bauthor{\binits{D.M.} \bsnm{{Zarro}}}, \bauthor{\binits{A.C.}
  \bsnm{{Sterling}}}, \bauthor{\binits{B.J.} \bsnm{{Thompson}}},
  \bauthor{\binits{H.S.} \bsnm{{Hudson}}}, \bauthor{\binits{N.}
  \bsnm{{Nitta}}},
\batitle{{SOHO EIT Observations of Extreme-Ultraviolet ``Dimming'' Associated
  with a Halo Coronal Mass Ejection}}.
\bjtitle{\apjl}
\bvolume{520},
\bfpage{139}--\blpage{142}
(\byear{1999}).
doi:\doiurl{10.1086/312150}
\end{barticle}
\endbibitem

\bibitem[\protect\citeauthoryear{{Zhang} et~al.}{2001}]{2001ApJ...559..452Z}
\begin{barticle}
\bauthor{\binits{J.} \bsnm{{Zhang}}}, \bauthor{\binits{K.P.} \bsnm{{Dere}}},
  \bauthor{\binits{R.A.} \bsnm{{Howard}}}, \bauthor{\binits{M.R.}
  \bsnm{{Kundu}}}, \bauthor{\binits{S.M.} \bsnm{{White}}},
\batitle{{On the Temporal Relationship between Coronal Mass Ejections and
  Flares}}.
\bjtitle{\apj}
\bvolume{559},
\bfpage{452}--\blpage{462}
(\byear{2001}).
doi:\doiurl{10.1086/322405}
\end{barticle}
\endbibitem

\end{thebibliography}

%
%



\end{document}